\documentclass[aps,preprintnumbers,preprint,nofootinbib,tightenlines,12pt]{revtex4}
\usepackage{graphicx,amsfonts,amssymb}
\usepackage[mathscr]{euscript}
\usepackage{graphics,epsfig, psfrag}
\usepackage{graphics,epsfig, psfrag}
\newcommand{\HT}{\widetilde{\mathcal H}}

\makeatletter       
\renewcommand{\thesection}{\Roman{section}}
\renewcommand{\p@subsection}{}

\renewcommand{\p@subsubsection}{}

\makeatother
\makeatletter
\@addtoreset{equation}{section}
\makeatother
\renewcommand{\theequation}{\arabic{section}.\arabic{equation}}
\begin{document}

\pagestyle{plain}

\title{\vspace{.5cm}
\Large Hilbert Space Structure in Classical Mechanics: (I)}

\author{\large E. Deotto}
\email[email: ]{deotto@mitlns.mit.edu}
\affiliation{Center for Theoretical Physics,
             Massachusetts Institute of Technology, \\
	     Cambridge, MA 02139, USA}

\author{\large E. Gozzi}
\email[email: ]{gozzi@ts.infn.it}

\author{\large D. Mauro}
\email[email: ]{mauro@ts.infn.it}
\affiliation{Dipartimento di Fisica Teorica, \\
	     Universit\`a di Trieste, 
	     Strada Costiera 11, P.O.Box 586, Trieste, Italy,
	     and INFN, Sezione di Trieste, Italy}

\date[]{7 August 2002}
\preprint{MIT-CTP-3293}

\begin{abstract}
In this paper we study the Hilbert space structure underlying the Koopman-von
Neumann (KvN) operatorial formulation
of classical mechanics. KvN limited themselves to study the Hilbert space 
of zero-forms that are the square integrable functions on phase space.
They proved that in this Hilbert space the evolution is unitary for every
system.
In this paper we extend the KvN Hilbert space to higher forms which are
basically
functions of the phase space points and the differentials on phase space. We
prove that if we 
equip this space with a positive definite scalar product the evolution can turn 
out to be non-unitary for some systems. Vice versa if we insist in having a
unitary
evolution for every system then the scalar product cannot be positive definite.
Identifying
the one-forms with the Jacobi fields we provide a physical explanation of these
phenomena. We also prove that the unitary/non unitary character of the evolution
is invariant under canonical transformations.

\end{abstract}
\maketitle

\section{Introduction}
\noindent
In the 30's Koopman and von Neumann (KvN) \cite{Koopman} gave an operatorial
formulation of {\it classical
mechanics} (CM). They first introduced square-integrable functions $\psi(\varphi^a)$ on the phase space ${\cal
M}$ of a classical system with Hamiltonian $H(\varphi)$ (with $\varphi^a$ we indicate the $2n$ phase-space
coordinates of the system $\varphi^a=q^1\ldots q^n,p^1\ldots p^n$). According to KvN the Liouville phase-space
distributions $\rho(\varphi)$ are obtained from $\psi(\varphi)$ as 
	\begin{equation}
	\rho(\varphi)=|\psi(\varphi)|^2 . \label{uno-uno}
	\end{equation}
The introduction of the $\psi(\varphi)$ 
is an acceptable assumption considering that $\rho(\varphi)$, having the meaning of a probability density,
is always positive semidefinite $\rho(\varphi)\ge 0$, and so one can always take its ``square root" and obtain
$\psi(\varphi)$. Moreover, as $\psi(\varphi)$ is  square integrable, i.e.
$\psi(\varphi)\in L^2$, it turns out that $\rho(\varphi)$ is integrable as it should be:
	\begin{equation}
	\displaystyle \int d^{2n}\varphi\,\psi^*(\varphi)\psi(\varphi)=\int d^{2n}\varphi\,\rho(\varphi) <\infty .
	\end{equation}
KvN {\it postulated} an evolution for $\psi(\varphi)$ via the Liouville operator
	\begin{equation}
	\displaystyle i\frac{\partial\psi(\varphi,t)}{\partial t}=\widehat{L}\psi(\varphi,t) \label{uno-due}
	\end{equation}
where 
	\begin{equation}
	\displaystyle \widehat{L}=i\frac{\partial H}{\partial q^{i}}\frac{\partial}{\partial p^i}-
	i\frac{\partial H}{\partial p^i}\frac{\partial}{\partial q^i}\,. \label{uno-tre}
	\end{equation}
This equation of motion for $\psi(\varphi)$ and (\ref{uno-uno}) 
lead to the same evolution  for $\rho(\varphi)$
	\begin{equation}
	\displaystyle i\frac{\partial\rho(\varphi,t)}{\partial t}=\widehat{L}\rho(\varphi,t).
	\end{equation}
This is is the well-known Liouville equation satisfied by the classical probability densities. 
Note that $\rho(\varphi)$ obeys
the same equation as  $\psi(\varphi)$
because $\widehat{L}$ is first order in the derivatives. The same does not happen in quantum mechanics where
the analog of (\ref{uno-due}) is the Schr\"odinger equation whose evolution operator is second order in the
derivatives. We will not spend more time here in explaining the interplay  between the quantum mechanical
wave functions  $\psi(q)$  and these classical ``wave functions" $\psi(\varphi)$. The interested reader  can
consult Ref. \cite{Mauro} where many details have been worked out. 

In order to have a true Hilbert structure a scalar product has to be introduced and KvN used the following one
	\begin{equation}
	\displaystyle \langle \psi|\Phi\rangle =\int d^{2n}\varphi \;\psi^*(\varphi)\Phi(\varphi). \label{uno-quattro}
	\end{equation}
We have introduced an abstract $\langle$bra$|$ and $|$ket$\rangle$ notation and we have used a particular
representation for the wave function. It is the $\varphi$-representation which is the analog of the $x$-representation 
in quantum mechanics. We could have used other representations and some of them are analyzed in Ref.
\cite{Mauro}. 

Sticking anyhow to the representation (\ref{uno-quattro}) it is easy to see that the Liouville operator
(\ref{uno-tre}) is Hermitian
	\begin{equation}
	\langle \widehat{L}\psi|\Phi\rangle =\langle\psi|\widehat{L}\Phi\rangle. \label{uno-cinque}
	\end{equation}
In fact the LHS of (\ref{uno-cinque}) is
	\begin{eqnarray}
	\displaystyle 
	&&\int d^{2n}\varphi\biggl(i\frac{\partial H}{\partial q^i}\frac{\partial \psi}{\partial p^i}-i\frac{\partial H}{\partial
	p^i}\frac{\partial \psi}{\partial q^i}\biggr)^*\Phi(\varphi)=\nonumber\\
	&& =-i\int d^{2n}\varphi\biggl(\frac{\partial H}{\partial q^i}\frac{\partial \psi^*}{\partial p^i}-
	\frac{\partial H}{\partial p^i}\frac{\partial \psi^*}{\partial q^i}\biggr)\Phi(\varphi)=\nonumber\\
	&&=i\int d^{2n}\varphi\biggl(\frac{\partial H}{\partial q^i}\frac{\partial \Phi}{\partial p^i}\psi^*-\frac{\partial
	H}{\partial p^i}\frac{\partial \Phi}{\partial q^i}\psi^*\biggr); \label{uno-sei}
	\end{eqnarray}
\noindent
where in the last step we have integrated by parts neglecting surface terms because both $\psi$ and $\Phi$ go to
zero at infinity. The RHS of (\ref{uno-cinque}) is exactly the last expression
in (\ref{uno-sei}) and this proves (\ref{uno-cinque}). The hermiticity of
$\widehat{L}$ is necessary in order to guarantee the unitarity of
the evolution operator which, in its infinitesimal form, is:
	\begin{equation}
	\displaystyle U(\Delta t)=e^{-i\widehat{L}\Delta t}.
	\end{equation}
The unitarity of the evolution, on the other hand,  is crucial in order to guarantee, via (\ref{uno-uno}), 
the conservation of
the total probability. 

In differential geometry \cite{Abraham} the operator $\widehat{L}$ is known as the Hamiltonian vector field
$(dH)^{\#}$ associated to the time evolution. It can be extended to an object ${\mathscr
L}_{(dH)^{\#}}$, known as the Lie-derivative along the Hamiltonian flow \cite{Abraham}. 
This is the operator which makes the
evolution of the higher forms:
	\begin{equation}
	\psi(\varphi,d\varphi)=\psi_0(\varphi)+\psi_a(\varphi)d\varphi^a+\psi_{ab}(\varphi)d\varphi^a\wedge
	d\varphi^b+\ldots \label{uno-otto}
	\end{equation}
The first term $\psi_0(\varphi)$ in the RHS of (\ref{uno-otto}) is the zero-form whose
evolution is given by the Liouvillian
$\widehat{L}$. We say that ${\mathscr L}_{(dH)^{\#}}$ is an extension of $\widehat{L}$ just because 
it makes the
evolution of quantities $\psi(\varphi,d\varphi)$ which are extensions of the
$\psi_0(\varphi)$. Moreover it is 
possible to prove \cite{Abraham} that
	\begin{equation}
	{\mathscr L}_{(dH)^{\#}}\bigg|_{d\varphi=0}=i\widehat{L}. \label{uno-nove}
	\end{equation}
\noindent
The differential geometry associated to classical mechanics is an old subject \cite{Abraham}
but it keeps arousing the interest of physicists \cite{little}.
The question we want to address in this paper is whether the space of higher forms (\ref{uno-otto}) can be turned
into a Hilbert space like KvN did for the zero-forms. This basically means that we want to see if it is possible
to introduce a positive definite scalar product in the space of the higher forms $\psi(\varphi,d\varphi)$. At the
same time we want to check if under this scalar product the Lie-derivative ${\mathscr L}_{(dH)^{\#}}$ is a
Hermitian operator. Surprisingly we will see that this is not possible, that means we will prove that both
conditions, positive definiteness of the scalar product and hermiticity of ${\mathscr L}_{(dH)^{\#}}$,
cannot hold at the same time. 

Instead of working in the abstract differential-geometric framework outlined above, 
we will use a more physical one derived from
a path integral formulation of CM \cite{Gozzi}. This  formulation
(from now on we will call it CPI as acronym for \underline{C}lassical
\underline{P}ath \underline{I}ntegral) is the functional counterpart of the operatorial method of KvN and
it generates some extra structures
which generalize the KvN approach. These structures are exactly the higher forms and the Lie derivative we
mentioned above. The CPI is basically defined as follows (for more details consult Ref. \cite{Gozzi}). Let
us build the following generating functional 
	\begin{equation}
	Z_{cl}[J]=\int {\mathscr D}\varphi \;\widetilde{\delta}[\varphi-\varphi_{cl}]\;\exp\left[\int J\varphi\right]
	\label{uno-dieci}
	\end{equation}
where $\varphi_{cl}$ are the classical solutions of the equations of motion \cite{Abraham}
	\begin{equation}
	\displaystyle 
	\dot{\varphi}^a=\omega^{ab}\frac{\partial H}{\partial \varphi^b}
	\end{equation}
with $\omega^{ab}$ the symplectic matrix. In (\ref{uno-dieci}) we gave 
weight one to the classical trajectories and weight zero to all the others
differently than what is done in the quantum mechanical path integral. As $\varphi^a_{cl}$ in
(\ref{uno-dieci}) are the zeroes of the following function: $\displaystyle 
\biggl(\dot{\varphi}^a-\omega^{ab}\frac{\partial H}{\partial
\varphi^b}\biggr)$, we can rewrite (\ref{uno-dieci}) as 
	\begin{equation}
	\displaystyle Z_{cl}[J]=\int {\mathscr D}\varphi \;\widetilde{\delta}\biggl[\dot{\varphi}^a-\omega^{ab}\frac{\partial H}
	{\partial\varphi^b}\biggr]\bigg|
	\delta_b^a\partial_t-\omega^{ad}\frac{\partial^2H}{\partial\varphi^d\partial\varphi^b}\bigg|\;\exp\;\int J\varphi
	\label{uno-dodici}
	\end{equation}
and using $6n$ auxiliary variables $(\lambda_a,c^a,\bar{c}_a)$ (where 
$c^a,\bar{c}_a$ have Grassmannian character) we can rewrite (\ref{uno-dodici}) in the following form
	\begin{equation}
	\displaystyle Z_{cl}[J]=\int {\mathscr D}\varphi^a{\mathscr D}\lambda_a{\mathscr D}c^a{\mathscr D}\bar{c}_a\;
	e^{i\int dt\widetilde{\cal L} +\int dt J\varphi } \label{uno-dodici-b}
	\end{equation}
where
	\begin{equation}
	\widetilde{\cal
	L}=\lambda_a\biggl[\dot{\varphi}^a-\omega^{ab}\frac{\partial H}{\partial\varphi^b}\biggr]+i\bar{c}_a\biggl[
	\dot{c}^a-\omega^{ad}\frac{\partial^2H}{\partial\varphi^d\partial\varphi^b}\biggr]c^b .\label{uno-tredici}
	\end{equation}
The Hamiltonian associated to $\widetilde{\cal L}$ is:
	\begin{equation}
	\displaystyle 
	\widetilde{\cal H}=\lambda_a\omega^{ab}\frac{\partial
	H}{\partial\varphi^b}+i\bar{c}_a\omega^{ad}\frac{\partial^2H}{\partial\varphi^d\partial\varphi^b}c^b .
	\label{uno-quattordici}
	\end{equation}
The contact with differential geometry was first established in Ref.
\cite{Gozzi} and further developed in
\cite{geometry}. In those references it was shown that the Grassmannian variables $c^a$ can be identified with
the basis $d\varphi^a$ of the space of forms and their associated wedge product is naturally taken into account
by the Grassmannian character of the variables $c^a$. The functions $\psi(\varphi,c)$ can
then be put into
correspondence with the inhomogeneous forms 
	\begin{equation}
	\psi(\varphi,c)=\psi_{0}(\varphi)+\psi_a(\varphi)c^a+\psi_{ab}(\varphi)c^ac^b+\ldots\;\Rightarrow\;
	\psi_{0}(\varphi)+\psi_a(\varphi)d\varphi^a+\psi_{ab}(\varphi)d\varphi^a\wedge d\varphi^b+\ldots
	\label{correspondence}
	\end{equation}
From the path integral (\ref{uno-dodici-b}) one can derive an operatorial formalism and a commutator
structure \cite{footnote1}, which is
given by
	\begin{equation}
	[\varphi^a,\lambda_b]_-=i\delta_b^a\;\;\;\;\;\;\;\;[c^a,\bar{c}_b]_+=\delta^a_b . \label{uno-quattordici-b}
	\end{equation}
To implement (\ref{uno-quattordici-b}) we can then realize  $\lambda_a$ and $\bar{c}_a$ as
	\begin{equation}
	\displaystyle \lambda_a=-i\frac{\partial}{\partial\varphi^a}\;\;\;\;\;\;\; 
	\bar{c}_a=\frac{\partial}{\partial c^a}. \label{uno-quindici}
	\end{equation}
In this representation  the ``wave functions" of the  theory are functions which
depend only on $\varphi$ and $c$, i.e. $\psi(\varphi,c)$~and these are precisely the inhomogeneous forms. 
The relations (\ref{uno-quindici}) turn the $\widetilde{\cal H}$ of Eq. (\ref{uno-quattordici})
into the following operator:
	\begin{equation}
	\displaystyle \widetilde{\cal H}=-i\omega^{ab}\frac{\partial H}{\partial \varphi^b}
	\frac{\partial}{\partial \varphi^a}-i\omega^{ad}\frac{\partial^2 H}{\partial\varphi^d\partial\varphi^b}
	c^b\frac{\partial}{\partial c^a} . \label{uno-sedici}
	\end{equation}
From now on we will use the same notation for the Hamiltonian (\ref{uno-quattordici}) and the associated operator
(\ref{uno-sedici}). 
The first term on the RHS of (\ref{uno-sedici}) is nothing else than the Liouville operator $\widehat{L}$
which acts only on the zero-forms $\psi_0(\varphi)$, while the second term acts also on higher forms
$\psi(\varphi,c)$. The combined action of these two terms identifies $\widetilde{\cal H}$ 
with the Lie-derivative along the Hamiltonian flow \cite{Gozzi}. Also all the other standard
differential geometric operations like exterior derivative, interior contraction with vector fields, Lie
brackets etc. can be rephrased in the language of the CPI. All the details can be found in Ref. \cite{Gozzi} and
\cite{geometry}. 

As we said previously, in this paper we would like to find out if the space of the higher forms, which now can be
identified with the functions $\psi(\varphi, c)$, can be endowed with a positive definite scalar product as KvN did for
the zero-forms and if the operator $\widetilde{\cal H}$ is Hermitian under this scalar product. 
The reader may wonder which are the {\it physical} reasons which motivate us to study this enlarged Hilbert space. The 
reasons are the following. It was well-known \cite{avez} that such features like the ergodicity of a dynamical 
system are indicated by the {\it spectral} properties of the Liouvillian. In particular a system is ergodic 
if the eigenstate associated to the zero eigenvalue of the Liouvillian is non degenerate. Later on 
it was discovered \cite{vari} that the {\it spectral} properties of $\widetilde{\cal H}$ (or of its analog for stochastic 
systems), which is the object that generalizes the Liouvillian, gives us further physical information
such as the Lyapunov exponents or the dynamical and topological entropies of the system. So the {\it spectrum}
of $\widetilde{\cal H}$ seems to be the central object encapsulating the most important physical features
of a dynamical system. To get the spectrum it is necessary first to study the Hilbert space 
in which $\widetilde{\cal H}$ is defined and to find out if it is a Hermitian operator or not. The previous studies 
\cite{vari} did not rigorously explore these mathematical features but derived the properties mentioned above 
in a rather formal way by functional techniques. 

The first scalar product we shall explore in {\bf Sec. II} is the one proposed by Salomonson and van Holten in Ref.
\cite{Salomonson} for a Hamiltonian which is {\it similar} to $\widetilde{\cal H}$ but describes 
supersymmetric quantum mechanics (susy QM) \cite{Witten}. The similarity with our Hamiltonian is in the fact
that both have ``bosonic" degrees
of freedom ($\varphi^a$) and Grassmannian ones ($c^a$), which are turned into one another by a kind of
supersymmetry invariance \cite{Witten}. The differences instead are mainly in
the fact that while the Hamiltonian of
susy QM contains second order derivatives in the kinetic term, like a good quantum Schr\"odinger-like operator,
ours contains only first order derivatives like a good classical
Liouville-like operator. In {\bf Sec. II} we
will show that also in the  space of the higher forms the scalar product introduced by Salomonson and
van Holten (SvH) is a positive definite one but we shall also show that our $\widetilde{\cal H}$ is not
Hermitian under this scalar product differently than what happens in susy QM. 

In {\bf Sec. III} we will turn to another scalar product which is the one used mostly in gauge theory and summarized
in Ref. \cite{Henneaux}. Under this scalar product we shall show that our $\widetilde{\cal H}$ is Hermitian but
not all states have positive norm. We shall anyhow prove that under both
scalar products (SvH  and gauge)
the operatorial formalism leads to the same path integral formulation given in (\ref{uno-dodici-b}).

In {\bf Sec.  IV} we will introduce a third type of scalar product that we call the symplectic one for the relevant
role played in it by the symplectic matrix. M. Reuter had already 
introduced the same scalar product \cite{Reuter1} in a different context. 
Unfortunately also the symplectic scalar product
 is not positive-definite but, like the gauge-theory-one of {\bf Sec. III}, the Hamiltonian turns out to be
 Hermitian under it. These two scalar products (the
symplectic and the gauge one) are actually very similar. This similarity is 
better revealed by performing a
transformation on the Grassmannian variables which turns them into a sort of
holomorphic and anti-holomorphic
variables. In {\bf Sec.  V} we will search for the subspace (of the full Hilbert
space) made of
positive norm states under either the symplectic or the gauge scalar product. We
shall also explore the issue of which
is the subspace where $\widetilde{\cal H}$ is Hermitian under the SvH scalar product. Our
conclusion will be that, in all three cases,  it is the space made of the zero
forms and of a set of higher ones
isomorphic to the zero-forms. So we conclude that, both in the case of the SvH scalar product and of the
symplectic (gauge) one, the ``physical states"  are basically the zero-forms or other forms {\it ``isomorphic"} to
them. By ``{\it physical}" we mean those which have both a positive norm and on which $\widetilde{\cal H}$ is
Hermitian, and by ``{\it isomorphic}" we mean those which transform in the same manner as the zero-forms under the
Lie-derivative along the Hamiltonian flow.

The reader may suspect at this point that these three scalar products are not the only possible ones. This is
correct and in {\bf Sec.  VI} we will generalize them introducing a sort of
metric. In this same section we shall prove that
even these most general scalar products have the same problems as the previous three above. That means either the
scalar product is positive definite but $\widetilde{\cal H}$ is not Hermitian or  
$\widetilde{\cal H}$ is Hermitian but the scalar product is not positive definite. This feature seems
to be a rather general one and it can be ascribed to the presence of the symplectic matrix and to the Grassmannian
nature of the variables $c^a$. In the conclusions ({\bf Sec.  VII}) we will stress that the kind of behavior that
we have analyzed above (positive definite scalar product but Hamiltonian non-Hermitian or vice versa) can disappear for
particular systems like for example the harmonic oscillator. Instead for systems in which
this behavior does not disappear (like for
example chaotic Hamiltonians) we shall give  a tentative explanation of why this happens. The explanation is based on the
fact that for chaotic systems the Jacobi fields fly away exponentially but at the same time the Liouville theorem
must hold. The paper ends with few appendices where we
confine several mathematical details of this work. We have given all these
mathematical details because we think they might help the reader to understand the results contained in this paper which
could look quite unusual at first sight.

\section{The SvH Scalar Product}

\noindent The set of basic operators of our theory is
	\begin{equation}
	(\hat{q},\hat{\lambda}_q, \hat{c}^q,\hat{\bar{c}}_q)\;\;\; ; \;\;\;(\hat{p}, \hat{\lambda}_p,\hat{c}^p,\hat{\bar{c}}_p).
	\label{due-zero}
	\end{equation}
The operators in the first set commute with those in the second one
according to the graded commutators 
(\ref{uno-quattordici-b}). We will concentrate now  on the first set. 
The strategy \cite{Salomonson} we shall 
follow is {\it to choose} some hermiticity conditions for these operators and
the normalization of one of the basic
states. These conditions will be sufficient to build a complete basis and
the scalar product which will exactly reproduce the hermiticity
conditions we have chosen originally. For the bosonic variables
($\hat{\varphi}^a,\hat{\lambda}_a$) we choose, once and for
all, the following hermiticity conditions:
	\begin{equation}
	\label{due-uno}
	\left\{
		\begin{array}{l}
		\displaystyle
		\hat{\varphi}^{a\dagger}=\hat{\varphi}^a\smallskip\\
		\displaystyle \hat{\lambda}_a^{\dagger}=\hat{\lambda}_a .\\
		\end{array}
		\right. 
	\end{equation}
These conditions are consistent with the first of the commutators  (\ref{uno-quattordici-b}). For the Grassmannian variables, due to the
anticommutator present in (\ref{uno-quattordici-b}), we could choose various hermiticity properties. The one we choose here
is, for the variables $\hat{c}^q$ and $\hat{\bar{c}}_q$, the following one:
	\begin{equation}
	\label{due-due}
	\left\{
		\begin{array}{l}
		\displaystyle
		\hat{c}^{q\dagger}=\hat{\bar{c}}_q\smallskip\\
		\displaystyle \hat{\bar{c}}_q^{\dagger}=\hat{c}^q.\\
		\end{array}
		\right.
	\end{equation}
This is the analog \cite{footnote2}
of the one used by SvH \cite{Salomonson}. Next let us  start building a basis of our Hilbert space.
First we define the state $|0-\rangle$ as 
	\begin{equation}
	\hat{c}^q|0-\rangle=0 . \label{due-tre}
	\end{equation}
The reader may wonder whether there exists such a state. The answer is yes:   let us
in fact take a generic state $|s\rangle$ such that $\hat{c}^q|s\rangle\neq 0$, then the following state
$|0-\rangle \equiv \hat{c}^q|s\rangle$ has precisely the property (\ref{due-tre}):
	\begin{equation}
	\hat{c}^q|0-\rangle=\hat{c}^q\hat{c}^q|s\rangle=0.
	\end{equation}
Like for the coherent states, from $|0 -\rangle$ it is easy to build the complete set of eigenvectors of $\hat{c}^q$
which are:
	\begin{equation}
	\displaystyle |\alpha -\rangle\equiv e^{-\alpha\hat{\bar{c}}_q}|0-\rangle=|0-\rangle -\alpha|0+\rangle. \label{due-quattro}
	\end{equation}
Using the anticommutator $[\hat{c}^q,\hat{\bar{c}}_q]=1$, it is in fact easy to show that  
	\begin{equation}
	\hat{c}^q|\alpha -\rangle=\alpha|\alpha -\rangle
	\end{equation}
where $\alpha$, due to the ``Grassmannian" nature of $\hat{c}^q$, is a 
Grassmannian odd parameter. The state 
$|0+\rangle$ which
appears in (\ref{due-quattro}) is defined as: 
	\begin{equation}
	|0+\rangle\equiv\hat{\bar{c}}_q|0 -\rangle. \label{due-cinque}
	\end{equation}  
Note that $|0+\rangle$ cannot be zero, otherwise we would end up in a
contradiction; in fact, if $|0 +\rangle$ were zero, we
would get $\hat{c}^q|0+\rangle=0$, and this cannot be true
because  we know that $\hat{c}^q|0+\rangle$ can also be written  as
$\hat{c}^q\hat{\bar{c}}_q|0-\rangle=[\hat{c}^q,\hat{\bar{c}}_q]|0-\rangle=|0-\rangle$
and this contradicts the previous relation. Via $|0+\rangle$ it is easy to build the
eigenstates of $\hat{\bar{c}}_q$. They are:
	\begin{equation}
	|\beta +\rangle\equiv e^{-\beta \hat{c}^q}|0+\rangle=|0+\rangle -\beta|0-\rangle \label{due-sei}
	\end{equation}
and using the commutation relation (\ref{uno-quattordici-b}) it is easy to prove
that  they really are eigenvectors of $\hat{\bar{c}}_q$
	\begin{equation}
	\hat{\bar{c}}_q|\beta +\rangle=\beta|\beta +\rangle. \label{due-sette}
	\end{equation}
Among this plethora of states that we built we have to choose a basis of our Hilbert \cite{footnote3} space. 
To do that we should diagonalize a Hermitian operator. 
Thanks to the hermiticity
conditions (\ref{due-due}) that we have used, it is easy to see that the operator 
$\hat{N}_q\equiv\hat{c}^q\hat{\bar{c}}_q$ is Hermitian. Moreover it is a 
projection operator $\hat{N}^2_q=\hat{N}_q$ which implies that
$\hat{N}_q$ has only 1 and 0 as eigenvalues. The corresponding eigenstates are
$|0-\rangle$ and $|0+\rangle$: 
	\begin{equation}
	\label{due-otto}
	\left\{
		\begin{array}{l}
		\displaystyle
		\hat{N}_q|0-\rangle=\hat{c}^q\hat{\bar{c}}_q|0-\rangle=[\hat{c}^q,\hat{\bar{c}}_q]|0-\rangle=|0-\rangle\smallskip\\
		\displaystyle \hat{N}_q|0+\rangle=\hat{c}^q\hat{\bar{c}}_q|0+\rangle=0.\\
		\end{array}
		\right.
	\end{equation}
So $|0-\rangle$ and $|0+\rangle$ make a basis in our space. The careful reader may object that the eigenstates
$\bigl\{|0+\rangle,|0-\rangle\bigr\}$ of a Hermitian operator are a basis of the Hilbert space only if the scalar product is a positive
definite one. We shall see this to be the case for the SvH scalar product (see Eq. (\ref{due-venticinque})) and so we can
proceed. In principle, anyhow, we should not make any statement at this point on the states
$\bigl\{|0+\rangle,|0-\rangle\bigr\}$. The
vector space which we can build out of $|0+\rangle$ and $|0-\rangle$ is made of any linear combination of
$|0+\rangle$ and $|0-\rangle$ made with either complex or Grassmannian coefficients $\delta$ and $\gamma$:
	\begin{equation}
	|\psi\rangle=\delta|0+\rangle+\gamma|0-\rangle .\label{due-otto-b}
	\end{equation}
Of this vector space the states $|\alpha -\rangle$ and $|\beta +\rangle$ of (\ref{due-quattro}) and (\ref{due-sei}) make
only a subset analogous to the one of coherent states. This subset will be sufficient to give a decomposition of the
identity as indicated in (\ref{due-quindici}). Moreover any state like (\ref{due-otto-b}) can be built as linear combination of
states belonging to the subset $\bigl\{|\beta+\rangle,|\alpha -\rangle\bigr\}$ because $|0+\rangle$ and $|0-\rangle$ belong to this
subset. 

Up to now we have not given any scalar product but only some hermiticity conditions like (\ref{due-uno}) and
(\ref{due-due}). To obtain the  scalar product, which would reproduce those hermiticity rules, the only extra condition we
have to give is the normalization of $|0 -\rangle$, which we {\it choose} to be
	\begin{equation}
	\langle-0|0-\rangle=1 .\label{due-nove}
	\end{equation}
As a consequence of this we get 
	\begin{equation}
	\langle+0|0+\rangle=1 \label{due-dieci}
	\end{equation}
and 
	\begin{equation}
	\langle+0|0-\rangle=0 .\label{due-undici}
	\end{equation}
To prove (\ref{due-dieci}) and (\ref{due-undici}) we use the following von Neumann notation for the scalar product
	\begin{equation}
	\biggl(|\alpha \pm\rangle, |\beta\pm\rangle\biggr)\equiv \langle\pm\alpha|\beta\pm\rangle .\label{due-dodici}
	\end{equation}
Relation (\ref{due-dieci}) is obtained in the following manner
	\begin{eqnarray}
	&&\biggl(|0+\rangle,|0+\rangle\biggr)=\biggl(\hat{\bar{c}}_q|0-\rangle,\hat{\bar{c}}_q|0-\rangle\biggr)=
	\biggl(|0-\rangle,\hat{c}^q\hat{\bar{c}}_q|0-\rangle\biggr)
	\nonumber=\\
	&&=\biggl(|0-\rangle,[\hat{c}^q,\hat{\bar{c}}_q]_+|0-\rangle\biggr)=\biggl(|0-\rangle,|0-\rangle\biggr)=1
	\end{eqnarray}
while (\ref{due-undici}) can be proven in this way:
	\begin{equation}
	\biggl(|0+\rangle,|0-\rangle\biggr)=\biggl(\hat{\bar{c}}_q|0-\rangle,|0-\rangle\biggr)=
	\biggl(|0-\rangle, \hat{c}^q|0-\rangle\biggr)=\biggl(|0-\rangle,0\biggr)=0.
	\end{equation}
So $\bigl\{|0+\rangle$, $|0-\rangle\bigr\}$ are an orthonormal basis. As $\hat{c}^q$ and $\hat{\bar{c}}_q$ are not Hermitian operators
(see Eq. (\ref{due-due})), we cannot say that their eigenstates $|\alpha-\rangle$  
and $|\beta +\rangle$, given in (\ref{due-quattro}) and (\ref{due-sei}), make up an
orthonormal basis for our Hilbert space. Nevertheless it is interesting to study the relations among these states. For
example it is easy to prove \cite{Appendix A} that
	\begin{equation}
	\left\{
		\begin{array}{l}
		\displaystyle
		|\beta +\rangle=-\int d\alpha \,e^{\alpha\beta}|\alpha-\rangle\smallskip \label{due-tredici-a}\\
		\displaystyle |\gamma -\rangle=-\int d\beta \,e^{\beta\gamma}|\beta+\rangle \label{due-tredici-b}
		\end{array}
		\right.
	\end{equation}
and that the scalar product among them  is:
	\begin{equation}
	\left\{
		\begin{array}{l}
		\displaystyle
		\biggl(|\alpha \pm\rangle,|\beta\pm\rangle\biggr)=e^{\alpha^*\beta}\smallskip \label{due-quattordici-a}\\
		\displaystyle \biggl(|\alpha\pm\rangle, |\beta\mp\rangle\biggr)=\pm\delta(\alpha^*-\beta). \label{due-quattordici-b}
		\end{array}
		\right.
	\end{equation}
With $|\alpha -\rangle$ and $|\beta +\rangle$ it is possible to build a sort of
``twisted" completeness relation which is 
	\begin{equation}
	-\int d\alpha|\alpha\pm\rangle\langle\mp \alpha^*|={\bf 1} .\label{due-quindici}
	\end{equation}
This is a sort of {\it resolution of the identity} analogous to the one given
by the coherent states. We call it ``twisted"
completeness relation because the bras and kets have the ``$+$" and ``$-$" signs interchanged. 

Since $|0+\rangle$ and $|0-\rangle$ form a basis of our Hilbert space, the completeness relation 
can also be written as:
	\begin{equation}
	|0+\rangle\langle+0|+|0-\rangle\langle-0|={\bf 1}. \label{due-sedici}
	\end{equation}
It is easy to prove that (\ref{due-quindici}) and  (\ref{due-sedici}) are
equivalent. In fact let us remember the relations:
	\begin{equation}
	\left\{
		\begin{array}{l}
		\displaystyle \label{due-diciassette}
		|\alpha -\rangle=|0-\rangle-\alpha|0+\rangle\smallskip \\
		\displaystyle |\beta +\rangle=|0+\rangle-\beta|0-\rangle 
		\end{array}
		\right.
	\end{equation}
	\begin{equation}
	\left\{
		\begin{array}{l}
		\displaystyle \label{due-diciotto}
		\langle -\alpha|=\langle  -0|-\alpha^*\langle+0| \smallskip \\
		\displaystyle \langle +\beta|=\langle  +0|+\beta^*\langle-0|
		\end{array}
		\right.
	\end{equation}
and let us insert them into the LHS of (\ref{due-quindici}). What we get is
	\begin{eqnarray}
	&& {\bf 1}=-\int d\alpha |\alpha +\rangle\langle -\alpha^*|=-\int d\alpha\biggl\{\biggl(|0+\rangle-\alpha|0-\rangle\biggr)
	\biggl(\langle -0|-\alpha\langle+0|\biggr)\biggr\}=\nonumber\\
	&&=-\int d\alpha\biggl\{|0+\rangle\langle-0|-\alpha |0-\rangle\langle-0|-\alpha |0+\rangle\langle+0|\biggr\}=
	|0-\rangle\langle -0|+|0+\rangle\langle +0| \nonumber\\
	\end{eqnarray}
which is exactly (\ref{due-sedici}). The careful reader may object to this proof  because, as we
said previously, we cannot state that $\bigl\{|0+\rangle,|0-\rangle\bigr\}$ are a basis before proving that the scalar
product is positive definite. However, as we already claimed, this is precisely the case as shown in 
Eq. (\ref{due-venticinque}).  

For our completeness relations, or better resolutions of the identity, we can either use (\ref{due-sedici}) or
(\ref{due-quindici}) and it will be this last one we will use most. For example we could use it to express a generic state
$|\psi\rangle$ in the following manner:
	\begin{eqnarray}
	|\psi\rangle&=&-\int d\alpha |\alpha-\rangle\langle+\alpha^*|\psi\rangle=-\int
	d\alpha|\alpha-\rangle(\psi_0+\psi_1\alpha)=\nonumber\\
	&=&-\int d\alpha\biggl(|0-\rangle-\alpha|0+\rangle\biggr)(\psi_0+\alpha\psi_1)=\psi_0|0+\rangle+\psi_1|0-\rangle.
	\label{aggiunta-1}
	\end{eqnarray}
Similarly to what happens in ordinary quantum mechanics, where $\langle
q|\psi\rangle=\psi(q)$ is a function of $q$, the projection of 
the state $|\psi\rangle$ onto the basis of the eigenstates $\langle+\alpha^*|$
is given by a function of 
the Grassmannian odd number $\alpha$:
	\begin{equation}
	\langle+\alpha^*|\psi\rangle=\psi(\alpha)=\psi_0+\psi_1\alpha .\label{aggiunta-2}
	\end{equation}
As we have to reproduce, via Eq. (\ref{aggiunta-2}), the forms (\ref{correspondence}) whose coefficients $\psi_0(\varphi)$,
$\psi_a(\varphi),\ldots$ are complex functions we will make the choice that $\psi_0$ and $\psi_1$ in (\ref{aggiunta-2}) are
complex numbers. Note anyhow that the most generic states $|\alpha\pm\rangle$ which enter the resolution of the identity
are of the form (\ref{due-otto-b}) whose coefficients $\delta$ and $\gamma$ could be both complex and Grassmannian. 

Making use of Eq. (\ref{aggiunta-1}) and of the scalar products
(\ref{due-nove})-(\ref{due-undici}) among the states $\bigl\{|0+\rangle, |0-\rangle\bigr\}$,
 it is very easy to derive the expression of the scalar product between two generic
states $|\psi\rangle, \;|\Phi\rangle$ of the theory:
	\begin{eqnarray}
	&&\langle\Phi|\psi\rangle=\biggl(|\Phi\rangle,|\psi\rangle\biggr)=\biggl(\Phi_0|0+\rangle+\Phi_1|0-\rangle,\psi_0|0+\rangle+\psi_1|0-
	\rangle\biggr)=\nonumber\\
	&&=\biggl(|0+\rangle,\Phi_0^*\psi_0|0+\rangle\biggr)+\biggl(|0-\rangle,\Phi^*_1\psi_0|0+\rangle\biggr)+\biggl(|0+\rangle,
	\Phi_0^*\psi_1|0-\rangle\biggr)+\biggl(|0-\rangle,\Phi_1^*\psi_1|0-\rangle\biggr)=\nonumber\\
	&&=\Phi_0^*\psi_0+\Phi_1^*\psi_1 .\label{scprod}
	\end{eqnarray}
We can also express the scalar product (\ref{scprod}) in terms of integrations
over Grassmannian variables:
	\begin{equation}
	\displaystyle \langle\Phi|\psi\rangle=\int d\eta d\alpha^* e^{\alpha^*\eta} \Phi^*(\alpha)\psi(\eta).
	\end{equation}
From the previous relation it is also easy to obtain the expression of the norm of a generic state $|\psi\rangle$:
	\begin{equation}
	\langle\psi|\psi\rangle=|\psi_0|^2+|\psi_1|^2\ge 0 .\label{due-venticinque}
	\end{equation}

Up to now, in all expressions, we have explicitly indicated the dependence only on the Grassmannian
variable $c^q$, but we know that the basic operators (\ref{due-zero}) were many more. Considering for example those
contained in the first set of (\ref{due-zero}), i.e. 
$(\hat{q},\hat{\lambda}_q,\hat{c}^q,\hat{\bar{c}}_q)$, we see that our basic states should have
included also a dependence on $q$ (if we choose the $q$ representation)
	\begin{eqnarray}
	&& |q,\alpha +\rangle\equiv |q\rangle\otimes |\alpha +\rangle\nonumber\\
	&& |q,\beta -\rangle\equiv |q\rangle\otimes|\beta -\rangle .
	\end{eqnarray}
The variables $\lambda_q$ and $\bar{c}_q$ are the momenta conjugated to $q$ and $c^q$ 
and so they would appear in the wave function only in the momentum
representation that for the moment we do not consider.
Also the completeness relations (\ref{due-quindici}) should include $q$ and be
	\begin{equation}
	-\int d\alpha dq\,|q,\alpha\pm\rangle\langle\mp\alpha^*,q|={\bf 1} .\label{due-ventisette}
	\end{equation}
For the same reason, the expansion (\ref{aggiunta-1}) should be
	\begin{eqnarray}
	|\psi\rangle&=&-\int d\alpha dq\biggl(|q, 0-\rangle-\alpha|q, 0+\rangle\biggr)
	\biggl(\psi_0(q)+\alpha\psi_1(q)\biggr)=\nonumber\\
	&=&\int dq\biggl[\psi_0(q)|q, 0+\rangle+ \psi_1(q)|q, 0-\rangle\biggr]
	\end{eqnarray}
and as a consequence the scalar product (\ref{due-venticinque}) turns out to be
	\begin{equation}
	\langle\psi|\psi\rangle=\int dq\biggl[|\psi_0(q)|^2+|\psi_1(q)|^2\biggr]\ge 0. \label{due-ventinove}
	\end{equation}
Next we should start including the variables of the second set in (\ref{due-zero}), i.e. $(\hat{p},
\hat{\lambda}_p,\hat{c}^p,\hat{\bar{c}}_p)$.
As it is a set which commutes with the first one things are not difficult.
Regarding the bosonic variables, the ``wave
functions" depend not only on $q$ but also on $p$: i.e. $\psi(q,p)$. Regarding instead the Grassmannian
variables, we should build the analog of the state (\ref{due-tre}) also for
$\hat{c}^p$ on which we impose the hermiticity conditions:
	\begin{equation}
	\label{hermiticityp}
	\left\{
		\begin{array}{l}
		\displaystyle
		\hat{c}^{p\dagger}=\hat{\bar{c}}_p\smallskip\\
		\displaystyle \hat{\bar{c}}_p^{\dagger}=\hat{c}^p.\\
		\end{array}
		\right.
	\end{equation}
We define a new state
$|0-,0-\rangle$ as
	\begin{equation}
	\left\{
		\begin{array}{l}
		\displaystyle 
		\hat{c}^q|0-,0-\rangle=0 \smallskip \label{due-trenta-a}\\
		\displaystyle \hat{c}^p|0-,0-\rangle=0 \label{due-trenta-b}
		\end{array}
		\right.
	\end{equation}
where the first ``$0-$" in $|0-,0-\rangle$ is associated to $c^q$ and the second one to $c^p$. Analogously to
(\ref{due-cinque}) we can define also the following other states 
	\begin{equation}
	\left\{
		\begin{array}{l}
		\displaystyle 
		|0+,0-\rangle=\hat{\bar{c}}_q|0-,0-\rangle \smallskip \label{due-trentuno-a}\\
		\displaystyle |0-,0+\rangle=-\hat{\bar{c}}_p|0-,0-\rangle. \label{due-trentuno-b}
		\end{array}
		\right.
	\end{equation}
The reason for the ``$-$" sign in front of the $\hat{\bar{c}}_p$ on the RHS of
(\ref{due-trentuno-b}) is because of the Grassmannian odd
nature of the first ``$0-$" in the state $|0-,0-\rangle$. These signs are the only things we should be careful about.
Analogously to (\ref{due-quattro}) it is also natural to construct the eigenstates of $\hat{c}^q$ and $\hat{c}^p$ which are 
	\begin{equation}
	\displaystyle |\alpha_q-,\alpha_p-\rangle\equiv e^{-\alpha_q\hat{\bar{c}}_q-\alpha_p\hat{\bar{c}}_p}|
	0-,0-\rangle .\label{due-trentadue}
	\end{equation}
If instead of one degree of freedom we have $n$ of them $(q^1\ldots q^n)$,  we will have $n$ $p$-variables
$(p^1\ldots p^n)$ and $2n$ Grassmannian ones $(c^{q_1}\ldots c^{q_n},c^{p_1}\ldots c^{p_n})$. The analog of
(\ref{due-trenta-a}) turns out to be
	\begin{equation}
	\label{due-trentatre}
	\left\{
		\begin{array}{l}
		\displaystyle 
		\hat{c}^{q_i}|\underbrace{0-,0-,\ldots, 0-}_q;\underbrace{0-,0-,\ldots, 0-}_p
		\rangle=0 \qquad\qquad \forall i=1\ldots 2n\smallskip\\
		\displaystyle \hat{c}^{p_i}|\underbrace{0-,0-,\ldots, 0-}_q;\underbrace{0-,0-,\ldots, 0-}_p
		\rangle=0 \qquad\qquad \forall i=1\ldots 2n.
		\end{array}
		\right.
	\end{equation}
Via the $\hat{\bar{c}}$ variables (either $\hat{\bar{c}}_q$ or $\hat{\bar{c}}_p$) 
we can then build the states analogous to those in (\ref{due-trentuno-a}) which are 
	\begin{equation}
	|0-,0-,\ldots,\underbrace{0+}_{i-th \; position},
	\ldots, 0-,\ldots, 0-\rangle=(-1)^{i-1}\hat{\bar{c}}_i|0-,0-,\ldots ,0-\rangle \label{due-trentaquattro}
	\end{equation}
where the exponent ``$i-1$" indicates how many ``$0-$" are present before the ``$0+$" 
on the LHS of (\ref{due-trentaquattro}). The variables $\hat{\bar{c}}_i$ can be
either a $\hat{\bar{c}}_q$ (if $i\leq n$) or a $\hat{\bar{c}}_p$ (if $i> n$). 
Generalizing (\ref{due-trentaquattro}) to the
case with two ``$0+$" in the ket on the LHS, one in the $i$-th position and another
in the $j$-th position, we get
	\begin{equation}
	|0-,\ldots, 0+,\ldots, 0+,\ldots ,0-,0-\rangle=(-1)^{i+j}\hat{\bar{c}}_i\hat{\bar{c}}_j|0-, 0-, \ldots, 0-\rangle.
	\end{equation}
When all entries on the LHS are ``$0+$"  the formula is 
	\begin{equation}
	|0+,\ldots, 0+,\ldots, 0+,\ldots ,0+,0+\rangle=(-1)^{n}\hat{\bar{c}}_1\hat{\bar{c}}_2\ldots \hat{\bar{c}}_{2n}|0-, 0-, \ldots,
	0-\rangle .
	\end{equation}
The eigenstates of the various $\hat{c}_i$ are built in a manner analogous to (\ref{due-trentadue})
	\begin{equation}
	\displaystyle |\alpha_1-,\ldots,\alpha_{2n} -\rangle=\exp\biggl[-\sum_{i=1}^{2n}\alpha_i
	\hat{\bar{c}}_i\biggr]|0-,\ldots, 0-\rangle
	\label{due-trentacinque}
	\end{equation}
and the eigenstates of $\hat{\bar{c}}_i$ as
	\begin{equation}
	\displaystyle |\beta_1+,\ldots,\beta_{2n} +\rangle=\exp\biggl[-\sum_{i=1}^{2n}\beta_i\hat{c}^i\biggr]|0+,\ldots, 0+\rangle.
	\label{due-trentasei}
	\end{equation}
Like we did in the case of one Grassmannian variable, imposing the normalization condition
	\begin{equation}
	\langle -0,-0,\ldots, -0|0-,0-,\ldots, 0-\rangle=1
	\end{equation}
and using the hermiticity conditions we get the scalar products among the states. 
So for example we get \cite{Appendix A}
	\begin{equation}
	\biggl(|0+, 0-,\ldots, 0-\rangle, |0+, 0-,\ldots, 0-\rangle\biggr)=1 \label{due-trentasette-a}
	\end{equation}
	\begin{equation}
	\biggl(|0-, 0+,\ldots, 0-\rangle, |0-, 0+,\ldots, 0-\rangle\biggr)=1 \label{due-trentasette-b}
	\end{equation}
	\begin{equation}
	\biggl(|0+, 0-,\ldots, 0-\rangle, |0-, 0+,\ldots, 0-\rangle\biggr)=0 \label{due-trentasette-c}
	\end{equation}
and similarly for any other combination of entries of ``$0+,0-$". We can also easily
\cite{Appendix A} derive the analog of Eq. (\ref{due-quattordici-a})
	\begin{equation}
	\biggl(|\alpha_q -,\alpha_p-\rangle,|\beta_q-,\beta_p-\rangle\biggr)=\exp[\alpha_q^*\beta_q+\alpha_p^*\beta_p]
	\label{due-trentotto-a}
	\end{equation}
and in general, for $n$ degrees of freedom, we have
	\begin{eqnarray}
	\displaystyle &&\biggl(|\alpha_{q_1}-,\alpha_{q_2}-,\ldots\alpha_{q_n}-,\alpha_{p_1}-,
	\ldots \alpha_{p_n}-\rangle,
	|\beta_{q_1}-,\beta_{q_2}-,\ldots \beta_{q_n}-,\beta_{p_1}-,\ldots \beta_{p_n}-\rangle\biggr)=\nonumber\\
	&&\qquad\quad =\exp\biggl[\sum_i\alpha_i^*\beta_i\biggr]. \label{due-trentotto-b}
	\end{eqnarray}
Also easy is the derivation of the following other relations
	\begin{equation}
	\label{due-trentotto-c}
	\left\{
		\begin{array}{l}
		\displaystyle 
		\biggl(|\alpha_q^*+,\alpha_p^*+\rangle,|\beta_q-,\beta_p-\rangle\biggr)=\delta(\alpha_q-\beta_q)\delta(\alpha_p-\beta_p) 
		\smallskip\\
		\displaystyle
		\biggl(|\alpha_q^*-,\alpha_p^*-\rangle,|\beta_q+,\beta_p+\rangle\biggr)=-\delta(\alpha_q-\beta_q)\delta(\alpha_p-\beta_p)
		\end{array}
		\right.
	\end{equation}
which, for $n$ degrees of freedom, become
	\begin{eqnarray}
	&&\displaystyle \biggl(|\alpha_{q_1}^*+,\alpha_{q_2}^*+,\ldots \alpha_{q_n}^*+,\alpha_{p_1}^*+,
	\ldots \alpha_{p_n}^*+\rangle,
	|\beta_{q_1}-,\beta_{q_2}-,\ldots \beta_{q_n}-,\beta_{p_1}-,\ldots
	\beta_{p_n}-\rangle\biggr)=\nonumber\\
	&&\qquad\quad=\prod_{i=1}^{2n}\delta(\alpha_i-\beta_i) \label{due-trentotto-d}
	\end{eqnarray}
and 
	\begin{eqnarray}
	&&\displaystyle \biggl(|\alpha_{q_1}^*-,\alpha_{q_2}^*-,\ldots \alpha_{q_n}^*-,\alpha_{p_1}^*-,\ldots 
	\alpha_{p_n}^*-\rangle,
	|\beta_{q_1}+,\beta_{q_2}+,\ldots \beta_{q_n}+,\beta_{p_1}+,\ldots
	\beta_{p_n}+\rangle\biggr)=\nonumber\\
	&&\qquad\quad =-\prod_{i=1}^{2n}\delta(\alpha_i-\beta_i). \label{due-trentotto-e}
	\end{eqnarray}
We can also build {\it mixed} states like
	\begin{equation}
	\displaystyle
	|\alpha_q-,\alpha_p+\rangle\equiv e^{-\alpha_q\hat{\bar{c}}_q-\alpha_p\hat{c}^p}|0-,0+\rangle
	\label{due-trentanove}
	\end{equation}
which are eigenstates of $\hat{c}^q$ with eigenvalue $\alpha_q$ and of $\hat{\bar{c}}_p$ with eigenvalue $\alpha_p$. With
$n$ degrees of freedom we could have states where the $2n$ entries are arbitrary combinations of ``$+$" and ``$-$". The scalar
products among these states can be worked out case by case using their expression in terms of $|0\pm, 0\pm\rangle$.

In the case of more than one Grassmannian variable, the round brackets
$\biggl(|\;\rangle,|\;\rangle\biggr)$, which indicate the scalar product,
are turned into the Dirac notation as follows:
	\begin{equation}
	\biggl(|\alpha_q,\alpha_p\rangle,|\beta_q,\beta_p\rangle\biggr)\equiv\langle\alpha_p,\alpha_q|\beta_q,\beta_p\rangle .\label{due-quaranta}
	\end{equation}
This indicates that to pass from the bra to the ket it is necessary to invert the order of the entries for the $q$ and the $p$.
The explanation for this is provided in Appendix A.

We have now all the ingredients to write down the ``twisted" completeness relations analog to Eq. (\ref{due-quindici}). They are
	\begin{eqnarray}
	&&\int d\alpha_qd\alpha_p|\alpha_q+,\alpha_p+\rangle\langle -\alpha_p^*,-\alpha_q^*|={\bf 1}\nonumber\\
	&&\int d\alpha_pd\alpha_q|\alpha_q-,\alpha_p-\rangle\langle+\alpha_p^*,+\alpha_q^*|={\bf 1}.
	\label{due-quarantuno}
	\end{eqnarray}
It is easy to prove (see Appendix A) that the LHS of (\ref{due-quarantuno}) turns out to be equal to
	\begin{equation}
	|0+,0+\rangle\langle+0,+0|+|0-,0+\rangle\langle+0,-0|+
	|0+,0-\rangle\langle-0,+0|+|0-,0-\rangle\langle-0,-0| \label{due-quarantadue}
	\end{equation}
and this is clearly equal to ${\bf 1}$ because it is made of the projectors on the 4 states:
	\begin{equation}
	|0+,0+\rangle, \;|0-,0+\rangle,\; |0+,0-\rangle,\; |0-,0-\rangle \label{due-quarantatre}
	\end{equation}
which are a complete basis in the case of 2 Grassmannian variables $c^p$, $c^q$. The
proof that they are a basis is analogous to the one presented in (\ref{due-otto}):
in the case of a 2-dimensional phase space we
can build two commuting Hermitian operators:
	\begin{equation}
	\label{due-quarantaquattro}
	\left\{
		\begin{array}{l}
		\displaystyle 
		\hat{N}_q=\hat{c}^q\hat{\bar{c}}_q
		\smallskip\\
		\displaystyle
		\hat{N}_p=\hat{c}^p\hat{\bar{c}}_p .
		\end{array}
		\right.
	\end{equation}
The states (\ref{due-quarantatre}) are eigenstates of $\hat{N}_q,\hat{N}_p$ with eigenvalues respectively
$(0,0)$, $(1,0)$, $(0,1)$, $(1,1)$:
	\begin{equation}
	\label{due-quarantacinque}
	\left\{
		\begin{array}{l}
		\displaystyle 
	 \hat{N}_q|0+,0+\rangle=0,\;\;\;\;\;\qquad\qquad\quad\; \hat{N}_p|0+,0+\rangle=0
		\smallskip\\
		\displaystyle
		\hat{N}_q|0-,0+\rangle=|0-,0+\rangle,\;\;\;\;\;\qquad\hat{N}_p|0-,0+\rangle=0
	 \smallskip\\
	 \displaystyle 
	 \hat{N}_q|0+,0-\rangle=0,\;\;\;\;\;\qquad\qquad\;\;\;\;\hat{N}_p|0+,0-\rangle=|0+,0-\rangle
	 \smallskip\\
	 \hat{N}_q|0-,0-\rangle=|0-,0-\rangle,\;\;\;\;\;\qquad\hat{N}_p|0-,0-\rangle=|0-,0-\rangle .
		\end{array}
		\right.
	\end{equation}
If we had chosen only $\hat{N}_q$ this would not have been a complete set of operators because the states $|0-,0-\rangle$ and
$|0-,0+\rangle$ have the same eigenvalue with respect to $\hat{N}_q$. The degeneracy is removed by diagonalizing another Hermitian
and commuting operator like $\hat{N}_p$. This removes completely the degeneracy and so $\hat{N}_q$ and $\hat{N}_p$ make a complete
set of operators. In the general case of a $2n$ dimensional phase space the complete set of Hermitian and commuting operators is
made of the following $2n$ operators:
	\begin{equation}
	 \label{due-quarantasei}
	 \left\{
		\begin{array}{l}
		\displaystyle 
		\hat{N}_{q_i}=\hat{c}^{q_i}\hat{\bar{c}}_{q_i}
		\smallskip\\
		\displaystyle
		\hat{N}_{p_i}=\hat{c}^{p_i}\hat{\bar{c}}_{p_i}
		\end{array}
		\right.
	\end{equation}
and the diagonalization will produce $2^{2n}$ states $|(\cdot)(\cdot)\ldots
(\cdot)\rangle$  with $2n$ entries which are
either ``$0+$" or ``$0-$". In this space of $2^{2n}$ states it is possible to give a matrix representation to the Grassmannian variables
$\hat{c}^q,\hat{c}^p,\hat{\bar{c}}_q,\hat{\bar{c}}_p$. The details of these
matrix representations are given in Ref. \cite{matrix}. The reader may have doubts about the fact that the four states (\ref{due-quarantatre}) really make a complete
basis. These doubts could have been generated by the fact that the Hermitian operators $\hat{N}_q$ and $\hat{N}_p$ have a
Grassmannian nature (an even one) and so the usual theorems on Hermitian
operators may not hold for them. To convince himself of that the reader can do the
long, but boring, calculation of checking Eq. (\ref{due-quarantuno}) on all possible states.

The next thing we are going to prove is the positivity of the SvH scalar
product for these systems with $2n$ Grassmannian
variables like we did in Eq. (\ref{due-venticinque}) for the case of one variable.
 Let us write the second of the
completeness relations (\ref{due-quarantuno}) as follows (for $n=1$)
	\begin{equation}
	\int dqdpdc^pdc^q|q,p,c^q-,c^p-\rangle\langle+c^{p*},+c^{q*},p,q|={\bf 1} \label{due-quarantasette}
	\end{equation}
where we have denoted with $c^q$, $c^p$, rather than $\alpha^q$, $\alpha^p$, the eigenvalues of $\hat{c}^q$ and $\hat{c}^p$. 
We have also added an integration over the $(p,q)$ variables because in general the wave functions depend also on $p$ and
$q$. Inserting this completeness relation into the scalar product $\langle\Phi|\psi\rangle$ between the states, we get
	\begin{eqnarray}
	\langle\Phi|\psi\rangle &=& \int dp dq dc^pdc^q\langle\Phi|q,p,c^q-,c^p-\rangle\langle+c^{p*},+c^{q*},p,q|\psi\rangle=\nonumber\\
	&=& \int dp dq dc^pdc^q\,\Phi_+^*(q,p,c^q,c^p)\psi_-(q,p,c^q,c^p) \label{due-quarantotto}
	\end{eqnarray}
where
	\begin{equation}
	 \label{due-quarantanove}
	 \left\{
		\begin{array}{l}
		\displaystyle 
		\Phi_+(q,p,c^q,c^p)\equiv\langle -c^p,-c^q,q,p|\Phi\rangle
		\smallskip \nonumber\\
		\displaystyle
		\psi_-(q,p,c^q,c^p)\equiv\langle +c^{p*},+c^{q*},p,q|\psi\rangle .
		\end{array}
		\right.
	\end{equation}
The function $\psi_-$ depends on $(q,p,c^p,c^q)$ and, because
$(c^p,c^q)$ are Grassmannian, it can only have the following form
	\begin{eqnarray}
	&&\psi_-(q,p,c^q,c^p)=\langle +c^{p*},+c^{q*},p,q|\psi\rangle=\nonumber\\
	&&=\psi_0(q,p)+\psi_q(q,p)c^q+\psi_p(q,p)c^p+\psi_2(q,p)c^qc^p
	\label{due-cinquanta-a}
	\end{eqnarray}
where we have indicated with $\psi_0$ the zero-form, with $\psi_q$ the coefficient of $c^q$, with $\psi_p$ the coefficient of $c^p$
and with $\psi_2$ the coefficient of the 2 form.
$\Phi_+(q,p,c^q,c^p)$ instead is
	\begin{eqnarray}
	&&\Phi_+(q,p,c^q,c^p)=\langle -c^p,-c^q,q,p|\Phi\rangle=\nonumber\\
	&&=\int dq^{\prime}dp^{\prime}dc^{p\prime}dc^{q\prime}\langle -c^p,-c^q,q,p|q^{\prime},p^{\prime},c^{q\prime}-,c^{p\prime}-\rangle
	\langle +c^{p\prime *},+c^{q\prime *},p^{\prime},q^{\prime}|\Phi\rangle=\nonumber\\
	&&=\int
	dq^{\prime}dp^{\prime}dc^{p\prime}dc^{q\prime}\delta(q-q^{\prime})
	\delta(p-p^{\prime})exp[c^{q*}c^{q\prime}+c^{p*}c^{p\prime}]
	\cdot (\Phi_0+\Phi_qc^{q\prime}+\Phi_pc^{p\prime}+\Phi_2c^{q\prime}c^{p\prime})=\nonumber\\
	&&=\int dc^{p\prime}dc^{q\prime}(1+c^{q*}c^{q\prime}+c^{p*}c^{p\prime}+c^{q*}c^{q\prime}c^{p*}c^{p\prime})
	\cdot(\Phi_0+\Phi_qc^{q\prime}+\Phi_pc^{p\prime}+\Phi_2c^{q\prime}c^{p\prime})=\nonumber\\
	&&=\Phi_2(q,p)+\Phi_p(q,p)c^{q*}-\Phi_q(q,p)c^{p*}-\Phi_0(q,p)c^{q*}c^{p*} \label{due-cinquanta-b}
	\end{eqnarray}
where we have made use of the completeness relation (\ref{due-quarantasette}) in the second step above and of the 
scalar product (\ref{due-trentotto-a}) in the third step.
Inserting now the equations (\ref{due-cinquanta-a})-(\ref{due-cinquanta-b}) into (\ref{due-quarantotto}) we get
	\begin{eqnarray}
	&&\langle\Phi|\psi\rangle =\int dpdqdc^pdc^q\,\Phi_+^*\psi_-=\nonumber\\
	&&=\int dpdq dc^pdc^q(\Phi_2^*+\Phi_p^*c^q 
	-\Phi_q^*c^p-\Phi_0^*c^pc^q)\cdot (\psi_0+\psi_qc^q+\psi_pc^p+\psi_2c^qc^p)=\nonumber\\
	&&=\int dp dq \,[\Phi_0^*\psi_0+\Phi^*_q\psi_q+\Phi^*_p\psi_p+\Phi_2^*\psi_2] .\label{svhprod}
	\end{eqnarray}
Using this equation we see that the norm of a generic state $|\psi\rangle$ 
is:
	\begin{equation}
	\langle\psi|\psi\rangle=\int dp dq\biggl[|\psi_0|^2+|\psi_q|^2+|\psi_p|^2+|\psi_2|^2\biggr]. \label{due-cinquantuno}
	\end{equation}
{\it This confirms that the SvH scalar product is positive definite}. 
Note that for the zero-forms the SvH scalar product is reduced to the
KvN one of Eq. (\ref{uno-quattro}). 

The  derivation that we have presented here can be repeated for any number of degrees of freedom. 
The state in that case is of the form 
	\begin{equation}
	\displaystyle \psi=\frac{1}{m!}\sum_{m=0}^{2n}\psi_{a_1\ldots a_m}c^{a_1}c^{a_2}\ldots c^{a_m}
	\end{equation}
and the SvH norm turns out to be 
	\begin{equation}
	\displaystyle \langle\psi|\psi\rangle=K\sum_{\{a_i\}}\sum_{m=0}^{2n}|\psi_{a_1\ldots a_m}|^2
	\end{equation}
where $K$ is a positive number. The derivation is long but straightforward.

All the construction we have done here is very similar to the one of SvH \cite{Salomonson} but there is a {\it crucial} difference.
The model which SvH examined is supersymmetric quantum mechanics and its
Hamiltonian is Hermitian \cite{Salomonson} under the
SvH scalar product. The crucial difference is that the Hamiltonian of our model (\ref{uno-sedici}) is not
Hermitian under the same scalar product as we shall show below. The Hamiltonian 
$\widetilde{\cal H}$ of (\ref{uno-quattordici}) or
(\ref{uno-sedici}) can be written as
	\begin{equation}
	\widetilde{\cal H}=\widetilde{\cal H}_{bos}+\widetilde{\cal H}_{ferm} \label{due-cinquantuno-x}
	\end{equation}
where 
	\begin{equation}
	 \label{due-cinquantuno-xx}
	 \left\{
		\begin{array}{l}
		\displaystyle 
		\widetilde{\cal H}_{bos}=\lambda_a\omega^{ab}\partial_bH
		\smallskip \nonumber\\
		\displaystyle
		\widetilde{\cal H}_{ferm}=i\bar{c}_a\omega^{ab}\partial_b\partial_dHc^d.
		\end{array}
		\right.
	\end{equation}
Let us check the Hermitian nature of each piece:
	\begin{eqnarray}
	\widetilde{\cal
	H}_{bos}^{\dagger}&=&(\lambda_a\omega^{ab}\partial_bH)^{\dagger}=\partial_bH(\omega^{\dagger})^{ba}\lambda_a^{\dagger}=\nonumber\\
	&=& \partial_bH\omega^{ab}\lambda_a=\lambda_a\omega^{ab}\partial_bH=\widetilde{\cal H}_{bos} \label{due-cinquantuno-b}
	\end{eqnarray}
where we have used the fact that $\lambda_a^{\dagger}=\lambda_a$ according to (\ref{due-uno}).
Now let us analyze the fermionic part $\widetilde{\cal H}_{ferm}$ of the
Hamiltonian. The SvH hermiticity conditions (\ref{due-due})-(\ref{hermiticityp}) 
for the Grassmannian variables are the following ones:
	\begin{equation}
	 \left\{
		\begin{array}{l}
		\displaystyle 
		(c^a)^{\dagger}=\bar{c}_a
		\smallskip \nonumber\\
		\displaystyle
		(\bar{c}_a)^{\dagger}=c^a .\label{due-cinquantadue}
		\end{array}
		\right.
	\end{equation}
Next let us write $\widetilde{\cal H}_{ferm}$ as
	\begin{equation}
	\widetilde{\cal H}_{ferm}=i\bar{c}_a\omega^{ab}\partial_b\partial_dHc^d=i\bar{c}_a{\cal F}^a_dc^d
	\end{equation}
where ${\cal F}_d^a=\omega^{ab}\partial_b\partial_dH$. Then
	\begin{equation}
	(\widetilde{\cal H}_{ferm})^{\dagger}=(i\bar{c}_a{\cal F}^a_dc^d)^{\dagger}=
	(-i)(c^d)^{\dagger}({\cal F}^{\dagger})^d_a
	(\bar{c}_a)^{\dagger}= -i\bar{c}_d({\cal F}^{\dagger})^d_ac^a.
	\end{equation}
So $\widetilde{\cal H}_{ferm}$ would be Hermitian if ${\cal F}^{\dagger}=-{\cal F}$. As ${\cal F}$ is real,
the relation ${\cal F}^{\dagger}=-{\cal F}$ implies that ${\cal F}^{T}=-{\cal F}$. 
Let us see if this happens by taking \cite{footnote4} 
	\begin{equation}
	{\cal F}^q_p=\omega^{qb}\partial_b\partial_pH=\omega^{qp}\partial_p^2H=\partial_p^2H \label{due-cinquantatre}
	\end{equation}
and comparing it with its transposed element
	\begin{equation}
	{\cal F}^p_q=\omega^{pb}\partial_b\partial_qH=\omega^{pq}\partial_q^2H=-\partial_q^2H .\label{due-cinquantaquattro}
	\end{equation}
We see that (\ref{due-cinquantatre}) and (\ref{due-cinquantaquattro}) are not the opposite of each other like it should
be for $\widetilde{\cal H}_{ferm}$ to be Hermitian. Note anyhow that for the harmonic oscillator with 
$\displaystyle H=\frac{1}{2}p^2+\frac{1}{2}q^2$, ${\cal F}^q_p=-{\cal F}^p_q$ and so in that case $\widetilde{\cal H}$
would be Hermitian. This of course would not happen for a generic potential and 
this concludes the proof that $\widetilde{\cal H}$ {\it is not in general Hermitian under the SvH scalar product}. 

Even if the CPI \cite{Gozzi} has been proposed 16 years ago, few people have really appreciated it. Most of the others have not carefully read the
papers and dismissed it as ``the same model as susy QM". The fact that now we have proved that, differently than susy QM, the CPI
Hamiltonian $\widetilde{\cal H}$ is not Hermitian may induce those people to
study the papers of Ref. \cite{Gozzi} more carefully.

Originally the CPI model was formulated directly via path integrals without deriving it explicitly from the operatorial formalism.
In quantum mechanics instead the path integral was derived \cite{Hibbs} by assembling infinitesimal time evolutions in operatorial
form and inserting completeness relations between those infinitesimal evolutions. We shall now do the same for the CPI and as
completeness relations we shall use the ones associated to the SvH scalar
product. Before proceeding we should remember that, besides
the ``configuration" \cite{footnote5} representation (\ref{uno-quindici})
in which $\hat{\varphi}^a$ is a multiplicative operator while 
$\displaystyle \lambda_a=-i\frac{\partial}{\partial\varphi^a}$ is a
derivative one, 
we can also have a sort of ``momentum" representation
in which $\hat{\lambda}_a$ is a multiplicative operator while $\hat{\varphi}^a$
is a derivative one and the same  for the Grassmannian variables. So in this momentum representation we have:
	\begin{equation}
	\label{due-cinquantacinque}
	 \left\{
		\begin{array}{l}
		\displaystyle 
		\varphi^a=i\frac{\partial}{\partial\lambda_a}
		\smallskip \nonumber\\
		\displaystyle
		c^a=\frac{\partial}{\partial\bar{c}_a}.
		\end{array}
		\right.
	\end{equation}
The ``wave functions" in this representation would depend on the $\lambda_a,\bar{c}_a$ and the completeness relations would be
	\begin{equation}
	\displaystyle
	\int d\lambda_qd\lambda_pd\bar{c}_qd\bar{c}_p|\lambda_q,\lambda_p,\bar{c}_q+,\bar{c}_p+\rangle
	\langle -\bar{c}^*_p, -\bar{c}^*_q,\lambda_p,\lambda_q|={\bf 1} \label{due-cinquantasei-a}
	\end{equation}
to be contrasted with the ``configuration" one
	\begin{equation}
	\int dqdpdc^pdc^q|q,p,c^q-,c^p-\rangle\langle +c^{p*},+c^{q*},p,q|={\bf 1}. \label{due-cinquantasei-b}
	\end{equation}
Using the two completeness relations above and the SvH scalar product we will prove that
the following transition amplitude:
	\begin{equation}
	\displaystyle K(f|i)=\langle+c_{p_f}^*,+c^*_{q_f},\varphi_f|e^{-i\widetilde{\cal H}
	(t_f-t_i)}|\varphi_i,c_{q_i}-,c_{p_i}-\rangle \label{due-cinquantasette}
	\end{equation}
is the transition amplitude to go from $(\varphi_i,c_{q_i},c_{p_i})$ to $(\varphi_f,c_{q_f},c_{p_f})$ and that it has the
same path integral structure which led to the CPI \cite{Gozzi} and to the expression (\ref{uno-dodici-b}) for the
generating functional. The reason we do this is to see whether the Hilbert space structure of SvH
leads to the CPI path integral or to something else. Later on we will do the same also for the other type of scalar
products which we will introduce in later sections. The reader may wonder why (\ref{due-cinquantasette}) is really the
transition amplitude to go from $(\varphi_i,c_{q_i},c_{p_i})$ to $(\varphi_f,c_{q_f},c_{p_f})$ since the final bra in 
(\ref{due-cinquantasette}) is the state $\langle +c_{p_f}^*,+c_{q_f}^*,\varphi_f|$. The reason is
that this is the bra which is eigenstate of $\hat{c}_q$ and $\hat{c}_p$ with eigenvalues $c_{q_f},c_{p_f}$. The proof goes
as follows. Let us start from the following relations which are the
analog of (\ref{due-sette}) 
	\begin{equation}
	\label{due-cinquantotto-a}
	 \left\{
		\begin{array}{l}
		\displaystyle 
		\hat{\bar{c}}_q|\varphi_f,c_{q_f}^*+,c_{p_f}^*+\rangle=c_{q_f}^*|\varphi_f,c_{q_f}^*+,c_{p_f}^*+\rangle
		\smallskip \nonumber\\
		\displaystyle
		\hat{\bar{c}}_p|\varphi_f,c_{q_f}^*+,c_{p_f}^*+\rangle=c_{p_f}^*|\varphi_f,c_{q_f}^*+,c_{p_f}^*+\rangle
		\end{array}
		\right.
	\end{equation}
and let us do the Hermitian conjugation of (\ref{due-cinquantotto-a}) with the
rules of the SvH scalar product. We get
	\begin{equation}
	\label{due-cinquantotto-b}
	 \left\{
		\begin{array}{l}
		\displaystyle 
		\langle+c_{p_f}^*,+c_{q_f}^*,\varphi_f|\hat{c}_q=c_{q_f}\langle+c_{p_f}^*,+c_{q_f}^*,\varphi_f|
		\smallskip \nonumber\\
		\displaystyle
		\langle+c_{p_f}^*,+c_{q_f}^*,\varphi_f|\hat{c}_p=c_{p_f}\langle+c_{p_f}^*,+c_{q_f}^*,\varphi_f|
		\end{array}
		\right.
	\end{equation}
that is what we wanted to prove. 

Now, as it is usually done in QM, let us divide the interval $t_f-t_i$ in (\ref{due-cinquantasette}) into $N$ intervals of
length $\epsilon$, so that $N\epsilon=t_f-t_i$. The amplitude $K(f|i)$ of (\ref{due-cinquantasette}) can then be written as
	\begin{equation}
	K(f|i)=\langle +c_{p_f}^*,+c_{q_f}^*,\varphi_f\biggl|\exp\bigl[-i\epsilon\widetilde{\cal H}\bigr]
	\ldots \exp\bigl[-i\epsilon\widetilde{\cal H}\bigr]\biggr|\varphi_i,c_{q_i}-,c_{p_i}-\rangle .\label{due-cinquantotto}
	\end{equation}
Let us then insert a ``momentum" completeness (\ref{due-cinquantasei-a}) in front of each exponential in 
(\ref{due-cinquantotto}) and a ``configuration"
completeness (\ref{due-cinquantasei-b}) behind each exponential. We get the following expression:
	\begin{eqnarray}
	\displaystyle
	&&K(f|i)=\langle
	+c_{p_f}^*,+c_{q_f}^*,p_f,q_f|\biggl\{\int d\lambda_{q_N} d\lambda_{p_N} 
	d\bar{c}_{q_N} d\bar{c}_{p_N}|
	\lambda_{q_N},\lambda_{p_N},\bar{c}_{q_N}+,\bar{c}_{p_N}+\rangle\cdot\nonumber\\
	&&\cdot\langle -\bar{c}_{p_N}^*,-\bar{c}_{q_N}^*,\lambda_{p_N},\lambda_{q_N}|\biggr\}\exp[-i\epsilon\widetilde{\cal
	H}]\biggl\{
	\int dq_{\scriptscriptstyle N}dp_{\scriptscriptstyle N}dc_{\scriptscriptstyle N}^pdc_{\scriptscriptstyle N}^q
	|q_{\scriptscriptstyle N},p_{\scriptscriptstyle N},c^q_{\scriptscriptstyle N}-,c^p_{\scriptscriptstyle N}-
	\rangle\nonumber\\
	&&\cdot\langle +c^*_{p_N},+c^*_{q_N},p_{\scriptscriptstyle N},q_{\scriptscriptstyle N}|\biggr\}\cdot\biggl\{\int
	d\lambda_{q_{N-1}}d\lambda_{p_{N-1}}d\bar{c}_{q_{N-1}}d\bar{c}_{p_{N-1}}
	|\lambda_{q_{N-1}},\lambda_{p_{N-1}},\bar{c}_{q_{N-1}}+,\bar{c}_{p_{N-1}}+\rangle\nonumber\\
	&&\cdot\langle -\bar{c}_{p_{N-1}}^*,
	-\bar{c}^*_{q_{N-1}},\lambda_{p_{N-1}},\lambda_{q_{N-1}}|\biggr\}\exp[-i\epsilon\widetilde{\cal H}]\biggl\{
	\int dq_{\scriptscriptstyle{N-1}}dp_{\scriptscriptstyle{N-1}}dc^p_{\scriptscriptstyle {N-1}}dc^q_{\scriptscriptstyle {N-1}}
	\ldots\biggr\}\ldots\nonumber\\
	&&\cdot \biggl\{\int
	dq_{\scriptscriptstyle 1}dp_{\scriptscriptstyle 1}dc^p_{\scriptscriptstyle 1}dc^q_{\scriptscriptstyle 1}|
	q_{\scriptscriptstyle 1},p_{\scriptscriptstyle 1},c^q_{\scriptscriptstyle 1}-,c^p_{\scriptscriptstyle 1}-
	\rangle\langle+c^{*}_{p_1},+c^{*}_{q_1},p_{\scriptscriptstyle 1},q_{\scriptscriptstyle 1}|\biggr\}|q_i,p_i,c_{q_i}-,
	c_{p_i}-\rangle .\label{due-cinquantotto-bis}
	\end{eqnarray}
The subindex $(1,2,\ldots,i,\ldots,N)$ on $(q,p,\lambda_q,\lambda_p,c^q,c^p,\bar{c}_q,\bar{c}_p)$ is the time
label in the subdivision of the interval $(t_f-t_i)$ in $N$ subintervals. 
There are various elements to evaluate in the expression above. We can start
from the last one which is easy to evaluate:
	\begin{equation}
	\langle +c_{p_1}^{*},+c_{q_1}^{*},p_{\scriptscriptstyle 1},q_{\scriptscriptstyle 1}
	|q_i,p_i,c_{q_i}-,c_{p_i}-\rangle=\delta(p_{\scriptscriptstyle 1}-p_i)\delta(q_{\scriptscriptstyle 1}-q_i)
	\delta(c_{q_1}-c_{q_i})\delta(c_{p_1}-c_{p_i}) .\label{due-cinquantanove}
	\end{equation}
Another element is 
	\begin{eqnarray}
	&&\langle
	+c_{p_j}^*,+c_{q_j}^*,p_j,q_j|\lambda_{q_{j-1}},\lambda_{p_{j-1}},\bar{c}_{q_{j-1}}+,\bar{c}_{p_{j-1}}+\rangle=\nonumber\\
	&&=\exp[ip_j\lambda_{p_{j-1}}+iq_j\lambda_{q_{j-1}}]\langle
	+c^{*}_{p_j},+c_{q_j}^{*}|\bar{c}_{q_{j-1}}+,\bar{c}_{p_{j-1}}+\rangle=\nonumber\\
	&&=\exp[ip_j\lambda_{p_{j-1}}+iq_j\lambda_{q_{j-1}}]\,\exp[c^q_j\bar{c}_{q_{j-1}}+c^p_j\bar{c}_{p_{j-1}}].
	\label{due-sessanta}
	\end{eqnarray}
These expressions are the analog of the QM ones which link the momentum and space eigenstates and are derived in
the same manner using the operatorial expression (\ref{uno-quindici}) for $\lambda_a$ and $\bar{c}_a$.
The last element that we need in (\ref{due-cinquantotto-bis}) is
	\begin{eqnarray}
	&& \langle -\bar{c}_{p_j}^*,-\bar{c}_{q_j}^*,\lambda_{p_j},\lambda_{q_j}|\exp\biggl[-i\epsilon\widetilde{\cal
	H}\biggr]|q_j,p_j,c^q_j-,c^p_j-\rangle=\nonumber\\
	&&=\exp\biggl[-i\epsilon\widetilde{\cal H}(\varphi_j,\lambda_j,c_j,\bar{c}_j)\biggr]\exp
	\biggl[\bar{c}_{q_j}c^q_j+\bar{c}_{p_j}c^p_j\biggr]
	\cdot \exp\biggl[-i\lambda_{q_j}q_j-i\lambda_{p_j}p_j\biggr]. \label{due-sessantuno}
	\end{eqnarray}
Its derivation is presented in Appendix A. Inserting (\ref{due-cinquantanove})-(\ref{due-sessanta})-(\ref{due-sessantuno})
into (\ref{due-cinquantotto-bis}) we get \cite{footnote6} 
	\begin{eqnarray}
	\displaystyle 
	&&K(f|i)=\int \biggl(\prod_{j=2}^Nd\varphi_jd\lambda_jd\bar{c}_{q_j}d\bar{c}_{p_j}dc^p_jdc^q_j\biggr)
	d\lambda_1d\bar{c}_{q_1}d\bar{c}_{p_1}\cdot \nonumber\\
	&&\cdot \exp\biggl[i\lambda_{\scriptscriptstyle N}(\varphi_{f}-\varphi_{\scriptscriptstyle N})+
	i\lambda_{\scriptscriptstyle{N-1}}(\varphi_{\scriptscriptstyle N}-\varphi_{\scriptscriptstyle{N-1}})\ldots
	+i\lambda_1(\varphi_2-\varphi_i)\biggr]\cdot \nonumber\\
	&&\cdot \exp\biggl[-\bar{c}_{\scriptscriptstyle N}(c_f-c_{\scriptscriptstyle N})-\bar{c}_{\scriptscriptstyle{N-1}}
	(c_{\scriptscriptstyle N}-c_{\scriptscriptstyle{N-1}})\ldots -\bar{c}_{\scriptscriptstyle 1}
	(c_{\scriptscriptstyle 2}-c_i)-i\epsilon\sum_{j=1}^N\widetilde{\cal
	H}(\varphi_j,\lambda_j,c_j,\bar{c}_j)\biggr]=\nonumber\\ 
	&&=\int {\mathscr D}\mu \,\exp\biggl[i\epsilon
	\sum_{j=1}^N\lambda_j\frac{(\varphi_{j+1}-\varphi_j)}{\epsilon}
	-\epsilon\sum_{j=1}^N\bar{c}_j\frac{(c_{j+1}-c_j)}{\epsilon}-i\epsilon \sum_{j=1}^N\widetilde{\cal
	H}(j)\biggr]
	\label{due-sessantadue}
	\end{eqnarray}
where the boundary conditions are:
	\begin{equation}
	\varphi_0=\varphi_i,\;\;\;\;\;\;\varphi_{\scriptscriptstyle N+1}=\varphi_f,\;\;\;\;\;\;c_0=c_i,\;\;\;\;\;\;
	c_{\scriptscriptstyle{N+1}}=c_f
	\end{equation}
and where the measure is:
	\begin{equation}
	\displaystyle {\mathscr D}\mu= \biggl(\prod_{j=2}^Nd\varphi_jd\lambda_jd\bar{c}_{q_j}d\bar{c}_{p_j}dc^p_jdc^q_j\biggr)
	d\lambda_1d\bar{c}_{q_1}d\bar{c}_{p_1}. \label{due-sessantatre}
	\end{equation}
This measure indicates that the initial and final $(\varphi,c)$ are not integrated over. The continuum limit of
(\ref{due-sessantadue}) can be easily derived:
	\begin{equation}
	\displaystyle K(f|i)=\int_{\varphi_ic_i}^{\varphi_fc_f}{\mathscr D}\mu \;e^{i\int dt\widetilde{\cal L}}
	\end{equation}
and $\widetilde{\cal L}$ turns out to be the Lagrangian in (\ref{uno-tredici}). This confirms that, via the scalar products and
completeness relations of SvH (\ref{due-cinquantasei-a})-(\ref{due-cinquantasei-b}), one gets the path integral 
(\ref{uno-dodici-b}) of the CPI.

We can summarize this long section by saying that with the SvH scalar product the Hilbert space is a true Hilbert space in
the sense that the scalar product is positive definite but unfortunately the Hamiltonian is not Hermitian even if the standard
path integral for classical mechanics can be reproduced. 

\section{The Gauge Scalar Product}
\noindent
In this section we will study a new scalar product which is the one typically used in gauge theory \cite{Henneaux}. That is
why we call it ``gauge scalar product". Proceeding as we did
in (\ref{due-uno})-(\ref{due-due}) for the SvH scalar product,
we shall first ``postulate" some hermiticity conditions and then
derive the scalar product and the completeness relations. 

The hermiticity conditions for the bosonic variables are chosen to be the same as in the SvH case
	 \begin{equation}
	 \label{tre-uno}
	 \left\{
		\begin{array}{l}
		\displaystyle 
		\hat{\varphi}^{a\dagger}=\hat{\varphi}^a
		\smallskip \nonumber\\
		\displaystyle
		\hat{\lambda}_a^{\dagger}=\hat{\lambda}_a
		\end{array}
		\right.
	\end{equation}
while for the Grassmannian variables we choose:
	 \begin{equation}
	 \label{tre-due}
	 \left\{
		\begin{array}{l}
		\displaystyle 
		\hat{c}^{a\dagger}=\hat{c}^a
		\smallskip \nonumber\\
		\displaystyle
		\hat{\bar{c}}_a^{\dagger}=\hat{\bar{c}}_a.
		\end{array}
		\right.
	\end{equation}
This means that the Grassmannian variables are Hermitian, differently 
from the SvH case (\ref{due-cinquantadue}). Thanks to the
hermiticity structure (\ref{tre-uno})-(\ref{tre-due}), the Hamiltonian of the
CPI now turns  out to be Hermitian. In fact
from
	\begin{equation}
	\widetilde{\cal H}=\widetilde{\cal H}_{bos}+\widetilde{\cal H}_{ferm} \label{tre-due-b}
	\end{equation}
we have that $\widetilde{\cal H}_{bos}$ is Hermitian as it was in the SvH case (\ref{due-cinquantuno-b}) because it
involves only the bosonic variables which have the same hermiticity structure both in the SvH (\ref{due-uno}) and in
the gauge case (\ref{tre-uno}). On the other hand,
doing the Hermitian conjugation of $\widetilde{\cal H}_{ferm}$ (\ref{due-cinquantuno-xx}), we get
	\begin{equation}
	(\widetilde{\cal H}_{ferm})^{\dagger}=-ic^c\omega^{ab}\partial_b\partial_cH\bar{c}_a=
	i\bar{c}_a\omega^{ab}\partial_b\partial_cHc^c=\widetilde{\cal H}_{ferm} \label{tre-tre}
	\end{equation}
where in the second step of (\ref{tre-tre}) we used the anticommutation relations
	\begin{equation}
	[c^c,\bar{c}_a]=\delta_a^c \label{tre-tre-a}
	\end{equation}
together with the fact that $\omega^{ab}\partial_b\partial_aH=0$. Eq.
(\ref{tre-tre}) is the proof that the overall Hamiltonian $\widetilde{\cal H}$ 
is Hermitian under the gauge scalar product. So far we have
established a first important difference from the SvH case where $\widetilde{\cal H}$ was not Hermitian. We
will next check whether the scalar product underlying the gauge hermiticity conditions (\ref{tre-uno})-(\ref{tre-due}) is
positive definite as in the SvH case. 

Let us now do all the necessary steps for the case of a single Grassmannian variable $c$ 
and its conjugate $\bar{c}$. Their algebra and hermiticity
relations can be summarized in the following table
	\begin{equation}
	 \label{tre-quattro}
	 \left\{
		\begin{array}{l}
		\displaystyle 
		[\hat{c},\hat{\bar{c}}]_+=1
		\smallskip \nonumber\\
		\displaystyle
		\hat{c}^2=\hat{\bar{c}}^2=0
		\smallskip \nonumber\\
		\hat{c}^{\dagger}=\hat{c}
		\smallskip \nonumber\\
		\hat{\bar{c}}^{\dagger}=\hat{\bar{c}}.
		\end{array}
		\right.
	\end{equation}
As we did in the SvH case, let us define a state $|0-\rangle$ as
	\begin{equation}
	\hat{c}|0-\rangle=0 \label{tre-cinque}
	\end{equation}
and a state $|0+\rangle$ by
	\begin{equation}
	|0+\rangle\equiv \hat{\bar{c}}|0-\rangle. \label{tre-sei}
	\end{equation}
If we now calculate the norm of $|0+\rangle$, we get:
	\begin{eqnarray}
	\biggl(|0+\rangle,|0+\rangle\biggr)&=&\biggl(\hat{\bar{c}}|0-\rangle,\;\hat{\bar{c}}|0-\rangle\biggr)=
	\biggl(|0-\rangle,\;\hat{\bar{c}}^{\dagger}\;\hat{\bar{c}}|0-\rangle\biggr)=\nonumber\\ &=&
	\biggl(|0-\rangle,\;\hat{\bar{c}}\;\hat{\bar{c}}|0-\rangle\biggr)=\biggl(|0-\rangle,0\cdot |0-\rangle\biggr)=0 ,\label{tre-sette}
	\end{eqnarray}
and the same we can do for $|0-\rangle$;
	\begin{eqnarray}
	\biggl(|0-\rangle,|0-\rangle\biggr)&=&\biggl(\hat{c}|0+\rangle,\hat{c}|0+\rangle\biggr)=
	\biggl(|0+\rangle,\hat{c}^2|0+\rangle\biggr)=\nonumber\\
	&=& \biggl(|0+\rangle,0\cdot |0+\rangle\biggr)=0 .\label{tre-otto}
	\end{eqnarray}
First we notice that, differently from the SvH case, neither of these norms
can be chosen as we like but they are determined by the algebra given in
(\ref{tre-quattro}); second  we notice that these states turn out to be  of zero norm; 
this is the first sign that the gauge scalar product is not positive
definite.
Another problem arises when we evaluate the following scalar product
	\begin{eqnarray}	
	\langle +0|0-\rangle &=& \biggl(|0+\rangle,|0-\rangle\biggr)=\biggl(\hat{\bar{c}}|0-\rangle,|0-\rangle\biggr)=
	\biggl(|0-\rangle,\hat{\bar{c}}^{\dagger}|0-\rangle\biggr)=\nonumber\\
	&=& \biggl(|0-\rangle,\hat{\bar{c}}|0-\rangle\biggr)=\biggl(|0-\rangle,|0+\rangle\biggr)=\langle -0|0+\rangle. \label{tre-nove}
	\end{eqnarray}
This means that $\langle +0|0-\rangle$ and $\langle -0|0+\rangle$ are not determined by the algebra (\ref{tre-quattro})
and so we could choose them to be 1:
	\begin{equation}
	\langle +0|0-\rangle=\langle -0|0+\rangle=1 .\label{tre-dieci}
	\end{equation}
In this way the complete set of scalar products is the following one:
	\begin{equation}
	 \label{tre-quindici}
	 \left\{
		\begin{array}{l}
		\displaystyle 
		\biggl(|0+\rangle,|0+\rangle\biggr)=0
		\smallskip \nonumber\\
		\displaystyle
		\biggl(|0-\rangle,|0-\rangle\biggr)=0
		\smallskip \nonumber\\
		\displaystyle 
		\biggl(|0+\rangle,|0-\rangle\biggr)=1
		\smallskip \nonumber\\
		\biggl(|0-\rangle,|0+\rangle\biggr)=1.
		\end{array}
		\right.
	\end{equation}
Analogously to what we did in the SvH case let us now build the eigenstates 
\cite{footnote7} of $\hat{c}$ and $\hat{\bar{c}}$ which are respectively:
	\begin{equation}
	 \label{tre-sedici}
	 \left\{
		\begin{array}{l}
		\displaystyle 
		|\alpha -\rangle=e^{-\alpha\hat{\bar{c}}}|0-\rangle=|0-\rangle-\alpha|0+\rangle
		\smallskip \nonumber\\
		\displaystyle
		|\beta +\rangle=e^{-\beta \hat{c}}|0+\rangle=|0+\rangle-\beta|0-\rangle.
		\end{array}
		\right.
	\end{equation}
The gauge scalar product among these states can be easily worked out using
 (\ref{tre-quindici}):
	\begin{equation}
	 \label{tre-diciassette}
	 \left\{
		\begin{array}{l}
		\displaystyle 
		\biggl(|\alpha\pm\rangle,|\beta\pm\rangle\biggr)=\mp\delta(\alpha^*-\beta)
		\smallskip \nonumber\\
		\displaystyle
		\biggl(|\alpha\pm\rangle,|\beta\mp\rangle\biggr)=e^{\alpha^*\beta}.
		\end{array}
		\right.
	\end{equation}
These relations should be compared with those of the SvH case which 
are given in Eq. (\ref{due-quattordici-a}).

\par 
Next we should check whether we can find a resolution of the identity analogous to the SvH ones which were given in
Eq. (\ref{due-quindici}). In the gauge case the resolution of the identity is
	\begin{equation}
	-\int d\alpha|\alpha\pm\rangle\langle\pm\alpha^*|={\bf 1}. \label{tre-diciotto}
	\end{equation}
Note that the signs in the bra and ket are no longer ``twisted" as they were
in the SvH case.
The reader may be tempted to prove (\ref{tre-diciotto}) in the following manner. The $|\alpha +\rangle$ are eigenstates of
$\hat{\bar{c}}$ which, in the gauge scalar product, are Hermitian operators (\ref{tre-due}), and the eigenstates of a
Hermitian operator usually make up a basis, so (\ref{tre-diciotto}) should just be the completeness relation associated to
that basis. The same for $|\alpha -\rangle$ which are eigenstates of another Hermitian operator which is $\hat{c}$.
Actually the eigenstates of a Hermitian operator make up a basis only if the scalar product is positive definite and
this is no longer the case here. In fact we have states, like $|0+\rangle$ or $|0-\rangle$, which have zero norm
(\ref{tre-sette})-(\ref{tre-otto}). 
So we have to prove (\ref{tre-diciotto}) by testing it on all the states we are interested in, i.e. $|\beta+\rangle$
and $|\beta-\rangle$. First let us start with $|\beta +\rangle$ and the relation
	\begin{equation}
	-\int d\alpha|\alpha +\rangle\langle +\alpha^*|={\bf 1} . \label{tre-venti}
	\end{equation}
Applying the LHS of (\ref{tre-venti}) to $|\beta +\rangle$ we have:
	\begin{eqnarray}
	&&\displaystyle -\int d\alpha|\alpha+\rangle\langle +\alpha^*|\beta+\rangle=-\int d\alpha|\alpha
	+\rangle(-)\delta(\alpha-\beta)=\nonumber\\
	&&=\int d\alpha |0+\rangle-\alpha|0-\rangle)(\alpha-\beta)=|0+\rangle-\beta|0-\rangle=|\beta +\rangle
	\end{eqnarray}
where in the first step above we used the scalar product (\ref{tre-diciassette}).
Let us now do the same for $|\beta-\rangle$ and we have:
	\begin{eqnarray}
	&&\displaystyle -\int d\alpha|\alpha+\rangle\langle+\alpha^*|\beta-\rangle=-\int d\alpha|\alpha+\rangle \exp(\alpha
	\beta)=\nonumber\\
	&&=-\int d\alpha(|0+\rangle-\alpha|0-\rangle)(1+\alpha\beta)=|0-\rangle-\beta|0+\rangle=|\beta-\rangle
	\end{eqnarray}
which proves (\ref{tre-venti}). The proof is the same for 
	\begin{equation}
	-\int d\alpha|\alpha-\rangle\langle-\alpha^*|={\bf 1} \label{tre-venti-b}
	\end{equation}
but we will leave it to the reader because it is quite simple.

Let us now find out, using the resolution of the identity (\ref{tre-venti-b}), which is the expression
of the gauge scalar product between two generic states $|\psi\rangle$ and $|\Phi\rangle$. 
Using (\ref{tre-venti-b}) we can write 
	\begin{equation}
	|\psi\rangle=-\int d\alpha|\alpha-\rangle\langle-\alpha^*|\psi\rangle. \label{tre-ventuno}
	\end{equation}
As $\langle -\alpha^*|\psi\rangle$ is a function of $\alpha$, we can write it as
$\langle -\alpha^*|\psi\rangle=\psi_1+\alpha\psi_2$ and let us insert in (\ref{tre-ventuno}) the expression
(\ref{tre-sedici}) for $|\alpha-\rangle$; what we get is
	\begin{eqnarray}
	|\psi\rangle&=&-\int d\alpha|\alpha-\rangle(\psi_1+\alpha\psi_2)=
	-\int d\alpha(|0-\rangle-\alpha|0+\rangle)(\psi_1+\alpha\psi_2)=\nonumber\\
	&=&\psi_1|0+\rangle+\psi_2|0-\rangle.
	\end{eqnarray}
Doing the same for $\langle\Phi|$ we get that the scalar product between two states $|\Phi\rangle$ and $|\psi\rangle$ 
is
	\begin{eqnarray}
	\langle\Phi|\psi\rangle&=&\biggl(|\Phi\rangle,|\psi\rangle\biggr)=\biggl(\Phi_1|0+\rangle+\Phi_2|0-\rangle,
	\psi_1|0+\rangle+\psi_2|0-\rangle\biggr)=\nonumber\\
	&=&\Phi_1^*\psi_1\biggl(|0+\rangle,|0+\rangle\biggr)+\Phi^*_1\psi_2\biggl(|0+\rangle,|0-\rangle\biggr)+
	\Phi_2^*\psi_1\biggl(|0-\rangle,|0+\rangle\biggr)+\Phi_2^*\psi_2\biggl(|0-\rangle,|0-\rangle\biggr)=\nonumber\\
	&=&\Phi_1^*\psi_2+\Phi_2^*\psi_1.
	\label{tre-ventidue}
	\end{eqnarray}
In the last step we have used the relations (\ref{tre-quindici}). 
The norm of a state $\psi_1+\alpha\psi_2$, according to (\ref{tre-ventidue}), is
	\begin{equation}
	\langle \psi|\psi\rangle=\psi_1^*\psi_2+\psi_2^*\psi_1=2 \,\text{Re}\;\psi_1^*\psi_2 \label{tre-ventitre}
	\end{equation}
and it should be compared with the SvH norm shown in (\ref{due-venticinque}).
While this one is positive definite, the gauge one given in (\ref{tre-ventitre}) is not. 
For example the zero-forms  (i.e.
states like $\psi=\psi_1$) have zero norm and the same happens for the one-forms $\psi=\alpha\psi_2$.
It is also easy to build negative norm states like for example:
	\begin{equation}
	\psi=\psi_1-\psi_1\alpha ,\label{tre-ventiquattro}
	\end{equation}
for which $\|\psi\|^2=-2|\psi_1|^2$. Having worked out all the details for the case of one Grassmannian
variable, we should now turn to two. This is the case of interest for the CPI. In fact, for
one degree of freedom, we have two Grassmannian variables ($c^q,c^p$) which become $2n$ for $n$ degrees of freedom.

We can proceed along the same lines we followed for the SvH case: we derive all the scalar products from 
one choice of normalization, for example $\biggl(|0-,0-\rangle,|0+,0+\rangle\biggr)=i$, 
plus the hermiticity conditions (\ref{tre-due}) and the commutation relations (\ref{tre-tre-a}) 
among the Grassmannian variables. For example: 
	\begin{eqnarray}
	&&\biggl(|0+,0+\rangle,|0-,0-\rangle\biggr)=\biggl(-\bar{c}_q\bar{c}_p|0-,0-\rangle,c^qc^p|0+,0+\rangle\biggr)=\nonumber\\
	&&=\biggl(|0-,0-\rangle,-\bar{c}_p\bar{c}_qc^qc^p|0+,0+\rangle\biggr)=
	\biggl(|0-,0-\rangle,-[\bar{c}_q,c^q][\bar{c}_p,c^p]|0+,0+\rangle\biggr)=\nonumber\\
	&&=-\biggl(|0-,0-\rangle,|0+,0+\rangle\biggr)=-i .
	\end{eqnarray}
In the steps above we used the fact that the states $|+\rangle$ are Grassmannian even while the 
states $|-\rangle$ are Grassmannian odd. Using the same kind of calculation it is possible to prove that the only non-zero 
scalar products are:
	\begin{equation}
	\label{tre-ventinove}
	 \left\{
		\begin{array}{l}
		\displaystyle 
		\biggl(|0-,0-\rangle,|0+,0+\rangle\biggr)=i
		\smallskip\nonumber\\
		\biggl(|0+,0+\rangle,|0-,0-\rangle\biggr)=-i
		\smallskip\nonumber\\
		\biggl(|0-,0+\rangle,|0+,0-\rangle\biggr)=-i
		\smallskip \nonumber\\
		\biggl(|0+,0-\rangle,|0-,0+\rangle\biggr)=i.
		\end{array}
		\right.
	\end{equation}
Then, as we did in the SvH case, we can build the following states 
	\begin{equation}
	|\alpha_q-,\alpha_p-\rangle\equiv \exp[-\alpha_q\hat{\bar{c}}_q-\alpha_p\hat{\bar{c}}_p]|0-,0-\rangle \label{tre-trenta}
	\end{equation}
which, like in the SvH case, are eigenstates of $\hat{c}^q$ and $\hat{c}^p$ with eigenvalues $\alpha_q,\alpha_p$
	\begin{equation}
	 \label{tre-trentuno}
	 \left\{
		\begin{array}{l}
		\displaystyle 
		\hat{c}^q|\alpha_q-,\alpha_p-\rangle=\alpha_q|\alpha_q-,\alpha_p-\rangle
		\smallskip \nonumber\\
		\displaystyle
		\hat{c}^p|\alpha_q-,\alpha_p-\rangle=\alpha_p|\alpha_q-,\alpha_p-\rangle.
		\end{array}
		\right.
	\end{equation}
The proof is the same as in the SvH case because it is based only on the commutation relations which are the same in both
cases. 

Analogously we can build the states 
	\begin{equation}
	|\beta_q+,\beta_p+\rangle=\exp[-\beta_q\hat{c}^q-\beta_p\hat{c}^p]|0+,0+\rangle \label{tre-trentadue}
	\end{equation}
which are eigenstates of $\hat{\bar{c}}_q,\hat{\bar{c}}_p$
	\begin{equation}
	 \label{tre-trentatre}
	 \left\{
		\begin{array}{l}
		\displaystyle 
		\hat{\bar{c}}_q|\beta_q+,\beta_p+\rangle=\beta_q|\beta_q+,\beta_p+\rangle
		\smallskip \nonumber\\
		\displaystyle
		\hat{\bar{c}}_p|\beta_q+,\beta_p+\rangle=\beta_p|\beta_q+,\beta_p+\rangle
		\end{array}
		\right.
	\end{equation}
and also the states 
	\begin{equation}
	 \label{tre-trentaquattro}
	 \left\{
		\begin{array}{l}
		\displaystyle 
		|\alpha_q-,\beta_p+\rangle=\exp[-\alpha_q\hat{\bar{c}}_q-\beta_p\hat{c}^p]|0-,0+\rangle
		\smallskip \nonumber\\
		\displaystyle
		|\beta_q+,\alpha_p-\rangle=\exp[-\beta_q\hat{c}^q-\alpha_p\hat{\bar{c}}_p]|0+,0-\rangle
		\end{array}
		\right.
	\end{equation}
which are respectively eigenstates of the following operators
	\begin{equation}
	 \label{tre-trentasei}
	 \left\{
		\begin{array}{l}
		\displaystyle 
		\hat{c}^q|\alpha_q-,\beta_p+\rangle=\alpha_q|\alpha_q-,\beta_p+\rangle
		\smallskip \nonumber\\
		\displaystyle
		\hat{\bar{c}}_p|\alpha_q-,\beta_p+\rangle=\beta_p|\alpha_q-,\beta_p+\rangle
		\end{array}
		\right.
	\end{equation}
	\begin{equation}
	 \label{tre-trentasette}
	 \left\{
		\begin{array}{l}
		\displaystyle 
		\hat{\bar{c}}_q|\beta_q+,\alpha_p-\rangle=\beta_q|\beta_q+,\alpha_p-\rangle
		\smallskip \nonumber\\
		\displaystyle
		\hat{c}^p|\beta_q+,\alpha_p-\rangle=\alpha_p|\beta_q+,\alpha_p-\rangle.
		\end{array}
		\right.
	\end{equation}
Next, we can calculate the gauge scalar products among these states using (\ref{tre-ventinove}) and 
the hermiticity relations (\ref{tre-due}). The results \cite{Appendix B} are
	\begin{eqnarray}
	\label{tre-trentotto}
	&&\biggl(|\alpha_q-,\alpha_p-\rangle,|\beta_q+,\beta_p+\rangle\biggr)=i\cdot \exp(\alpha_q^*\beta_q+\alpha_p^*\beta_p)\nonumber\\
	&&\biggl(|\alpha_q-,\alpha_p-\rangle,|\alpha_q^{\prime}-,\beta_p+\rangle\biggr)=i\cdot \delta(\alpha_q^*-\alpha_q^{\prime})
	\exp(\alpha_p^*\beta_p)\nonumber\\
	&&\biggl(|\alpha_q-,\alpha_p-\rangle,|\beta_q+,\alpha_p^{\prime}-\rangle\biggr)=i\cdot \delta(\alpha_p^*-\alpha_p^{\prime})
	\exp(\alpha_q^*\beta_q)\nonumber\\
	&&\biggl(|\alpha_q-,\alpha_p-\rangle,|\alpha_q^{\prime}-,\alpha_p^{\prime}-\rangle\biggr)=i\cdot
	\delta(\alpha_q^*-\alpha_q^{\prime})\delta
	(\alpha_p^*-\alpha_p^{\prime})\\
	&&\biggl(|\beta_q+,\beta_p+\rangle,|\alpha_q-,\beta_p^{\prime}+\rangle\biggr)=i\cdot\delta(\beta_p^*-\beta_p^{\prime})\exp(\beta_q^*\alpha_q)
	\nonumber\\
	&&\biggl(|\beta_q+,\beta_p+\rangle,|\beta_q^{\prime}+,\alpha_p-\rangle\biggr)=i\cdot\delta(\beta_q^{\prime}-\beta_q^*)\exp(\beta_p^*\alpha_p)
	\nonumber\\
	&&\biggl(|\beta_q+,\beta_p+\rangle,
	|\beta_q^{\prime}+,\beta_p^{\prime}+\rangle\biggr)=i\cdot\delta(\beta_q^*-\beta_q^{\prime})\delta(\beta_p^*-\beta_p^{\prime})\nonumber\\
	&&\biggl(|\beta_q+,\beta_p+\rangle,|\alpha_q-,\alpha_p-\rangle\biggr)=-i\cdot
	\exp(\beta_q^*\alpha_q+\beta_p^*\alpha_p)\nonumber.
	\end{eqnarray}
Via these scalar products it is easy \cite{Appendix B} to prove the following resolutions of the identity:
	\begin{equation}
	i\int d\alpha_q d\alpha_p|\alpha_q\pm,\alpha_p\pm\rangle\langle\pm\alpha_p^*,\pm\alpha_q^*|={\bf 1} \label{tre-trentanove}
	\end{equation}
which should be compared with the SvH ones (\ref{due-quarantuno}). 

From (\ref{tre-trentanove})  we can derive the scalar product between two
generic states as we have done in the SvH case (see 
Eqs. (\ref{due-quarantotto})-(\ref{svhprod})). Let us first extend
(\ref{tre-trentanove}) to include the variables $\varphi$ or $\lambda$ as in
(\ref{due-cinquantasei-a})-(\ref{due-cinquantasei-b}). 
Written in terms of $c$ and $\bar{c}$ these relations are \cite{footnote8} 
	\begin{eqnarray}
	&&i\int dqdpdc^qdc^p|q,p,c^q-,c^p-\rangle\langle-c^{p*},-c^{q*},p,q|={\bf 1} \label{tre-quaranta-a}\\
	&&\displaystyle i\int d\lambda_q d\lambda_p d\bar{c}_qd\bar{c}_p|\lambda_q,\lambda_p,\bar{c}_q+,\bar{c}_p+\rangle
	\langle+\bar{c}_p^*,+\bar{c}_q^*,\lambda_p,\lambda_q|={\bf 1} \label{tre-quaranta-b}
	\end{eqnarray}
Inserting the completeness relation (\ref{tre-quaranta-a}) into the scalar product
$\langle\psi|\Phi\rangle$ we obtain:
	\begin{eqnarray}
	\label{reality-1}
	\langle\psi|\Phi\rangle&=&i\int dqdpdc^qdc^p\langle\psi|q,p,c^q-,c^p-\rangle
	\cdot \langle-c^{p*},-c^{q*},p,q|\Phi\rangle=\nonumber\\
	&=&i\int dq dp dc^qdc^p\psi_+^*(q,p,c^q,c^p)\Phi_-(q,p,c^q,c^p) 
	\end{eqnarray}
where 
	\begin{equation}
	 \left\{
		\begin{array}{l}
		\displaystyle 
		\psi_+(q,p,c^q,c^p)\equiv\langle -c^p,-c^q,p,q|\psi\rangle
		\smallskip \nonumber\\
		\displaystyle
		\Phi_-(q,p,c^q,c^p)\equiv\langle
		-c^{p*},-c^{q*},p,q|\Phi\rangle\equiv\Phi_0+\Phi_qc^q+\Phi_pc^p+\Phi_2c^qc^p\nonumber.
		\end{array}
		\right.
	\end{equation}
Note that:
	\begin{eqnarray}
	&&\psi_+(q,p,c^q,c^p)=\langle -c^p,-c^q, p,q|\psi\rangle=\nonumber\\
	&&=i\int dq^{\prime}dp^{\prime}dc^{q\prime}dc^{p\prime}\langle -c^p,-c^q,p,q|q^{\prime},p^{\prime},c^{q\prime}-,
	c^{p\prime}-\rangle\langle -c^{p*\prime},-c^{q*\prime},p^{\prime}, q^{\prime}|\psi\rangle=\nonumber\\
	&& =\int dc^{q\prime}dc^{p\prime}
	\delta(c^{p\prime}-c^{p*})\delta(c^{q\prime}-c^{q*})
	\cdot(\psi_0+\psi_qc^{q\prime}+\psi_pc^{p\prime}+\psi_2c^{q\prime}c^{p\prime})=\nonumber\\
	&&=\psi_0+\psi_qc^{q*}+\psi_pc^{p*}-\psi_2c^{p*}c^{q*}.
	\end{eqnarray}
Therefore the complex conjugate of $\psi_+(q,p,c^q,c^p)$ is
	\begin{equation}
	\psi^*_+(q,p,c^q,c^p)=\psi_0^*+\psi_q^*c^{q}+\psi_p^*c^{p}-\psi_2^*c^{q}c^{p} \label{reality0}
	\end{equation}
and inserting (\ref{reality0}) into (\ref{reality-1}) we obtain:
	\begin{eqnarray}
	\langle \psi|\Phi\rangle&=&i\int d\varphi dc^qdc^p[\psi_0^*+c^q\psi_q^*+c^p\psi_p^*-c^qc^p\psi_2^*]\cdot
	[\Phi_0+\Phi_qc^q+\Phi_pc^p+\Phi_2c^qc^p]=\nonumber\\ 
	&=&i\int d\varphi\int
	dc^qdc^p[\psi_0^*\Phi_2c^qc^p+c^q\psi_q^*\Phi_pc^p+c^p\psi_p^*\Phi_qc^q-c^qc^p\psi_2^*\Phi_0]=\nonumber\\
	&=&i\int d\varphi[\psi_2^*\Phi_0-\Phi_2\psi^*_0+\psi_p^*\Phi_q-\psi_q^*\Phi_p] .\label{tre-ventotto-x}
	\end{eqnarray}
The presence of the factor $i$ in (\ref{tre-ventotto-x}) is crucial in order to have a real norm. In fact:
	\begin{eqnarray}
	\langle\psi|\psi\rangle&=&i\int d\varphi\,[\psi_2^*\psi_0-\psi_2\psi_0^*+\psi_p^*\psi_q-\psi_q^*\psi_p]=\nonumber\\
	&=&2\,\text{Im}\int d\varphi[\psi_2\psi_0^*+\psi_q^*\psi_p] .\label{mart}
	\end{eqnarray}
From (\ref{mart}) we see
that this is not a positive definite scalar product. Moreover the zero-forms have zero-norm. This means that the scalar
product of
KvN for the zero-forms is definitely not of the gauge type because the zero-forms in KvN had positive
norm. 

The resolutions of the identity (\ref{tre-quaranta-a})-(\ref{tre-quaranta-b}) are also useful in the derivation of the path
integral as we did in (\ref{due-sessantadue}). There we proved that the SvH
resolutions of the identity (\ref{due-cinquantasei-a}) and (\ref{due-cinquantasei-b}) 
turned the operatorial formalism into the CPI path
integral. Let us now see what happens for the gauge scalar product. 
In this case the transition amplitude $K(f|i)$ between an initial configuration
$(\varphi_i,c_{q_i},c_{p_i})$ and a final one $(\varphi_f,c_{q_f},c_{p_f})$ is
	\begin{equation}
	\displaystyle K(f|i)=\langle -c_{p_f}^{*},-c_{q_f}^{*}, \varphi_f|e^{-i\widetilde{\cal H}(t_f-t_i)}
	|\varphi_i,c_{q_i}-,c_{p_i}-\rangle.
	\label{tre-quarantuno}
	\end{equation}
Note that the difference with respect to (\ref{due-cinquantasette}) is that the final bra is 
$\langle-c_{p_f}^{*},-c_{q_f}^{*},\varphi_f|$
and not $\langle +c_{p_f}^{*},+c_{q_f}^{*},\varphi_f|$. The reason is that the dual of a ket $|c_q-,c_p-\rangle$ 
in the SvH case 
is a bra of the form $\langle +c_p^*,+c_q^*|$ while in the gauge case is a bra of the form $\langle
-c^*_p,-c^*_q|$ as it is clear from the relations (\ref{due-cinquantasei-b}) and (\ref{tre-quaranta-a}).
Another more elementary explanation is that the bra $\langle-c_{p_f}^*,-c_{q_f}^*,\varphi_f|$ is an eigenstate of 
$\hat{c}^q,\hat{c}^p$ with eigenvalues
$c_{q_f},c_{p_f}$ as it is easy to see from the following. Let us start from the relations:
	\begin{equation}
	 \label{tre-quarantuno-a}
	 \left\{
		\begin{array}{l}
		\displaystyle 
		\hat{c}^q|\varphi_f,c_{q_f}^*-,c_{p_f}^*-\rangle=c^*_{q_f}|\varphi_f,c^*_{q_f}-,c^*_{p_f}-\rangle
		\smallskip \nonumber\\
		\displaystyle
		\hat{c}^p|\varphi_f,c_{q_f}^*-,c_{p_f}^*-\rangle=c^*_{p_f}|\varphi_f,c^*_{q_f}-,c^*_{p_f}-\rangle
		\end{array}
		\right.
	\end{equation}
and let us do the Hermitian conjugation of (\ref{tre-quarantuno-a}) according to the gauge scalar product. 
We get
	\begin{equation}
	 \label{tre-quarantuno-b}
	 \left\{
		\begin{array}{l}
		\displaystyle 
		\langle-c^*_{p_f},-c^*_{q_f},\varphi_f|\hat{c}^q=\langle-c^*_{p_f},-c^*_{q_f},\varphi_f|c_{q_f}
		\smallskip \nonumber\\
		\displaystyle
		\langle-c^*_{p_f},-c^*_{q_f},\varphi_f|\hat{c}^p=\langle-c^*_{p_f},-c^*_{q_f},\varphi_f|c_{p_f}
		\end{array}
		\right.
	\end{equation}
that is what we wanted to prove. 

\par
Let us now proceed by slicing the time interval in (\ref{tre-quarantuno}) into $N$ intervals
of length $\epsilon$ like in (\ref{due-cinquantotto}) and inserting a relation like (\ref{tre-quaranta-b}) on the left of 
each operator
$\displaystyle e^{-i\widetilde{\cal H}\epsilon}$ and one like (\ref{tre-quaranta-a}) on the right. What we obtain is
	\begin{eqnarray}
	\displaystyle
	\label{tre-quarantadue}
	&&K(f|i)=\langle
	-c_{p_f}^*,-c_{q_f}^*,p_f,q_f|\biggl\{i\int d\lambda_{q_N} d\lambda_{p_N} 
	d\bar{c}_{q_N} d\bar{c}_{p_N}|
	\lambda_{q_N},\lambda_{p_N},\bar{c}_{q_N}+,\bar{c}_{p_N}+\rangle\cdot\nonumber\\
	&&\cdot\langle +\bar{c}_{p_N}^*,+\bar{c}_{q_N}^*,\lambda_{p_N},\lambda_{q_N}|\biggr\}\exp[-i\epsilon\widetilde{\cal
	H}]\biggl\{i
	\int dq_{\scriptscriptstyle N}dp_{\scriptscriptstyle N}dc_{\scriptscriptstyle N}^qdc_{\scriptscriptstyle N}^p
	|q_{\scriptscriptstyle N},p_{\scriptscriptstyle N},c^q_{\scriptscriptstyle N}-,c^p_{\scriptscriptstyle N}-
	\rangle\nonumber\\
	&&\cdot\langle -c^*_{p_N},-c^*_{q_N},p_{\scriptscriptstyle N},q_{\scriptscriptstyle N}|\biggr\}\cdot\biggl\{i\int
	d\lambda_{q_{N-1}}d\lambda_{p_{N-1}}d\bar{c}_{q_{N-1}}d\bar{c}_{p_{N-1}}
	|\lambda_{q_{N-1}},\lambda_{p_{N-1}},\bar{c}_{q_{N-1}}+,\bar{c}_{p_{N-1}}+\rangle\nonumber\\
	&&\cdot\langle +\bar{c}_{p_{N-1}}^*,
	+\bar{c}^*_{q_{N-1}},\lambda_{p_{N-1}},\lambda_{q_{N-1}}|\biggr\}\exp[-i\epsilon\widetilde{\cal H}]\biggl\{
	i\int dq_{\scriptscriptstyle{N-1}}dp_{\scriptscriptstyle{N-1}}dc^q_{\scriptscriptstyle
	{N-1}}dc^p_{\scriptscriptstyle {N-1}}
	\ldots\biggr\}\ldots\nonumber\\
	&&\cdot \biggl\{i\int
	dq_{\scriptscriptstyle 1}dp_{\scriptscriptstyle 1}dc^q_{\scriptscriptstyle 1}dc^p_{\scriptscriptstyle 1}|
	q_{\scriptscriptstyle 1},p_{\scriptscriptstyle 1},c^q_{\scriptscriptstyle 1}-,c^p_{\scriptscriptstyle 1}-
	\rangle\langle-c^{*}_{p_1},-c^{*}_{q_1},p_{\scriptscriptstyle 1},q_{\scriptscriptstyle
	1}|\biggr\}|q_i,p_i,c_{q_i}-,
	c_{p_i}-\rangle .
	\end{eqnarray}
In the expression above the last term, using (\ref{tre-trentotto}), gives
	\begin{equation}
	\langle
	-c^{*}_{p_1},-c^{*}_{q_1},
	p_{\scriptscriptstyle 1},q_{\scriptscriptstyle 1}|q_i,p_i,c_{q_i}-,c_{p_i}-\rangle=
	i\delta(q_{\scriptscriptstyle 1}-q_i)\delta(p_{\scriptscriptstyle 1}-p_i)\delta(c_{q_1}-
	c_{q_i})\delta(c_{p_1}-c_{p_i}).
	\label{en1}
	\end{equation}
The Dirac deltas above, via the integration in $q_1,p_1,c_{q_1},c_{p_1}$,
identify $q_{\scriptscriptstyle 1}=q_i$, $p_{\scriptscriptstyle 1}=p_i$, $c_{q_1}=c_{q_i}$ and
$c_{p_1}=c_{p_i}$. The other scalar products in (\ref{tre-quarantadue}) are of the type
	\begin{eqnarray}
	&&\langle -c^{*}_{p_j},-c^{*}_{q_j},p_j,q_j|\lambda_{q_{j-1}},\lambda_{p_{j-1}},\bar{c}_{q_{j-1}}+,
	\bar{c}_{p_{j-1}}+\rangle=\nonumber\\
	&&=i \,\exp\bigl[ip_j\lambda_{p_{j-1}}+iq_j\lambda_{q_{j-1}}\bigr]\cdot \exp\bigl[c^q_j\bar{c}_{q_{j-1}}+c^p_j\bar{c}_{p_{j-1}}\bigr]
	\label{en2}
	\end{eqnarray}
where again we used the gauge scalar product (\ref{tre-trentotto}). The last terms we have to evaluate in
(\ref{tre-quarantadue}) have the form
	\begin{equation}
	\displaystyle \langle +\bar{c}^{*}_{p_j},+\bar{c}^{*}_{q_j},\lambda_{p_j},\lambda_{q_j}|
	e^{-i\epsilon\widetilde{\cal H}}|q_j,p_j,c^q_j-,c^p_j-\rangle.
	\end{equation}
for which it is necessary to work out $\displaystyle \langle
+\bar{c}^{*}_{p_j},+\bar{c}^{*}_{q_j},\lambda_{p_j},\lambda_{q_j}|
\widetilde{\cal H}|q_j,p_j,c^q_j-,c^p_j-\rangle$.
For the bosonic part of $\widetilde{\cal H}$, via (\ref{tre-trentotto}), we get: 
	\begin{eqnarray}
	&&\langle+\bar{c}_{p_j}^{*},+\bar{c}_{q_j}^{*}+,\lambda_{p_j},\lambda_{q_j}|\hat{\lambda}_q\hat{p}-\hat{\lambda}_pV^{\prime}
	(\hat{q})|
	q_j,p_j,c^q_j-,c^p_j-\rangle=\nonumber\\
	&&=-i(\lambda_{q_j}p_j-\lambda_{p_j}V^{\prime}(q_j))\exp[-i\lambda_{q_j}q_j-
	i\lambda_{p_j}p_j+\bar{c}_{q_j}c^q_j+\bar{c}_{p_j}c^{p}_j]
	\end{eqnarray}
and for the Grassmannian part of $\widetilde{\cal H}$  we obtain
	\begin{eqnarray}
	&&\langle
	+\bar{c}_{p_j}^{*},+\bar{c}_{q_j}^{*},\lambda_{p_j},\lambda_{q_j}|i\hat{\bar{c}}_q\hat{c}^p-i\hat{\bar{c}}_pV^{\prime\prime}
	\hat{c}^q|
	q_j,p_j,c_j^q-,c^p_j-\rangle=\nonumber\\
	&&=(-i)(i\bar{c}_{q_j}c_j^p-i\bar{c}_{p_j}V^{\prime\prime}c^q_j)\exp[-i\lambda_{q_j}q_j-i\lambda_{p_j}p_j+
	\bar{c}_{q_j}c^q_j+\bar{c}_{p_j}c^p_j].
	\end{eqnarray}
Combining the bosonic and fermionic part we end up with 
	\begin{eqnarray}
	\displaystyle &&\langle +\bar{c}_{p_j}^{*},+\bar{c}_{q_j}^{*},\lambda_{p_j},\lambda_{q_j}|e^{-i\epsilon\widetilde{\cal
	H}}|q_j,p_j,c_j^q-,c_j^p-\rangle=\nonumber\\
	&&=-i\cdot \exp\bigl[-i\epsilon\widetilde{\cal H}(j)\bigr]\exp[-i\lambda_{q_j}q_j-i\lambda_{p_j}p_j+
	\bar{c}_{q_j}c^q_j+\bar{c}_{p_j}c^p_j].
	\end{eqnarray}
The $N$ factors ``$-i$" which appear in the expression above will get compensated by the $N$ factors ``$i$"
which appear in the completeness relations (\ref{tre-quaranta-b})
we used. The $N$ factors ``$+i$"  of the completeness relations (\ref{tre-quaranta-a}) can be combined with the $N$ factors ``$+i$"
contained in the scalar products (\ref{en2}). The $N$ ``$-$" signs which one obtains in this way  can be
absorbed by turning around the integration in $c^q$ and $c^p$:
	\begin{equation}
	-\int dq_jdp_jdc^q_jdc^p_j=\int dq_jdp_jdc^p_jdc^q_j.
	\end{equation}
So, combining all the pieces, the final result is \cite{footnote8bis}
	\begin{equation}
	K(f|i)=i \int {\mathscr D}\mu \;\exp\biggl[i\epsilon\biggl(\sum_{j=1}^N\lambda_j\cdot\frac{\varphi_{j+1}-\varphi_j}{\epsilon}+
	i\bar{c}_j\frac{c_{j+1}-c_j}{\epsilon}-\widetilde{\cal H}(j)\biggr)\biggr] \label{tre-quarantadue-b}
	\end{equation}
where the boundary conditions are:
	\begin{equation}
	\varphi_0=\varphi_i,\;\;\;\;\;\;\varphi_{\scriptscriptstyle N+1}=\varphi_f,\;\;\;\;\;\;c_0=c_i,\;\;\;\;\;\;
	c_{\scriptscriptstyle{N+1}}=c_f
	\end{equation}
and the measure is
	\begin{equation}
	\displaystyle {\mathscr D}\mu=\biggl(\prod_{j=2}^Nd\varphi_jd\lambda_jd\bar{c}_{q_j}d\bar{c}_{p_j}dc^p_jdc^q_j\biggr)
	d\lambda_1d\bar{c}_{q_1}d\bar{c}_{p_1}
	\end{equation}
which is identical to the measure of the SvH case in (\ref{due-sessantatre}). In that case the exchange between 
$c^q$ and $c^p$ in the
integration measure took place already in the completeness relation (\ref{due-cinquantasei-b}). Both measures can be turned, without any sign
change, into the following one:
	\begin{equation}
	\displaystyle {\mathscr
	D}\mu=d\lambda_{q_1}d\lambda_{p_1}d\bar{c}_{q_1}d\bar{c}_{p_1}
	\prod_{j=2}^Ndq_jdp_jd\lambda_{q_j}d\lambda_{p_j}dc_j^qd\bar{c}_{q_j}
	dc^p_jd\bar{c}_{p_j}
	\end{equation}
which is the one which originally appeared in the CPI \cite{Gozzi}. 
The discretized Lagrangian appearing in (\ref{tre-quarantadue-b}) goes into the usual $\widetilde{\cal L}$ 
of (\ref{uno-tredici}) in the continuum limit.

All this confirms that also via the gauge scalar product we
can reproduce the CPI path integral. This concludes this section which we can summarize by saying that, 
in the gauge scalar
product, $\widetilde{\cal H}$ {\it is Hermitian}, the path integral \cite{Gozzi} can be obtained from the operatorial formalism but {\it the
scalar product is not positive definite}. 

\section{The Symplectic Scalar Product}
\noindent
The gauge scalar product which we explored in the previous section is not the only one under which $\widetilde{\cal H}$
is Hermitian. In this section we will explore another one which has the same feature and whose hermiticity conditions are:
	\begin{equation}
	 \label{quattro-uno}
	 \left\{
		\begin{array}{l}
		\displaystyle 
		(\hat{c}^a)^{\dagger}=i\omega^{ab}\hat{\bar{c}}_b
		\smallskip \nonumber\\
		\displaystyle
		(\hat{\bar{c}}_d)^{\dagger}=i\omega_{df}\hat{c}^f
		\end{array}
		\right.
	\end{equation}
	\begin{equation}
	 \label{quattro-due}
	 \left\{
		\begin{array}{l}
		\displaystyle 
		\hat{\varphi}^{a\dagger}=\hat{\varphi}^a
		\smallskip \nonumber\\
		\displaystyle
		\hat{\lambda}_a^{\dagger}=\hat{\lambda}_a.
		\end{array}
		\right.
	\end{equation}
Under these conditions the bosonic part of $\widetilde{\cal H}$ turns out to be Hermitian as it was in the SvH case
(\ref{due-cinquantuno-b}). If we take $\displaystyle H=\frac{p^2}{2}+V(q)$, 
the fermionic part (\ref{due-cinquantuno-xx}) can be written
as\break $\widetilde{\cal H}_{ferm}=i\bar{c}_qc^p-i\bar{c}_pV^{\prime\prime}c^q$ where $\displaystyle
V^{\prime\prime}=\frac{\partial^2V}{\partial q^2}$. Applying the hermiticity conditions (\ref{quattro-uno}) we get
	\begin{eqnarray}
	\widetilde{\cal
	H}^{\dagger}_{ferm}&=&(i\bar{c}_qc^p)^{\dagger}-(i\bar{c}_pV^{\prime\prime}c^q)^{\dagger}=
	-ic^{p\dagger}\bar{c}_q^{\dagger}+i{c^q}^{\dagger}V^{\prime\prime}\bar{c}_p^{\dagger}=\nonumber\\
	&=&-i(-i\bar{c}_q)(-ic^p)+i\cdot i\bar{c}_pV^{\prime\prime}ic^q=
	i\bar{c}_qc^p-i\bar{c}_pV^{\prime\prime}c^q=\widetilde{\cal H}_{ferm}.
	\label{quattro-tre}
	\end{eqnarray}
We call ``symplectic" the  scalar product which produces the hermiticity condition above. The name is due
to the presence of the symplectic matrix $\omega^{ab}$ in (\ref{quattro-uno}).

As we did in the SvH case, let us now proceed by constructing a decomposition of the identity. This will help us
find out the expression of the symplectic scalar product in terms of the components of the wave functions, as we did in
(\ref{due-cinquantuno}) for the SvH case and in (\ref{tre-ventotto-x}) for the gauge case. The
vector space is the same as in the SvH and the gauge cases and is spanned by
the four states $\bigl\{|0-,0-\rangle$, $|0-,0+\rangle$, $|0+,0-\rangle$, $|0+,0+\rangle\bigr\}$ defined in the usual way:
	\begin{equation}
	 \label{quattro-quattro}
	 \left\{
		\begin{array}{l}
		\displaystyle 
		c^q|0-,0-\rangle=c^p|0-,0-\rangle=0
		\smallskip \nonumber\\
		\displaystyle
		|0+,0-\rangle=\bar{c}_q|0-,0-\rangle
	 \smallskip \nonumber\\
	 \displaystyle 
	 |0-,0+\rangle=-\bar{c}_p|0-,0-\rangle
	 \smallskip \nonumber\\
	 |0+,0+\rangle =\bar{c}_q|0-,0+\rangle=\bar{c}_p\bar{c}_q|0-,0-\rangle.
		\end{array}
		\right.
	\end{equation}
As in the previous cases we can choose how to normalize one of 
these states. This time the normalization we choose is
	\begin{equation}
	\biggl(|0+,0+\rangle,|0+,0+\rangle\biggr)=1 , \label{quattro-quattro-a}
	\end{equation}
and from this and (\ref{quattro-quattro}) one can easily \cite{Appendix C}  obtain that the only non-zero scalar
products are
	 \begin{equation}
	 \label{quattro-cinque}
	 \left\{
		\begin{array}{l}
		\displaystyle
		\biggl(|0-,0+\rangle,|0+,0-\rangle\biggr)=i
	 \smallskip \nonumber\\
	 \displaystyle 
	 \biggl(|0+,0-\rangle,|0-,0+\rangle\biggr)=-i
	 \smallskip \nonumber\\
	 \biggl(|0-,0-\rangle,|0-,0-\rangle\biggr)=-1.
		\end{array}
		\right.
	\end{equation}
In (\ref{quattro-cinque}) we notice the presence of negative norm states like $|0-,0-\rangle$.
The states which are needed in the decomposition of the identity are the usual ones we have already built 
in the SvH and gauge cases:
	\begin{equation}
	 \label{quattro-sei}
	 \left\{
		\begin{array}{l}
		\displaystyle  
		|\alpha_q-,\alpha_p-\rangle\equiv e^{-\alpha_q\hat{\bar{c}}_q-\alpha_p\hat{\bar{c}}_p}|0-,0-\rangle
		\smallskip \nonumber\\
		\displaystyle
		|\beta_q+,\alpha_p-\rangle\equiv e^{-\beta_q\hat{c}^q-\alpha_p\hat{\bar{c}}_p}|0+,0-\rangle
	 \smallskip \nonumber\\
	 \displaystyle 
	 |\alpha_q-,\beta_p+\rangle\equiv e^{-\alpha_q\hat{\bar{c}}_q-\beta_p\hat{c}^p}|0-,0+\rangle
	 \smallskip \nonumber\\
	 |\beta_q+,\beta_p+\rangle\equiv e^{-\beta_q\hat{c}^q-\beta_p\hat{c}^p}|0+,0+\rangle.
		\end{array}
		\right.
	\end{equation}
Using (\ref{quattro-quattro-a})-(\ref{quattro-cinque}) and the commutation relations (\ref{uno-quattordici-b}) we can easily
\cite{Appendix C} derive the scalar products among the states (\ref{quattro-sei}). They are
	\begin{eqnarray}
	\label{quattro-sette}
	&&\biggl(|\alpha_q-,\alpha_p-\rangle,|\alpha_q^{\prime}-,\alpha_p^{\prime}-\rangle\biggr)=
	-\exp(-i\alpha_q^*\alpha_p^{\prime}+i\alpha_p^*\alpha_q^{\prime})\nonumber\\
	&&\biggl(|\alpha_q-,\beta_p+\rangle,|\beta_q+,\beta_p^{\prime}+\rangle\biggr)=
	i\delta(\beta_p^{\prime}+i\alpha_q^*)\exp(i\beta_q\beta_p^*)\nonumber\\
	&&\biggl(|\alpha_q-,\alpha_p-\rangle,|\beta_q+,\alpha_p^{\prime}-\rangle\biggr)=
	\delta(\beta_q-i\alpha_p^*)\exp(i\alpha_p^{\prime}\alpha_q^*)\nonumber\\
	&&\biggl(|\alpha_q-,\alpha_p-\rangle,|\beta_q+,\beta_p+\rangle\biggr)=\delta(\beta_q-i\alpha^*_p)
	\delta(\beta_p+i\alpha_q^*)\nonumber\\
	&&\biggl(|\alpha_q-,\alpha_p-\rangle,|\alpha_q^{\prime}-,\beta_p+\rangle\biggr)=
	-\delta(\beta_p+i\alpha_q^*)\exp(-i\alpha_q^{\prime}\alpha_p^*)\nonumber\\
	&&\biggl(|\beta_q+,\alpha_p-\rangle,|\beta_q^{\prime}+,\beta_p+\rangle\biggr)=
	-i\delta(\beta_q^{\prime}-i\alpha_p^*)\exp(-i\beta_p\beta_q^*)\\
	&&\biggl(|\beta_q+,\alpha_p-\rangle,|\alpha_q-,\beta_p+\rangle\biggr)=
	-i\exp(i\beta_q^*\beta_p+i\alpha_p^*\alpha_q)\nonumber\\
	&&\biggl(|\beta_q+,\alpha_p-\rangle,|\beta_q^{\prime}+,\alpha_p^{\prime}-\rangle\biggr)=
	\delta(\alpha_p^*+i\beta_q^{\prime})\delta(\alpha_p^{\prime}-i\beta_q^*)\nonumber\\
	&&\biggl(|\alpha_q-,\beta_p+\rangle,|\alpha_q^{\prime}-,\beta_p^{\prime}+\rangle\biggr)=
	\delta(\alpha_q^{\prime}+i\beta_p^*)\delta(i\beta_p^{\prime}-\alpha_q^*)\nonumber\\
	&&\biggl(|\alpha_q^{\prime}-,\alpha_p^{\prime}-\rangle,|\alpha_q-,\alpha_p-\rangle\biggr)=
	-\exp(-i\alpha_q^{\prime*}\alpha_p+i\alpha_p^{\prime*}\alpha_q)\nonumber\\
	&&\biggl(|\beta_q+,\beta_p+\rangle,|\beta_q^{\prime}+,\beta_p^{\prime}+\rangle\biggr)=
	\exp(i\beta_q^*\beta_p^{\prime}-i\beta_p^*\beta_q^{\prime})\nonumber .
	\end{eqnarray}
These relations should be compared with their analogs in the SvH (\ref{due-trentotto-a})-(\ref{due-trentotto-e})
and in the gauge case (\ref{tre-trentotto}). Before proceeding let us remember what we discussed in Eq. (\ref{due-quaranta}). There
we proved that in passing from the $|ket\rangle$ to the $\langle bra|$ the order of the entries had to be reversed: if in the $|ket
\rangle$ the first entry was $q$, in the corresponding $\langle bra|$ it had to be $p$. This was the case for  the SvH and gauge scalar
products but it is no longer the case here. In fact let us remember the relation 
	\begin{equation}
	|0_q-,0_p-\rangle=\hat{c}^q\hat{c}^p|0_q+,0_p+\rangle \label{quattro-otto-a}
	\end{equation}
and performing now the Hermitian conjugation (\ref{quattro-uno}) on the operators appearing in (\ref{quattro-otto-a}), we see that
the following relation is perfectly consistent
	\begin{equation}
	\langle 0_q+,0_p+|\hat{\bar{c}}_q\hat{\bar{c}}_p=\langle 0_q-,0_p-| .\label{quattro-otto-b}
	\end{equation}
This consistency indicates that we can keep the same order in the entries of the
$\langle bra|$ as we had at the level of the $|ket\rangle$. 

The resolutions of the identity that we get for the symplectic scalar product are 
	\begin{eqnarray}
	\displaystyle \label{quattro-nove-a} 
	&&\int d\alpha_p d\alpha_q|\alpha_q-,\alpha_p-\rangle\langle i\alpha_p^*+,(-i\alpha_q^*)+|={\bf 1} \\
	\displaystyle \label{quattro-nove-b}
	&&\int d\alpha_p d\alpha_q|\alpha_q+,\alpha_p+\rangle\langle (-i\alpha_p^*)-,i\alpha_q^*-|={\bf 1}.
	\end{eqnarray}
In Appendix C we will derive them. These resolutions should be compared with those of the SvH case
(\ref{due-quarantuno}) and with those of the gauge case (\ref{tre-trentanove}). The strange form of
the $\langle bra|$ appearing in Eq. (\ref{quattro-nove-a}) is actually necessary in order to make it an eigenstate of
$\hat{c}^p$. In fact if we start from
	\begin{equation}
	\hat{\bar{c}}_q|i\alpha_p^*+,(-i\alpha_q^*)+\rangle=i\alpha_p^*|i\alpha_p^*+,(-i\alpha_q^*)+\rangle
	\end{equation}
and do the Hermitian conjugation, we get:
	\begin{equation}
	\langle i\alpha_p^*+,(-i\alpha_q^*)+|(\hat{\bar{c}}_q)^{\dagger}=\langle i\alpha_p^*+,(-i\alpha_q^*)+|(i\alpha_p^*)^*.
	\end{equation}
Using the hermiticity conditions (\ref{quattro-uno}), we obtain that 
	\begin{eqnarray}
	&&\langle i\alpha_p^*+,(-i\alpha_q^*)+|(-i\hat{c}^p)=\langle i\alpha_p^*+,(-i\alpha_q^*)+|(-i\alpha_p)\Rightarrow\nonumber\\
	&&\Rightarrow \langle i\alpha_p^*+, (-i\alpha_q^*)+|\hat{c}^p=\langle i\alpha_p^*+,(-i\alpha_q^*)+|\alpha_p.
	\label{quattro-nove-c}
	\end{eqnarray}
and this proves our claim. The same can be shown for $\hat{c}^q$. Let us start from 
	\begin{equation}
	\hat{\bar{c}}_p|i\alpha_p^*+,(-i\alpha_q^*)+\rangle=-i\alpha_q^*|i\alpha_p^*+,(-i\alpha_q^*)+\rangle
	\end{equation}
and, doing the Hermitian conjugation of it, we get 
	\begin{eqnarray}
	&&\langle i\alpha_p^*+,(-i\alpha_q^*)+|i\hat{c}^q=\langle i\alpha_p^*+, (-i\alpha_q^*)+|i\alpha_q \Rightarrow\nonumber\\
	&&\Rightarrow \langle i\alpha_p^*+,(-i\alpha_q^*)+|\hat{c}^q=\langle i\alpha_p^*+,(-i\alpha_q^*)+|\alpha_q.
	\label{quattro-nove-d}
	\end{eqnarray}
Via the relation (\ref{quattro-nove-a}) it is now easy to write the symplectic scalar product in terms of
the components of the states as we did in (\ref{due-cinquantuno}) for the SvH case and in (\ref{tre-ventotto-x}) for the gauge case. Let us
consider two states
$|\psi\rangle$ and $|\Phi\rangle$ and their scalar product $\langle\Phi|\psi\rangle$. By inserting the first resolution of the
identity (\ref{quattro-nove-a}), we obtain:
	\begin{eqnarray}
	\langle \Phi|\psi\rangle&=&\int d\alpha_p d\alpha_q\langle\Phi|\alpha_q-,\alpha_p-\rangle
	\langle i\alpha_p^*+,(-i\alpha_q^*)+|\psi\rangle=\nonumber\\
	&=&\int d\alpha_pd\alpha_q\Phi^*_+(\alpha_q,\alpha_p)\psi_-(\alpha_q,\alpha_p) \label{quattro-dieci-a}
	\end{eqnarray}
where:
	\begin{equation}
	\psi_-(\alpha_q,\alpha_p)\equiv\langle i\alpha_p^*+,(-i\alpha_q^*)+|\psi\rangle\label{quattro-dieci-b}
	\end{equation}
and 
	\begin{equation}
	\Phi_+(\alpha_q,\alpha_p)\equiv\langle\alpha_q-,\alpha_p-|\Phi\rangle .\label{quattro-undici}
	\end{equation}
Since $\psi_-(\alpha_q,\alpha_p)$ is a function of $\alpha_q,\alpha_p$, we write it as 
	\begin{equation}
	\psi_-(\alpha_q,\alpha_p)=\psi_0+\psi_q\alpha_q+\psi_p\alpha_p+\psi_2\alpha_q\alpha_p. \label{quattro-dodici}
	\end{equation}
As $\alpha_q,\alpha_p$ are Grassmannian variables, $\psi_-$ can only contain them to the first power. The freedom we have in the choice
of $\psi_-$ will only be in the factors of ``$i$" or ``$-i$" in front of the various $\psi_0,\psi_q,\psi_p,\psi_2$, and these
factors can easily be
reabsorbed in the definitions of the components. Let us now find out the expression of $\Phi_+$ in terms of components:
	\begin{eqnarray}
	&&\Phi_+(\alpha_q,\alpha_p)=\langle\alpha_q-,\alpha_p-|\Phi\rangle=\nonumber\\
	&&=\int d\alpha_p^{\prime}d\alpha_q^{\prime}\langle\alpha_q-,\alpha_p-|\alpha_q^{\prime}-,
	\alpha_p^{\prime}-\rangle\langle i\alpha_p^{\prime *}+,(-i\alpha_q^{\prime *})+|\Phi\rangle=\nonumber\\
	&&=\int d\alpha_p^{\prime}d\alpha_q^{\prime}\langle\alpha_q-,\alpha_p-|\alpha_q^{\prime}-,
	\alpha_p^{\prime}-\rangle\Phi_-(\alpha_q^{\prime},\alpha_p^{\prime})=\nonumber\\
	&&=\int
	d\alpha_p^{\prime}d\alpha_q^{\prime}(-1+i\alpha_q^*\alpha_p^{\prime}-i\alpha_p^*\alpha_q^{\prime}+\alpha_p^*\alpha_q^*
	\alpha_q^{\prime}\alpha_p^{\prime})\cdot
	(\Phi_0+\Phi_q\alpha_q^{\prime}+\Phi_p\alpha_p^{\prime}+\Phi_2\alpha_q^{\prime}\alpha_p^{\prime})\nonumber\\
	&&=-\Phi_2-i\Phi_p\alpha_p^*-i\Phi_q\alpha_q^*+\Phi_0\alpha_p^*\alpha_q^* .\label{quattro-tredici}
	\end{eqnarray}
Inserting the expressions (\ref{quattro-dodici}) and (\ref{quattro-tredici}) in (\ref{quattro-dieci-a}) we get:
	\begin{eqnarray}
	\langle \Phi|\psi\rangle&=&\int d\alpha_p d\alpha_q(-\Phi_2^*+i\Phi_p^*\alpha_p+i\Phi_q^*\alpha_q+\Phi_0^*\alpha_q\alpha_p)
	\cdot (\psi_0+\psi_q\alpha_q+\psi_p\alpha_p+\psi_2\alpha_q\alpha_p)=\nonumber\\
	&=&\Phi_0^*\psi_0+i(\Phi_q^*\psi_p-\Phi_p^*\psi_q)-\Phi_2^*\psi_2 .
	\end{eqnarray}
If we include also the bosonic variables $\varphi$ we obtain:
	\begin{equation}
	\displaystyle \langle \Phi|\psi\rangle
	=\int d\varphi\biggl[\Phi_0^*\psi_0+i(\Phi_q^*\psi_p-\Phi_p^*\psi_q)-\Phi_2^*\psi_2\biggr]. \label{quattro-quattordici}
	\end{equation}
The hermiticity conditions (\ref{quattro-uno}) now read
	\begin{equation}
	\label{quattro-quattordici-a}
	 \left\{
		\begin{array}{l}
		\displaystyle  
		\langle\Phi|\hat{c}^q\psi\rangle=\langle i\hat{\bar{c}}_p\Phi|\psi\rangle, \quad \;\;\;\;\;\;\;\;\;
	 \langle \Phi|i\hat{\bar{c}}_p\psi\rangle=\langle \hat{c}^q\Phi|\psi\rangle 
		\smallskip \nonumber\\
		\displaystyle
		\langle \Phi|\hat{c}^p\psi\rangle=\langle -i\hat{\bar{c}}_q\Phi|\psi\rangle, \;\;\;\;\;\;\;\;\;\;
	 \langle \Phi|-i\hat{\bar{c}}_q\psi\rangle=\langle \hat{c}^p\Phi|\psi\rangle.
	 \smallskip
		\end{array}
		\right.
	\end{equation}
The details of these derivations can be found in Appendix C.

Let us first notice that the scalar product (\ref{quattro-quattordici}) reproduces the one of KvN for the zero-forms,
	\begin{equation}
	\displaystyle \langle \Phi|\psi\rangle=\int d\varphi \;\Phi_0^*\psi_0 .\label{quattro-quattordici-a1}
	\end{equation}
This happened also in the SvH but not in the gauge case.

Unfortunately the symplectic scalar product is not
positive definite. In fact from (\ref{quattro-quattordici}) we notice that, for example, 
a 2-form $\Phi=\Phi_2\alpha_q\alpha_p$ has negative norm 
	\begin{equation}
	\displaystyle \langle \Phi|\Phi\rangle=-\int d\varphi|\Phi_2|^2 .\label{quattro-quattordici-b}
	\end{equation}
There are also states of zero norm like for example one-forms with real coefficients $\psi=\psi_q\alpha^q+\psi_p\alpha^p$:
	\begin{equation}
	\displaystyle \langle\psi|\psi\rangle=i\int d\varphi (\psi_q\psi_p-\psi_p\psi_q)=0 .\label{quattro-quattordici-c}
	\end{equation}
One last thing we should notice is that it was the choice (\ref{quattro-quattro-a}) for the normalization of $|0+,0+\rangle$ which
led to the result (\ref{quattro-quattordici}). If we had chosen the normalization 
$\displaystyle \||0-,0-\rangle\|^2=1$ we would have got $\displaystyle \| |0+,0+\rangle\|^2=-1$ and the following
scalar product between zero-forms \cite{Appendix C}: 
	\begin{equation}
	\displaystyle \langle\Phi|\psi\rangle=- \int d \varphi\;\Phi_0^*\psi_0 \label{quattro-quindici}
	\end{equation}
which is in contradiction with the KvN result (\ref{uno-quattro}). Remember that we wanted to have a scalar product which 
reproduces the KvN result for the zero-forms and provide an extension of it for higher forms. 

We have shown in this paper that there is more than one extension of the KvN scalar product, 
the SvH and the symplectic one, but both
of which have defects: for the first one $\widetilde{\cal H}$ is not Hermitian, while for the second one the scalar
product is not positive definite. The gauge scalar product instead was not even an extension of the KvN because the zero-forms
states had zero-norms (\ref{tre-ventotto-x}). 

We will now study the symplectic scalar product for $n$ degrees of freedom. 
Let us first note that the case (\ref{quattro-quattordici}) of one degree of freedom can be written in the following manner:
	\begin{equation}
	\displaystyle
	\langle\Phi|\psi\rangle=\int d\varphi\biggl[\Phi_0^*\psi_0
	+i\Phi_a^*\omega^{ab}\psi_b+\frac{i^2}{2!}\Phi_{a_1a_2}^*\omega^{a_1b_1}\omega^{a_2b_2}\psi_{b_1b_2}
	\biggr] \label{quattro-sedici}
	\end{equation}
where the notation has the following meaning:
	\begin{equation}
	\varphi^a=(q,p),\;\;\;\;\Phi_a=(\Phi_q,\Phi_p),\;\;\;\;\Phi_{qp}=\Phi_2,\;\;\;\;\Phi_{qq}=0,\;\;\;\;\Phi_{pp}=0. \nonumber
	\end{equation}
The last two equations are due to the antisymmetry of the two forms. It is a long but straightforward calculation to prove
that two generic states of the form:
	\begin{equation}
	\displaystyle
	\psi=\sum_{m=0}^{2n}\frac{1}{m!}\psi_{a_1\ldots a_m}c^{a_1}\ldots c^{a_m};\;\;\;\;\;
	\Phi=\sum_{m=0}^{2n}\frac{1}{m!}\Phi_{b_1\ldots b_m}c^{b_1}\ldots c^{b_m} \label{quattro-diciassette}
	\end{equation}
have the following symplectic scalar product:
	\begin{equation}
	\displaystyle
	\langle\Phi|\psi\rangle=\sum_{m=0}^{2n}\frac{i^m}{m!}\int d\varphi \;
	\Phi^*_{a_1\ldots a_m}\omega^{a_1b_1}\ldots
	\omega^{a_mb_m}\psi_{b_1\ldots b_m} .\label{quattro-diciotto}
	\end{equation}

The last issue we want to explore is whether we can reproduce, via the symplectic scalar product and its
decomposition of the identity, the CPI path-integral starting from the operatorial formalism,
as we did in the SvH and gauge cases. 
Let us first write for the symplectic scalar product the analog of the resolutions of the identity
(\ref{tre-quaranta-a})-(\ref{tre-quaranta-b}). Using (\ref{quattro-nove-a}) we get 
	\begin{eqnarray}
	&&\int d\varphi\, dc^pdc^q|\varphi,\,c^q-,c^p-\rangle\langle ic^{p*}+,(-ic^{q*})+,\varphi|={\bf 1}
	\label{quattro-diciannove-a}\\
	&&\int d\lambda\,d\bar{c}_p\,d\bar{c}_q|\lambda,\,\bar{c}_q+,\bar{c}_p+\rangle\langle(-i\bar{c}_p^*)-,
	i\bar{c}_q^*-,\lambda|={\bf 1}. \label{quattro-diciannove-b}
	\end{eqnarray}
The kernel $K(f|i)$, analogous to (\ref{due-cinquantasette}) and (\ref{tre-quarantuno}), in the symplectic case takes the
form:
	\begin{equation}
	\displaystyle 
	K(f|i)=\langle ic^{*}_{p_f}+,(-ic^{*}_{q_f})+,\varphi_f|\exp\bigl[-i\widetilde{\cal
	H}(t_f-t_i)\bigr]|\varphi_i,\,c_{q_i}-,c_{p_i}-\rangle.
	\label{quattro-venti}
	\end{equation}
Note that the difference with respect to (\ref{due-cinquantasette}) and (\ref{tre-quarantuno}) is in the initial $\langle bra|$.
Let us remember in fact that the $K(f|i)$ is the transition amplitude 
	\begin{equation}
	K(f|i)\equiv K(\varphi_f,c^q_f,c^p_f|\varphi_i,c^q_i,c^p_i)
	\end{equation}
between an initial configuration $(\varphi_i,c^q_i,c^p_i)$ and a final one $(\varphi_f,c^q_f,c^p_f)$ and so the $\langle bra|$
which appears in (\ref{quattro-venti}), i.e. $\langle ic_{p_f}^*+,(-ic_{q_f}^*)+,\varphi_f|$, 
must be an eigenstate of $\hat{c}^p$
and $\hat{c}^q$ with eigenvalues $c^p_f$ and $c^q_f$. This is actually the case as it has been proved in 
(\ref{quattro-nove-c})-(\ref{quattro-nove-d}). 
The procedure then continues, as in the previous cases, by splitting the time interval $(t_f-t_i)$ in
(\ref{quattro-venti}) into $N$ intervals of length $\epsilon$ 
	\begin{equation}
	\displaystyle 
	K(f|i)=\langle ic^{*}_{p_f}+,(-ic^{*}_{q_f})+,\varphi_f|\underbrace{e^{-i\epsilon\widetilde{\cal
	H}}e^{-i\epsilon\widetilde{\cal
	H}}\ldots e^{-i\epsilon\widetilde{\cal H}}}_{N\; terms}|\varphi_i,c_{q_i}-,c_{p_i}-\rangle.
	\end{equation}
Inserting (\ref{quattro-diciannove-b}) on the left and (\ref{quattro-diciannove-a}) on the right of each term
$e^{-i\epsilon\widetilde{\cal H}}$, we get the same discretized expression as in
(\ref{due-sessantadue}). We skip the details here because they are very similar to SvH and
gauge cases and the result is the same. That means also in the symplectic case we can reproduce the CPI 
path integral (\ref{uno-dodici-b}).

We can summarize this section by saying that in the symplectic scalar product
the $\HT$ is Hermitian but the Hilbert space is not positive definite.

\section{``Physical" Hilbert Space} 
\noindent
In this section we will address the issue of what is the {\it physical} subspace of the full Hilbert space.
With a little abuse of notation, we call {\it physical} subspace the one made of only positive norm states
and on which $\widetilde{\cal H}$ is Hermitian. We shall do this analysis for all the three scalar products
studied in the previous sections starting from the SvH one of
Sec. II. 
In this case all the states in the Hilbert space have positive definite norm but
$\widetilde{\cal H}$ is not Hermitian. This last one is an unacceptable feature because it would lead to the non-conservation of the
norm. This in turn would create difficulties in assigning the meaning of probability to the norm of a generic state 
as instead we could do in the
zero-form case  (see Eq. (\ref{uno-uno})). The linear subspace on which $\widetilde{\cal H}$ is Hermitian is defined
by the following condition
	\begin{equation}
	(\widetilde{\cal H}-\widetilde{\cal H}^{\dagger})|\psi\rangle=0. \label{cinque-uno}
	\end{equation}
In fact it is easy to see that, for the states obeying (\ref{cinque-uno}), we have 
	\begin{equation}
	\langle \Phi|\widetilde{\cal H}-\widetilde{\cal H}^{\dagger}|\psi\rangle=0 \label{cinque-due-a}
	\end{equation}
i.e.
	\begin{equation}
	\langle \Phi|\widetilde{\cal H}|\psi\rangle=\langle\Phi|\widetilde{\cal H}^{\dagger}|\psi\rangle
	\label{cinque-due-b}.
	\end{equation}
Considering that $\langle\Phi|$ is a generic state of the full Hilbert space, (\ref{cinque-due-b}) implies that
$\widetilde{\cal H}$ is Hermitian in the physical subspace. 
Obviously the states which are solutions of (\ref{cinque-uno}) make a linear vector space. 
The next thing we have to guarantee is that the subspace defined by (\ref{cinque-uno}) be closed
under time evolution. This means that  a state $|\psi^{\prime}\rangle$, obtained via an infinitesimal time evolution from a physical
state $|\psi\rangle$: 
	\begin{equation}
	\displaystyle |\psi^{\prime}\rangle=e^{-i\epsilon\widetilde{\cal H}}|\psi\rangle \label{cinque-tre},
	\end{equation}
must still be physical, that means it must still be a solution of (\ref{cinque-uno}):
	\begin{equation}
	(\widetilde{\cal H}-\widetilde{\cal H}^{\dagger})
	|\psi^{\prime}\rangle=0. \label{cinque-quattro}
	\end{equation}
Inserting (\ref{cinque-tre}) into (\ref{cinque-quattro}) we get
	\begin{eqnarray}
	&&(\widetilde{\cal H}-\widetilde{\cal H}^{\dagger})
	|\psi^{\prime}\rangle=(\widetilde{\cal
	H}-\widetilde{\cal H}^{\dagger})|
	\psi\rangle-i\epsilon (\widetilde{\cal H}-\widetilde{\cal
	H}^{\dagger})\widetilde{\cal H}|\psi\rangle=\nonumber\\ 
	&&=-i\epsilon \biggl[\widetilde{\cal H},(\widetilde{\cal H}-\widetilde{\cal H}^{\dagger})\biggr]|\psi\rangle=
	i\epsilon [\widetilde{\cal H},\widetilde{\cal H}^{\dagger}]|\psi\rangle
	\end{eqnarray}
that implies that for $|\psi^{\prime}\rangle$ to be physical the following condition must also be satisfied 
	\begin{equation}
	[\widetilde{\cal H},\widetilde{\cal H}^{\dagger}]|\psi\rangle=0. \label{cinque-cinque}
	\end{equation}
Let us analyze the commutator structure contained in (\ref{cinque-cinque}). If we write
$\widetilde{\cal H}$ as in (\ref{due-cinquantuno-x})
	\begin{equation}
	\widetilde{\cal H}=\widetilde{\cal H}_{bos}+\widetilde{\cal H}_{ferm} \label{cinque-sei}
	\end{equation}
we get that the commutator contained in the LHS of (\ref{cinque-cinque}) turns into the following expression:
	\begin{equation}
	[\widetilde{\cal H},\widetilde{\cal H}^{\dagger}]=
	[\widetilde{\cal H}_{ferm},\widetilde{\cal
	H}^{\dagger}_{ferm}]+[\widetilde{\cal H}_{bos},\widetilde{\cal H}_{ferm}^{\dagger}
	]+[\widetilde{\cal H}_{ferm},\widetilde{\cal H}_{bos}] .\label{cinque-sette}
	\end{equation}
Let us look at the first term on the RHS of (\ref{cinque-sette}). The general expression of $\widetilde{\cal
H}_{ferm}$ was given in (\ref{due-cinquantuno-xx}) and, choosing $H$ to be of the form $\displaystyle
H=\sum_{i=1}^np_i^2/2+V(q_1,\ldots,q_n)$, we get 
	\begin{equation}
	\label{cinque-sette-b} \widetilde{\cal H}_{ferm}=i\bar{c}_{q_i}c^{p_i}-i\bar{c}_{p_j}\partial_i\partial_jVc^{q_i} .
	\end{equation}
Using the SvH Hermitian conjugation rules (\ref{due-cinquantadue}), we obtain:
	\begin{equation}
	\widetilde{\cal H}^{\dagger}_{ferm}=-i\bar{c}_{p_i}c^{q_i}+i\bar{c}_{q_i}\partial_i\partial_jVc^{p_j}.
	\end{equation}
So the first term in (\ref{cinque-sette}) turns out to be 
	\begin{equation}
	[\widetilde{\cal H}_{ferm},\widetilde{\cal H}^{\dagger}_{ferm}]=\bar{c}_{q_i}c^{q_i}-
	\bar{c}_{p_i}c^{p_i}+(\partial_i\partial_jV)(\partial_l\partial_mV)[\bar{c}_{p_j}c^{p_m}\delta^i_l-\bar{c}_{q_l}c^{q_i}\delta_j^m].
	\label{cinque-otto}
	\end{equation}
The second and the third term in (\ref{cinque-sette}) contain third order derivatives in the potential $V$. To find solutions
$|\psi\rangle$ of Eq. (\ref{cinque-cinque}), whose form is independent of the potential, we should impose that 
$|\psi\rangle$ be annihilated separately by the terms in (\ref{cinque-sette}) which contain no derivative in $V,$ next by
those which contain first derivatives of $V$ and then by those which contain second derivatives and so on. By looking at
(\ref{cinque-sette}) and (\ref{cinque-otto}) the term with no derivative of $V$ is
$(\bar{c}_{q_i}c^{q_i}-\bar{c}_{p_i}c^{p_i})$ and imposing it on $|\psi\rangle$ we get
	\begin{equation}
	(\bar{c}_{q_i}c^{q_i}-\bar{c}_{p_i}c^{p_i})|\psi\rangle=0
	\end{equation}
which implies 
	\begin{equation}
	\displaystyle c^{q_i}\frac{\partial}{\partial c^{q_i}}|\psi\rangle=c^{p_i}\frac{\partial}{\partial c^{p_i}}|\psi\rangle.
	\label{cinque-nove}
	\end{equation}
If we write a representation of $|\psi\rangle$ as
	\begin{equation}
	\displaystyle \psi(\varphi,c)=\sum_{j=0}^{2n}\frac{1}{j!}\psi_{a,b,\ldots,j}(\varphi)c^ac^b\ldots c^j \label{cinque-nove-b}
	\end{equation}
then (\ref{cinque-nove}) is satisfied by those $\psi(\varphi,c)$ which contain the same number of $c^q$ and of 
$c^p$. Clearly these forms are Grassmannian even, which implies that odd forms are not physical.   
Before going on to check whether also the terms in (\ref{cinque-otto}) with second derivatives in $V$ 
annihilate these forms, let us remember that we must also satisfy the condition (\ref{cinque-uno}). 
The operator $\widetilde{\cal H}-\widetilde{\cal H}^{\dagger}$ with an $H$
of the form $\displaystyle H=\sum_{i=1}^n\frac{p_i^2}{2}+V(q_1,\ldots,q_n)$ has the expression 
	\begin{equation}
	\widetilde{\cal H}-\widetilde{\cal H}^{\dagger}=(i\bar{c}_{q_i}c^{p_i}+i\bar{c}_{p_i}c^{q_i})-i
	(\bar{c}_{p_j}\partial_i\partial_jVc^{q_i}+\bar{c}_{q_i}\partial_i\partial_jVc^{p_j}) .\label{cinque-dieci}
	\end{equation}
Again, a physical form must be annihilated separately by the terms independent of $V$ and by those depending on it. So,
using (\ref{cinque-dieci}), Eq. (\ref{cinque-uno}) gives the following two conditions:
	\begin{eqnarray}
	&&\label{cinque-undici-a} (\bar{c}_{q_i}c^{p_i}+\bar{c}_{p_i}c^{q_i})|\psi\rangle=0 \\
	&&\label{cinque-undici-b} (\bar{c}_{p_j}\partial_i\partial_jVc^{q_i}+\bar{c}_{q_i}
	\partial_i\partial_jVc^{p_j})|\psi\rangle=0.
	\end{eqnarray}
Let us now remember  that, because of Eq. (\ref{cinque-nove}), the state $|\psi\rangle$ must contain the same number of $c^q$
and $c^p$. Therefore it is easy to realize that (\ref{cinque-undici-a}) implies that the following two relations must
both hold:
	\begin{equation}
	\bar{c}_{q_i}c^{p_i}|\psi\rangle=0 \label{cinque-dodici-a}
	\end{equation}
and
	\begin{equation}
	\bar{c}_{p_i}c^{q_i}|\psi\rangle=0. \label{cinque-dodici-b}
	\end{equation}
Analogously, for (\ref{cinque-undici-b}) to hold, it must be that each
term in it separately annihilates $|\psi\rangle$:
	\begin{equation}
	 \left\{
		\begin{array}{l}
		\displaystyle 
	\label{cinque-tredici-a}	\bar{c}_{p_j}(\partial_i\partial_jV)c^{q_i}|\psi\rangle=0
		\smallskip \\
		\displaystyle
	\label{cinque-tredici-b}	\bar{c}_{q_i}(\partial_i\partial_jV)c^{p_j}|\psi\rangle=0.
		\end{array}
		\right.
	\end{equation}
The careful reader will at this point notice that our proof that the commutator (\ref{cinque-sette}) on our states is zero is
not yet complete. In fact, of Eq. (\ref{cinque-sette}) we checked only the first term (\ref{cinque-otto}) and, of this, only the part
which is independent of the potential. How can we be sure that our states are annihilated also by the part depending on the
potential in (\ref{cinque-otto}) and by the two last terms in (\ref{cinque-sette})? To complete the proof we proceed as follows.
Let us make a linear combination of (\ref{cinque-dodici-a}) and (\ref{cinque-tredici-a}) of the following form:
	\begin{equation}
	(i\bar{c}_{q_i}c^{p_i}-i\bar{c}_{p_j}\partial_i\partial_jVc^{q_i})|\psi\rangle=0.
	\end{equation}
It is easy to realize, looking at (\ref{cinque-sette-b}), that this is:
	\begin{equation}
	\widetilde{\cal H}_{ferm}|\psi\rangle=0. \label{cinque-diciotto}
	\end{equation}
Doing instead a linear combination of (\ref{cinque-dodici-b}) and (\ref{cinque-tredici-b}) of the form 
	\begin{equation}
	(-i\bar{c}_{p_i}c^{q_i}+i\bar{c}_{q_j}\partial_i\partial_jVc^{p_i})|\psi\rangle=0
	\end{equation}
we immediately realize that this is
	\begin{equation}
	\widetilde{\cal H}^{\dagger}_{ferm}|\psi\rangle=0. \label{cinque-diciannove}
	\end{equation}
These are the two relations which complete our proof. In fact, using
(\ref{cinque-diciotto})-(\ref{cinque-diciannove}), we have that (\ref{cinque-otto}) is automatically satisfied while 
in (\ref{cinque-sette}) we get
	\begin{eqnarray}
	[\widetilde{\cal H},\widetilde{\cal H}^{\dagger}]|\psi\rangle &=&
	[\widetilde{\cal H}_{ferm},\widetilde{\cal
	H}^{\dagger}_{ferm}]|\psi\rangle+
	[\widetilde{\cal H}_{bos},\widetilde{\cal H}_{ferm}^{\dagger}]|\psi\rangle+
	[\widetilde{\cal H}_{ferm},\widetilde{\cal H}_{bos}]|\psi\rangle=\nonumber\\
	&=&-\widetilde{\cal H}^{\dagger}_{ferm}\widetilde{\cal H}_{bos}|\psi\rangle+\widetilde{\cal
	H}_{ferm}\widetilde{\cal H}_{bos}|\psi\rangle=-\widetilde{\cal H}_{ferm}^{\dagger}|\psi^{\prime}
	\rangle+\widetilde{\cal H}_{ferm}|\psi^{\prime}\rangle=0.
	\end{eqnarray}
The last step is based on the fact that $|\psi^{\prime}\rangle\equiv\widetilde{\cal H}_{bos}|\psi\rangle$
is  still a physical state. In fact $\widetilde{\cal H}_{bos}$ acts only on the bosonic coefficients of the states 
and so it will leave the Grassmannian structure intact and it is this Grassmannian structure which
characterizes our physical states. 

Up to now we have proved that a state, to be physical, must be annihilated by the fermionic part of $\widetilde{\cal H}$
(\ref{cinque-diciotto}). 
The next step is to find the explicit form of such states. We want to start with an example. Let us take  a 
2-form with $n$ degrees of freedom. In order to satisfy (\ref{cinque-nove}) the 2-form must contain a $c^q$ and a $c^p$ and so it must be of the form:
	\begin{equation}
	\psi=\psi_{q_ip_k}c^{q_i}c^{p_k} .\label{gio1}
	\end{equation}
If we impose (\ref{cinque-dodici-b}) on the state (\ref{gio1}) we obtain:
	\begin{equation}
	\displaystyle c^{q_{\alpha}}\frac{\partial}{\partial c^{p_{\alpha}}}\psi=0 \Rightarrow \psi_{q_ip_{\alpha}}c^{q_{\alpha}}
	c^{q_i}=0 .\label{gio2}
	\end{equation}
For the properties of the Grassmannian variables the previous relation is satisfied if we take $\alpha=i$, i.e. if we take a 2-form
of the type:
	\begin{equation}
	\psi=\psi_{q_1p_1}c^{q_1}c^{p_1}+\psi_{q_2p_2}c^{q_2}c^{p_2}+\ldots+ \psi_{q_np_n}c^{q_n}c^{p_n} \label{cinque-quattordici-a}
	\end{equation} 
i.e. a form in which {\it each} $c^{q_i}$ is coupled with the relative $c^{p_i}$. 
Let us, for simplicity, indicate the various $\psi_{q_jp_j}$ as $\psi_{\scriptscriptstyle (j)}
(\varphi)$. Then (\ref{cinque-quattordici-a}) 
can be written as 
	\begin{equation}
	\displaystyle \psi=\sum_j\psi_{\scriptscriptstyle (j)}c^{q_j}c^{p_j} .\label{cinque-quattordici-b}
	\end{equation} 
Inserting (\ref{cinque-quattordici-b}) into Eq. (\ref{cinque-tredici-a}), 
it is easy to prove that it can be satisfied only if all the
coefficients $\psi_{\scriptscriptstyle (j)}$ in (\ref{cinque-quattordici-b}) are the same
	\begin{equation}
	\psi_{\scriptscriptstyle (j)}(\varphi)=K(\varphi) .\label{cinque-quindici-b}
	\end{equation}
So (\ref{cinque-quattordici-a}) turns out to be 
	\begin{equation}	
	\psi=K(\varphi)[c^{q_1}c^{p_1}+c^{q_2}c^{p_2}+\ldots+c^{q_n}c^{p_n}] .\label{cinque-sedici}
	\end{equation}
One sees that somehow the dependence on the indices of the coefficients $\psi_{ab\ldots j}$ has disappeared. The coefficients 
$K(\varphi)$ will be the same for forms of the same rank but will change with the rank. So an example of a ``{\it physical}"
inhomogeneous form of rank up to 4 will be:
	\begin{equation}
	\displaystyle \psi=\psi_0(\varphi)+K(\varphi)\sum_ic^{q_i}c^{p_i}+S(\varphi)\sum_{i,j}(c^{q_i}c^{p_i})(c^{q_j}c^{p_j})
	+\ldots
	\label{cinque-diciassette}
	\end{equation}
It is easy to realize that among these physical states we always have the zero-forms and the $2n$ or volume forms.
Anyhow all our construction proves only that states of the form
(\ref{cinque-diciassette}) are physical but not that they are the only ones. We feel anyhow very confident that they are actually
the only ones. From the physical point of view, if we take the homogeneous physical forms, like (\ref{cinque-sedici}), we see that
they are ``somehow" isomorphic to the zero-forms. In fact $\widetilde{\cal H}_{ferm}$ annihilates them (see
(\ref{cinque-diciotto})) and this is the same that happens on the zero-forms. Basically $\widetilde{\cal H}_{ferm}$
acts on the Grassmannian variables in (\ref{cinque-sedici}) annihilating them; so we are left with only
$K(\varphi)$ changing under the time evolution and this $K(\varphi)$ evolves like a zero-form. 
Instead an inhomogeneous state like
(\ref{cinque-diciassette}) is made of a sum of terms, each isomorphic to a zero-form; so we can say that it is 
like a linear superposition of zero-forms. These zero-forms in (\ref{cinque-diciassette}) are 
$\psi_0(\varphi),K(\varphi),S(\varphi)$.
Before concluding we should point out that the physical condition (\ref{cinque-uno}) limits the forms to be of the type
(\ref{cinque-diciassette}) only if we do not put any restriction on the potential $V$. If we put restriction instead, 
for example choosing a harmonic oscillator potential or a separable potential, then condition (\ref{cinque-uno}) is satisfied by a
wider class of forms than the ones in (\ref{cinque-diciassette}). This concludes the analysis of the SvH case.

Let us now turn to the other scalar products and in particular to the symplectic one of Sec. IV. The
Hamiltonian $\widetilde{\cal H}$ in that case is Hermitian but not all the states of the Hilbert space have positive
norm. So the ``physical" Hilbert space, which we will indicate with ${\bf{\cal H}}_{phys}$, should be a subspace of the full
Hilbert space and made only of positive norm states. 
Anyhow this subspace ${\bf{\cal H}}_{phys}$ cannot be the set of {\it all} positive
norm states, which we will indicate with ${\bf{\cal H}}^{\scriptscriptstyle (+)}$, contained in the original Hilbert space. In fact it is easy to realize that 
${\bf{\cal H}}^{\scriptscriptstyle (+)}$ is not a vector space because the linear combination of two states with positive norm,
$\psi_+^{(1)},\psi^{(2)}_+$:
	\begin{equation}
	\psi\equiv\alpha\psi_+^{(1)}+\beta\psi_+^{(2)} 
	\end{equation}
where $\alpha$ and $\beta$ are complex coefficients, does not necessarily belong to ${\bf{\cal H}}^{\scriptscriptstyle (+)}$. We will provide
an explicit example of this fact in (\ref{counter}). So
${\bf{\cal H}}_{phys}$ can only be a particular subspace of ${\bf{\cal H}}^{\scriptscriptstyle (+)}$. In order to build it, 
it is better to change the
variables, and pass from the $(q_i,p_i,\lambda_{q_i},\lambda_{p_i}, c^{q_i},c^{p_i},\bar{c}_{q_i},\bar{c}_{p_i})$ to the
following ones:
	\begin{equation}
	 \left\{
		\begin{array}{l}
		\displaystyle 
	\label{cinque-venti-a}	z_i\equiv\frac{1}{\sqrt{2}}(q_i+ip_i), \qquad\qquad\;\;\;\;\;\; \bar{z}_i\equiv\frac{1}{\sqrt{2}}(q_i-ip_i)
		\smallskip \\
		\displaystyle
	\label{cinque-venti-b}	l_i\equiv\frac{1}{\sqrt{2}}(\lambda_{q_i}-i\lambda_{p_i}), \qquad\qquad \;\;\;\;
	        \bar{l}_i\equiv\frac{1}{\sqrt{2}}(\lambda_{q_i}+i\lambda_{p_i}) 	
	        \smallskip \\
	        \displaystyle
	\label{cinque-venti-c}	\xi_i\equiv\frac{1}{\sqrt{2}}(c^{q_i}+ic^{p_i}), \qquad\qquad\;\;\;\;
	        \bar{\xi}_i\equiv\frac{1}{\sqrt{2}}(-\bar{c}_{q_i}+i\bar{c}_{p_i})
	        \smallskip \\
	        \displaystyle
	\label{cinque-venti-d}	\xi_i^*\equiv\frac{1}{\sqrt{2}}(c^{q_i}-ic^{p_i}), \qquad\qquad\;\;\;
	        \bar{\xi}^*_i\equiv\frac{1}{\sqrt{2}}(\bar{c}_{q_i}+i\bar{c}_{p_i}).
		\end{array}
		\right.
	\end{equation}
The operation which we have indicated with ``$*$" should not be identified with the complex conjugation. From
(\ref{uno-quattordici-b}) it is easy to work out the graded commutators among the new variables
(\ref{cinque-venti-a}). In particular we will be
interested in the following ones:
	\begin{equation}
	 \left\{
		\begin{array}{l}
		\displaystyle 
	\label{cinque-ventuno}	[\xi_i,\bar{\xi}_j]=-\delta_{ij}, \qquad\qquad\; [\xi_i,\bar{\xi}_j^*]=0
		\smallskip \\
		\displaystyle
		[\xi^*_i,\bar{\xi}^*_j]=+\delta_{ij}, \qquad\qquad [\xi_i^*,\bar{\xi}_j]=0.
		\end{array}
		\right.
	\end{equation}
Under the symplectic Hermitian conjugation (\ref{quattro-uno}), we get 
	\begin{equation}
	\xi_i^{\dagger}=\bar{\xi}_i,\qquad\qquad\qquad \xi_i^{*\dagger}=\bar{\xi}_i^*. \label{cinque-ventidue}
	\end{equation}
Note that this ``hermiticity" property is, for the Grassmannian variables $(\xi,\xi^*), (\bar{\xi},\bar{\xi}^*)$, the same
as the SvH one (\ref{due-cinquantadue}) for the variables $c^a,\bar{c}_a$. The crucial difference 
is in the anticommutator 
        \begin{equation}
        [\xi_i,\bar{\xi}_j]=-\delta_{ij}
        \end{equation}
which, for the analog SvH variables, had the opposite sign on the RHS: $[c_{q_i},\bar{c}_{q_j}]=\delta_{ij}$. 
We shall show  that this sign-difference gives rise to negative norm states in the symplectic case. 
Turning the  new variables $(\xi_i,\xi_j^*,\bar{\xi}_k,\bar{\xi}^*_l)$ into
operators, let us define the states:
	\begin{equation}
	\hat{\xi}_i|0-,0-,\ldots\rangle=\hat{\xi}_i^*|0-,0-,\ldots\rangle=0. \label{cinque-ventiquattro}
	\end{equation}
To start with $n=1$, let us take the following states:
	\begin{equation}
	 \left\{
		\begin{array}{l}
		\displaystyle 
	\label{cinque-venticinque-a}	|0+,0-\rangle=\hat{\bar{\xi}}|0-,0-\rangle
		\smallskip \\
		\displaystyle
	\label{cinque-venticinque-b}	|0-,0+\rangle=-\hat{\bar{\xi}}^*|0-,0-\rangle
	        \smallskip \\
	        \displaystyle 
	\label{cinque-venticinque-c}    |0+,0+\rangle=\hat{\bar{\xi}}^*\hat{\bar{\xi}}|0-,0-\rangle.
		\end{array}
		\right.
	\end{equation}
These states $|0\pm,0\pm\rangle$ will be different from those defined in (\ref{quattro-quattro}) via the $c$ variables. Besides 
the hermiticity conditions, let us make the choice of the normalization for one of the states $|0\pm,0\pm\rangle$,
as we did previously. We choose 
	\begin{equation}
	\biggl(|0-,0-\rangle,|0-,0-\rangle\biggr)=-1 \label{cinque-ventisei-a}
	\end{equation}
and, via the definitions (\ref{cinque-venticinque-a}) and the commutation relations (\ref{cinque-ventuno}), we obtain
the following normalization conditions:
	\begin{equation}
	 \left\{
		\begin{array}{l}
		\displaystyle 
	\label{cinque-ventisei}	\biggl(|0+,0-\rangle,|0+,0-\rangle\biggr)=1
		\smallskip \nonumber\\
		\displaystyle
		\biggl(|0-,0+\rangle,|0-,0+\rangle\biggr)=-1
	        \smallskip \nonumber\\
	        \displaystyle 
	        \biggl(|0+,0+\rangle,|0+,0+\rangle\biggr)=1.
		\end{array}
		\right.
	\end{equation}
Note the difference with (\ref{quattro-cinque}). From the definition (\ref{cinque-venticinque-a}) 
we could build a representation of the states as follows:
	\begin{equation}
	 \left\{
		\begin{array}{l}
		\displaystyle 
	\label{cinque-ventisette}	
		\displaystyle
		|0-,0-\rangle=\xi\xi^*,\qquad\qquad |0-,0+\rangle=\xi
	        \smallskip \nonumber\\
	        \displaystyle 
	        |0+,0-\rangle=\xi^*,\qquad\qquad \;\;|0+,0+\rangle=1.
		\end{array}
		\right.
	\end{equation}
From this representation one sees that $|0+,0+\rangle$ is the basis of the zero-forms. This explains why we made the
normalization choice $\biggl(|0-,0-\rangle,|0-,0-\rangle\biggr)=-1$; in this way the normalization of $|0+,0+\rangle$ would turn out to be +1
and as a consequence the zero-form-states would have positive norm as in the KvN case. The scalar products
(\ref{cinque-ventisei}) can be written as
	\begin{equation}
	 \left\{
		\begin{array}{l}
		\displaystyle 
	\label{cinque-ventisette-b}	
		\displaystyle
		(\xi\xi^*,\xi\xi^*)=-1,\qquad\qquad\; (\xi,\xi)=-1
	        \smallskip \nonumber\\
	        \displaystyle 
	        (\xi^*,\xi^*)=1,\qquad\qquad\qquad (1,1)=1
		\end{array}
		\right.
	\end{equation}
and so a generic state 
	\begin{equation}
	\psi(\xi,\xi^*)=\psi_0+\psi_{\xi}\xi+\psi_{\xi^*}\xi^*+\psi_2\xi\xi^*
	\end{equation}
has a norm 
	\begin{equation}
	\langle\psi|\psi\rangle=|\psi_0|^2-|\psi_{\xi}|^2+|\psi_{\xi^*}|^2-|\psi_2|^2 .\label{cinque-ventotto}
	\end{equation}
It is easy to derive (\ref{cinque-ventotto}) considering that, via the $\xi$ variables, the symplectic scalar product is like the SvH one. 
The sign-differences in (\ref{cinque-ventotto}) with respect to (\ref{due-cinquantuno}) 
are due to the minus signs appearing in (\ref{cinque-ventisei}) 
and (\ref{cinque-ventisette-b}) relative
to the plus signs appearing in (\ref{due-trentasette-a})-(\ref{due-trentasette-b}). 

We will now generalize our treatment to the case of
$n=2$. If we still want that $|0+0+0+0+\rangle$ be the basis of the zero-forms with positive norm like in the KvN
case, then we should not choose the analog of the normalization (\ref{cinque-ventisei-a}) but rather:
	\begin{equation}
	\biggl(|0-0-0-0-\rangle,|0-0-0-0-\rangle\biggr)=1 .\label{cinque-ventinove}
	\end{equation}
With this choice it is easy to prove \cite{Appendix C} that we get
	\begin{equation}
	\biggl(|0+0+0+0+\rangle,|0+0+0+0+\rangle\biggr)=1 \label{cinque-trenta}
	\end{equation}
and as a consequence the zero-forms have positive norm. In general, for $n$ degrees of freedom, in order to have positive norm
for the zero-forms  we should make the following choice for the normalization of the state
$|0-0-\ldots 0-0-\rangle$
	\begin{equation}
	\langle 0-0-\ldots 0-0-|0-0-\ldots 0-0-\rangle=(-1)^n. \label{cinque-trentuno}
	\end{equation}
We have now all the ingredients to start looking for the physical states. From the norms in (\ref{cinque-ventisette-b}) we see that in
general a homogeneous form $\displaystyle \psi=\frac{1}{l!}\psi_{ij\ldots l}\xi^i\xi^{*j}\ldots \xi^l$ 
has positive norm if the number of
$\xi$ variables is odd. This is also confirmed by relation (\ref{cinque-ventotto}) which contains four homogeneous forms
$\psi_0,\psi_{\xi},\psi_{\xi^*},\psi_{2}$ and each of them obeys the rule we have just stated . This rule  
holds not  only for $n=1$ but also for higher $n$. For example,
the representation of $|0-0-0-\ldots 0-\rangle$ in $n$ dimensions is 
	\begin{equation}
	|0-0-0-\ldots 0-\rangle=\xi_1\xi_1^*\xi_2\xi_2^*\ldots \xi_n\xi_n^*
	\end{equation}
and its norm is (\ref{cinque-trentuno}), i.e. $(-1)^n$, which is $+1$ if $n$ (number of $\xi$ contained) is
even and $-1$ if $n$ is odd. This confirms the rule above. The detailed proof is easy and it is given in Appendix C. 
So we have a criterion to look for  positive norm states: if a generic homogeneous state $\displaystyle 
\frac{1}{l!}\psi_{ab\ldots l}\cdot c^ ac^b\ldots
c^l$ is given, we first transform the $c^ a$ variables into $\xi^ i,\xi^{i*}$ variables via
(\ref{cinque-venti-a}):
	\begin{equation}
	\displaystyle \frac{1}{l!} \psi_{ab\ldots l}c^ac^b\ldots c^l\;\Longrightarrow\; 
	\frac{1}{l!}\widetilde{\psi}_{ij\ldots}\xi^i\xi^{j*}\ldots
	\end{equation}
and then we count the number of $\xi$; if they are even, the state has positive norm and if they are odd, it has negative norm. Of
course this is a sufficient and necessary condition only for homogeneous states but not for non-homogeneous ones. For
example the state 
	\begin{equation}
	\displaystyle \psi=\frac{1}{2!}\psi_{ab}\xi^a\xi^b+\psi_a\xi^a \label{cinque-trentadue}
	\end{equation}
is made of two parts, a 2-form $\psi_{ab}\xi^a\xi^b$, and a one-form $\psi_a\xi^a$. From what we said above the 2-form has positive
 norm because  it
contains two $\xi$, while the one-form has a negative one. Still the overall norm 
	\begin{equation}
	\displaystyle \|\psi\rangle\|^2=\sum_{a<b}\psi_{ab}\psi^*_{ab}-\sum_a\psi_a\psi_a^* \label{counter}
	\end{equation}
could well be positive. Indeed this is the statement that the subspace $\bf{\cal H}^{\scriptscriptstyle (+)}$ of 
$\bf{\cal H}$ 
is not a vector space. In fact  in the example (\ref{cinque-trentadue}) we have summed a vector in 
$\bf{\cal H}^{\scriptscriptstyle (+)}$
with one belonging to $\bf{\cal H}^{\scriptscriptstyle (-)}$ and ended up in one in
$\bf{\cal H}^{\scriptscriptstyle (+)}$.  So the criterion that a state must contain an even number of
$\xi$ to have a positive norm is true only for homogeneous forms. Anyhow it may happen that a subspace of
$\bf{\cal H}^{\scriptscriptstyle (+)}$ is a vector space and this is what we shall find. 

Let us first stick to the homogeneous positive forms and let us check what happens under time evolution. First we  rewrite
the Hamiltonian $\widetilde{\cal H}$ (\ref{uno-quattordici}) in terms of the new variables
(\ref{cinque-venti-a}):
	\begin{equation}
	\widetilde{\cal H}=i\partial_aHl_a-i\bar{\partial}_aH\bar{l}_a+(\hat{\xi}_k\hat{\bar{\xi}}_a+\hat{\xi}^{*}_a\hat{\bar{\xi}}_k^*)
	\partial_k\bar{\partial}_aH
	+\hat{\xi}^{*}_a\hat{\bar{\xi}}_k\bar{\partial}_a\bar{\partial}_kH+\hat{\xi}_a\hat{\bar{\xi}}_k^*\partial_a\partial_kH 
	\label{cinque-trentatre}
	\end{equation}
where $\displaystyle \bar{\partial}_i=\frac{\partial}{\partial \bar{z}_i}$ and $\displaystyle
\partial_i=\frac{\partial}{\partial z_i}$. If we now take a generic homogeneous state of positive norm, i.e., with an even
number of $\xi$:
	\begin{equation}
	\displaystyle \psi=\frac{1}{q!}\psi_{ij\ldots q}\xi_i\xi^{*}_j\ldots \xi_q, \label{cinque-trentaquattro}
	\end{equation}
in general the time evolution will turn it into a positive norm state because $\widetilde{\cal H}$ is Hermitian and the
evolution is unitary, but the final state will be the sum of two terms, the first with an even number of $\xi$ and the second with an odd
number. In fact, while the first four terms (\ref{cinque-trentatre}) in $\widetilde{\cal H}$ do not change the
number of $\xi$ and $\xi^*$ factors in the state (\ref{cinque-trentaquattro}), the last two terms in
(\ref{cinque-trentatre}) do. For example the first of these two terms removes a factor $\xi$ from
(\ref{cinque-trentaquattro}) and injects a factor $\xi^*$ into it. So the resulting state is an {\it inhomogeneous} 
form in $\xi$.
Our $\widetilde{\cal H}$ does not conserve the form number in $\xi_i$
and $\xi^{*}_i$ separately while it conserves the one in $c^a$. If we restrict our space of homogeneous positive norm states to those
which are annihilated by the last two terms (\ref{cinque-trentatre}), then the time evolution will occur only via the first
four terms of (\ref{cinque-trentatre}) and these, as we said previously, will not modify the number of $\xi$ and $\xi^*$
contained in the state. It is easy \cite{Appendix C} to check that states of the form:
	\begin{equation}
	\displaystyle \psi_{phys}\equiv\psi_0(\varphi)+B(\varphi)\sum_{i,j}\xi_i\xi_{i}^*\xi_j\xi_{j}^*
	+C(\varphi)\sum_{i,j,k,l}\xi_i\xi_{i}^*\xi_j\xi_{j}^*\xi_k\xi_{k}^*\xi_l\xi_{l}^*+\ldots \label{cinque-trentacinque}
	\end{equation}
are annihilated by the last two terms of $\widetilde{\cal H}$ (\ref{cinque-trentatre}). The features of these states 
are: \medskip
\newline
{\bf 1)} each homogeneous form contained in them is made of products of an even number of $\xi_i,\xi_{i}^*$;\medskip
\newline
{\bf 2)} all indices are summed over;\medskip
\newline {\bf 3)} in the homogeneous forms each term has the same coefficient. 
In our case for the terms of the
4-form the coefficient is $B(\varphi)$ and for the terms of the 8-form is $C(\varphi)$.
\medskip
\newline
These states have positive norm because they are the sum of orthogonal positive norm states. 
Moreover the time evolution turns them in states with
the same feature because the last two terms in $\widetilde{\cal H}$, which could break the pairs
$\xi_i\xi_{i}^*$, give zero on states of the form (\ref{cinque-trentacinque}). So this family of states is closed under time
evolution. Last but not least, differently than generic positive norm states, those of the form
(\ref{cinque-trentacinque}) make a vector space: the sum of two with arbitrary coefficients is still a form which has the
properties {\bf 1), 2), 3)} which define this family. 
These states have all the features to be physical states: they have positive norm, they are closed under time evolution and
they make a vector space. They may not be the only physical states but, after a long analysis, we have the strong
feeling they are. A further
feature they have is that, not only the last two terms of $\widetilde{\cal H}$
but also the previous two containing second derivatives of $H$ annihilate them \cite{Appendix C}:
	\begin{equation}
	[(\hat{\xi}_k\hat{\bar{\xi}}_a+\hat{\xi}_{a}^*\hat{\bar{\xi}}_k^*)\partial_k\bar{\partial}_aH]\psi_{phys}=0.
	\label{cinque-trentasei}
	\end{equation}
The four terms containing  second derivatives of $H$ are what we called $\widetilde{\cal H}_{ferm}$ 
in the first part of this section. So (\ref{cinque-trentasei}) implies that
	\begin{equation}
	\widetilde{\cal H}_{ferm}\psi_{phys}=0. \label{cinque-trentasette}
	\end{equation}
It is trivial to prove that this feature is preserved under time evolution because 

\noindent $[\widetilde{\cal H},
\widetilde{\cal H}_{ferm}]\psi_{phys}=0$. Note that Eq. (\ref{cinque-trentasette}) is the same as Eq. 
(\ref{cinque-diciotto}) obtained in the SvH case. It implies that also for the states (\ref{cinque-trentacinque}) there is no evolution
of the Grassmannian variables. Only $\widetilde{\cal H}_{bos}$ is left to evolve the states and it acts on the
coefficients $\psi_0(\varphi),B(\varphi),C(\varphi)$, which all evolve as zero-forms. In this sense the symplectic physical states,
like those for the SvH case, are ``isomorphic" to a set of zero-forms. Nevertheless the SvH physical states are many more than
the symplectic physical ones. 
In fact if we take for example a 4-form in (\ref{cinque-trentacinque}), by turning the $\xi_i,\xi_{i}^*$ into $c^q,c^p$
variables via (\ref{cinque-venti-c}), we get :
	\begin{eqnarray}
	\displaystyle &&
	A(z,\bar{z})\xi_i\xi_{i}^*\xi_j\xi_{j}^*=\frac{\widetilde{A}(\varphi)}{4}[(c^{q_i}+ic^{p_i})(c^{q_i}-ic^{p_i})
	(c^{q_j}+ic^{p_j})(c^{q_j}-ic^{p_j})]=\nonumber\\
	&&=\frac{\widetilde{A}(\varphi)}{4}[(-ic^{q_i}c^{p_i}+ic^{p_i}c^{q_i})(-ic^{q_j}c^{p_j}+ic^{p_j}c^{q_j})]=
	\frac{\widetilde{A}(\varphi)}{4}2ic^{p_i}c^{q_i}c^{p_j}c^{q_j}2i=\nonumber\\ 
	&&=-\widetilde{A}(\varphi)c^{p_i}c^{q_i}c^{p_j}c^{q_j}
	\label{cinque-trentotto}
	\end{eqnarray}
and this is the same as a physical 4-form in the SvH case. But if we take a 6-form an analogous calculation gives:
	\begin{equation}
	\displaystyle A(z,\bar{z})\:\xi_i\xi_{i}^*\xi_j\xi_{j}^*\xi_k\xi_{k}^*=
	-i\widetilde{A}(\varphi)\:c^{p_i}c^{q_i}c^{p_j}c^{q_j}c^{p_k}c^{q_k}.
	\end{equation}
A 6-form like this is physical in the  SvH case  because it is annihilated by $\widetilde{\cal H}_{ferm}$ and so
$\widetilde{\cal H}$ is Hermitian on it. Nevertheless in the symplectic case 
it cannot be a physical form since it has negative norm because it contains an odd number of $\xi$. 
So the class of physical states is wider in the SvH case than in the symplectic one. 

What about the gauge scalar product we analyzed in Section III? Actually we are less interested in it because the
zero-forms have zero norm violating in this way the main feature of the KvN scalar product which we wanted to
maintain. Nevertheless, if one wants  to find the physical Hilbert space also in this
case, the way to proceed is the following. Let us define the new Grassmannian variables:
	\begin{equation}
	 \left\{
		\begin{array}{l}
		\displaystyle 
	\label{cinque-trentanove}	
		\displaystyle
		\hat{\psi}^a\equiv\frac{\hat{c}^a+i\omega^{ab}\hat{\bar{c}}_b}{\sqrt{2}}
	        \smallskip \nonumber\\
	        \displaystyle 
	        \hat{\bar{\psi}}_a\equiv\frac{\hat{\bar{c}}_a+i\omega_{ab}\hat{c}^b}{\sqrt{2}}.
	        \end{array}
		\right.
	\end{equation}
Using the symplectic hermiticity conditions (\ref{quattro-uno}) for the variables
($\hat{c}^a,\hat{\bar{c}}_a$), we get that
	\begin{equation}
	\hat{\psi}^{a\dagger}=\hat{\psi}^a, \qquad,\qquad \hat{\bar{\psi}}_a^{\dagger}=\hat{\bar{\psi}}_a
	\label{cinque-trentanove-a}
	\end{equation}
which means that $\hat{\psi}^a$ and $\hat{\bar{\psi}}_a$ are Hermitian like the Grassmannian variables in the gauge
scalar product (\ref{tre-due}). This is the connection between the symplectic and the gauge scalar product.
It is also easy to prove that the commutation relations among the ($\hat{\psi}^a,\hat{\bar{\psi}}_a$) are the same as the
ones among the variables $\hat{c}^a$
	\begin{equation}
	[\hat{\psi}^a,\hat{\bar{\psi}}_b]=\delta_b^a,\;\;\;\; [\hat{\psi}^a,\hat{\psi}^b]=0,\;\;\;\; 
	[\hat{\bar{\psi}}_a,\hat{\bar{\psi}}_b]=0.
	\end{equation}
The inverse transformation of (\ref{cinque-trentanove}) is
	\begin{equation}
	 \left\{
		\begin{array}{l}
		\displaystyle 
	\label{cinque-quaranta}	
		\displaystyle
		\hat{c}^a=\frac{\hat{\psi}^a-i\omega^{ab}\hat{\bar{\psi}}_b}{\sqrt{2}}
	        \smallskip \nonumber\\
	        \displaystyle 
	        \hat{\bar{c}}_a=\frac{\hat{\bar{\psi}}_a-i\omega_{ab}\hat{\psi}^b}{\sqrt{2}}.
	        \end{array}
		\right.
	\end{equation}
Having proved all this we can introduce, as in (\ref{cinque-venti-c}), the Grassmannian variables: 
	\begin{eqnarray}
	&&\displaystyle \hat{\xi}_i=\frac{1}{\sqrt{2}}(\hat{c}^{q_i}+i\hat{c}^{p_i})=\frac{1}{2}(\hat{\psi}^{q_i}+i\hat{\psi}^{p_i}-i
	\hat{\bar{\psi}}_{p_i}-\hat{\bar{\psi}}_{q_i})\nonumber\\
	&&\displaystyle
	\hat{\xi}_i^*=\frac{1}{\sqrt{2}}(\hat{c}^{q_i}-i\hat{c}^{p_i})=\frac{1}{2}(\hat{\psi}^{q_i}-i\hat{\psi}^{p_i}-
	i\hat{\bar{\psi}}_{p_i}+\hat{\bar{\psi}}_{q_i})\nonumber\\
	&&\displaystyle \hat{\bar{\xi}}_i=\frac{1}{\sqrt{2}}(-\hat{\bar{c}}_{q_i}+i\hat{\bar{c}}_{p_i})=
	\frac{1}{2}(-\hat{\bar{\psi}}_{q_i}-i\hat{\psi}^{p_i}+i\hat{\bar{\psi}}_{p_i}+\hat{\psi}^{q_i})
	\nonumber\\
	&&\displaystyle \hat{\bar{\xi}}_i^*=\frac{1}{\sqrt{2}}(\hat{\bar{c}}_{q_i}+i\hat{\bar{c}}_{p_i})=
	\frac{1}{2}(\hat{\bar{\psi}}_{q_i}+i\hat{\psi}^{p_i}+i\hat{\bar{\psi}}_{p_i}+\hat{\psi}^{q_i}).
	\end{eqnarray}
It is easy to realize that, if $\hat{\psi}$ and $\hat{\bar{\psi}}$ satisfy the algebra and the commutation relations
of the gauge scalar product, then the set of operators $\hat{\xi}, \hat{\xi}^*, \hat{\bar{\xi}},
\hat{\bar{\xi}}^*$ satisfy exactly the Eqs. (\ref{cinque-ventuno})-(\ref{cinque-ventidue}). Therefore, starting from the
gauge scalar product, we can repeat the same kind of considerations made in the symplectic case in order to find
out what is the subset of physical states. 

\section{Generalized Scalar Products}
\noindent
From the previous sections we can conclude that all the three scalar products we have analyzed have the feature of having either
$\widetilde{\cal H}$ non-Hermitian or the scalar product non-positive definite. The reader may object that this is a
feature of these particular scalar products but that there may be other more general ones which have both $\widetilde{\cal
H}$ Hermitian and no negative-norm states. We shall show in this section that even the most general scalar product cannot
have both features.

Let us limit our analysis to the case of $n=1$ and $\displaystyle H=\frac{p^2}{2}+V(q)$. The logic of what we will do next
is to find out which are the most general hermiticity conditions for ($\hat{c}^a,\hat{\bar{c}}_a$) under which 
$\widetilde{\cal H}$ is Hermitian \cite{footnote9}. Having done this,
we will then analyze whether any of the scalar products associated to those most general hermiticity conditions is positive definite. 

The bosonic part of $\widetilde{\cal H}$ is always Hermitian and therefore we should only care about the fermionic part
which is
	\begin{equation}
	\widetilde{\cal H}_{ferm}=i\hat{\bar{c}}_q\hat{c}^p-i\hat{\bar{c}}_pV^{\prime\prime}(q)\hat{c}^q. \label{sei-uno}
	\end{equation}
For this to be Hermitian the two pieces on the RHS of (\ref{sei-uno}) must be separately Hermitian. In fact the
second one contains the potential $V$ differently from the first. So we must have:
	\begin{eqnarray}
	&& \label{sei-due} (i\hat{\bar{c}}_q\hat{c}^p)^{\dagger}=i\hat{\bar{c}}_q\hat{c}^p \\
	&& \label{sei-tre} (-i\hat{\bar{c}}_p\hat{c}^q)^{\dagger}=-i\hat{\bar{c}}_p\hat{c}^q.
	\end{eqnarray}
Let us now, for example, see which is the most general hermiticity condition on $\hat{\bar{c}}_q$ and $\hat{c}^p$ which 
satisfies (\ref{sei-due}). Imposing a general condition of the form:
	\begin{equation}
	 \left\{
		\begin{array}{l}
		\displaystyle 
	        \label{sei-quattro}	
		\displaystyle
		\hat{c}^{p\dagger}=\alpha\hat{c}^p+\beta\hat{\bar{c}}_q
	        \smallskip \nonumber\\
	        \displaystyle 
	        \hat{\bar{c}}_q^{\dagger}=\gamma\hat{c}^p+\delta\hat{\bar{c}}_q
	        \end{array}
		\right.
	\end{equation}
and inserting it in (\ref{sei-due}) we get
	\begin{equation}
	\alpha\delta-\beta\gamma=1. \label{sei-cinque}
	\end{equation}
Besides (\ref{sei-due}) we also have to satisfy the relations $\bigl((\hat{c}^p)^{\dagger}\bigr)^{\dagger}=\hat{c}^p,
\; \bigl((\hat{\bar{c}}_q)^{\dagger}\bigr)^{\dagger}=\hat{\bar{c}}_q$ and inserting (\ref{sei-quattro}) into them
we get the following conditions on the coefficients $\alpha,\beta,\gamma,\delta$:
	\begin{eqnarray}
	\label{sei-sei-a}	
		\displaystyle
		&&\alpha^*\alpha+\beta^*\gamma=1
	        \smallskip \\
	        \label{sei-sei-b}
	        \displaystyle 
	        &&\alpha^*\beta+\beta^*\delta=0
	\end{eqnarray}
	\begin{eqnarray}
	        \label{sei-sette-a}	
		\displaystyle
		&&\alpha\gamma^*+\delta^*\gamma=0
	        \smallskip\\
	        \label{sei-sette-b}
	        \displaystyle 
	        &&\gamma^*\beta+\delta^*\delta=1.
	\end{eqnarray}
From (\ref{sei-sei-a}) and (\ref{sei-sette-b}) we get 
	\begin{equation}
	|\alpha|^2-|\delta|^2=-2i \,\text{Im}(\beta^*\gamma) \label{sei-otto}
	\end{equation}
and, being the LHS of (\ref{sei-otto}) real and the RHS imaginary, we obtain 
	\begin{equation}
	 \left\{
		\begin{array}{l}
	        \label{sei-nove-a}	
		\displaystyle
		|\alpha|=|\delta|
	        \smallskip \\
	        \label{sei-nove-b}
	        \displaystyle 
	        \text{Im}(\beta^*\gamma)=0.
	        \end{array}
		\right.
	\end{equation}
So $\beta^*\gamma$ is a real number and from (\ref{sei-sei-a}) we get that 
	\begin{equation}
	\beta^*\gamma \leq 1. \label{sei-dieci}
	\end{equation}
In Appendix D we will analyze the three cases of (\ref{sei-dieci}) 
	\begin{equation}
	 \left\{
		\begin{array}{l}
	        \label{sei-undici}	
		\displaystyle
		\beta^*\gamma=1
	        \smallskip \nonumber\\
	        \displaystyle 
	        \beta^*\gamma=0
	        \smallskip \nonumber\\
	        \beta^*\gamma=z
	        \end{array}
		\right.
	\end{equation}
where $z$ is real and different from 0 and 1. We shall show that the hermiticity conditions
(\ref{sei-quattro}) which we will get in the three cases (\ref{sei-undici}) are respectively:
\begin{eqnarray}
	\beta^*\gamma=1 & \Rightarrow & \left\{
		\begin{array}{l}
	        \label{sei-dodici-1}	
		\displaystyle
		\hat{c}^{p\dagger}=ib\hat{\bar{c}}_q
	        \smallskip \\
	        \displaystyle 
	        \hat{\bar{c}}_q^{\dagger}=\frac{i}{b}\hat{c}^p
	        \end{array}
		\right. \\
		\medskip
	\beta^*\gamma=0 & \Rightarrow & \left\{
		\begin{array}{l}
	        \label{sei-dodici-2}	
		\displaystyle
		\hat{c}^{p\dagger}=e^{i\theta_{\alpha}}\hat{c}^p
	        \smallskip \\
	        \displaystyle 
	        \hat{\bar{c}}_q^{\dagger}=i\gamma_I\hat{c}^p+e^{-i\theta_{\alpha}}\hat{\bar{c}}_q
	        \end{array}
		\right. \\
		\medskip
	\beta^*\gamma=z & \Rightarrow & \left\{
		\begin{array}{l}
	        \label{sei-dodici-3}	
		\displaystyle
		\hat{c}^{p\dagger}=e^{i\theta_{\alpha}}\hat{c}^p+ib\hat{\bar{c}}_q
	        \smallskip \\
	        \displaystyle 
	        \hat{\bar{c}}_q^{\dagger}=e^{-i\theta_{\alpha}}\hat{\bar{c}}_q
	        \end{array}
		\right.
\end{eqnarray}
where $b$ and $\gamma_I$ are respectively the imaginary part of $\beta$ and $\gamma$. The variable 
$\theta_\alpha$ is the phase of $\alpha$ 
($\alpha=e^{i\theta_{\alpha}}$). Note that, if $b=-1$, Eq. (\ref{sei-dodici-1}) would give part of the hermiticity conditions of
the symplectic scalar product (\ref{quattro-uno}). The choice $\theta_{\alpha}=0$ and $\gamma_I=0$ for (\ref{sei-dodici-2})
would give part of the gauge scalar product hermiticity conditions (\ref{tre-due}). The choice $\theta_{\alpha}=0$ and
$b=0$ in (\ref{sei-dodici-3}) would also give part of the gauge scalar product hermiticity conditions
(\ref{tre-due}). So basically we got, via our procedure, generalizations of either the symplectic or the gauge scalar
product. We did not get generalizations of the SvH one because in that case $\widetilde{\cal H}$ was not
Hermitian while here our procedure was that of searching for all the scalar products under which $\widetilde{\cal H}$ is
Hermitian. Let us remember that up to now we have only satisfied the conditions (\ref{sei-due}). If we want to satisfy
(\ref{sei-tre}) we should take steps  similar to those performed up to now. Note that
(\ref{sei-tre}) can be obtained from (\ref{sei-due}), replacing $p$ with $q$ and vice versa. So the associated hermiticity
conditions can be derived from (\ref{sei-dodici-1}), (\ref{sei-dodici-2}),
(\ref{sei-dodici-3}), replacing $p$ with $q$ and vice versa: 
	\begin{eqnarray}
	 &&\left\{
		\begin{array}{l}
	        \label{sei-tredici-1}	
		\displaystyle
		\hat{c}^{q\dagger}=ia\hat{\bar{c}}_p
	        \smallskip \\
	        \displaystyle 
	        \hat{\bar{c}}_p^{\dagger}=\frac{i}{a}\hat{c}^q
	        \end{array}
		\right. \\
		\medskip
	&&\left\{
		\begin{array}{l}
	        \label{sei-tredici-2}	
		\displaystyle
		\hat{c}^{q\dagger}=e^{i\theta_{\beta}}\hat{c}^q
	        \smallskip \\
	        \displaystyle 
	        \hat{\bar{c}}_p^{\dagger}=i\gamma_I^{\prime}\hat{c}^q+e^{-i\theta_{\beta}}\hat{\bar{c}}_p
	        \end{array}
		\right. \\
		\medskip
	&&\left\{
		\begin{array}{l}
	        \label{sei-tredici-3}	
		\displaystyle
		\hat{c}^{q\dagger}=e^{i\theta_{\beta}}\hat{c}^q+ia\hat{\bar{c}}_p
	        \smallskip \\
	        \displaystyle 
	        \hat{\bar{c}}_p^{\dagger}=e^{-i\theta_{\beta}}\hat{\bar{c}}_p.
	        \end{array}
		\right.
	\end{eqnarray}
In the formulae above the variables $a,\theta_{\beta},\gamma_I^{\prime}$ are real parameters that can vary like 
$b,\theta_{\alpha},\gamma_I$ 
did in (\ref{sei-dodici-1}), (\ref{sei-dodici-2}), (\ref{sei-dodici-3}). Now, having three conditions
(\ref{sei-dodici-1}), (\ref{sei-dodici-2}), (\ref{sei-dodici-3}) which satisfy (\ref{sei-due}), and three
(\ref{sei-tredici-1}), (\ref{sei-tredici-2}), (\ref{sei-tredici-3}) which satisfy (\ref{sei-tre}), 
we can have nine combinations which satisfy
both (\ref{sei-due}) and (\ref{sei-tre}). 
The next step in our procedure is to see whether among the scalar products associated to the nine hermiticity conditions 
(\ref{sei-dodici-1})-(\ref{sei-tredici-3}) there is any which is positive definite. In order to do this analysis in a
neater form we shall introduce a metric $g^{ij}$ in each scalar product. This metric is defined for $n=1$ in the
following way. Let us write the states as
	\begin{equation}
	 \left\{
		\begin{array}{l}
	        \label{sei-quattordici}	
		\displaystyle
		\psi=\psi_0+\psi_1c^q+\psi_2c^p+\psi_3c^qc^p
	        \smallskip \nonumber\\
	        \displaystyle 
	        \Phi=\Phi_0+\Phi_1c^q+\Phi_2c^p+\Phi_3c^qc^p
	        \end{array}
		\right.
	\end{equation}
then we express the scalar product as 
	\begin{equation}
	\langle \Phi|\psi\rangle=\int d\varphi\,\Phi_i^*g^{ij}\psi_j  \label{sei-quindici}
	\end{equation}
where $g^{ij}$ is the metric mentioned above and $i,j$ can be $(0,1,2,3)$. It is easy to get convinced that all the three
scalar products, SvH (\ref{due-cinquantuno}), gauge (\ref{tre-ventotto-x}) and symplectic (\ref{quattro-quattordici}) one,
 can be
written in the form (\ref{sei-quindici}) with different metrics. 

Let us now see which metric we obtain out of the first set of hermiticity conditions presented in (\ref{sei-dodici-1})
and (\ref{sei-tredici-1}) which are:
	\begin{equation}
	 \left\{
		\begin{array}{l}
	        \label{sei-sedici}	
		\displaystyle
		\hat{c}^{p\dagger}=ib\hat{\bar{c}}_q
	        \smallskip \nonumber\\
	        \displaystyle 
	        \hat{\bar{c}}_q^{\dagger}=\frac{i}{b}\hat{c}^p
	        \smallskip \nonumber\\
	        \hat{c}^{q\dagger}=ia\hat{\bar{c}}_p
	        \smallskip \nonumber\\
	        \displaystyle \hat{\bar{c}}_p^{\dagger}=\frac{i}{a}\hat{c}^q.
	        \end{array}
		\right.
	\end{equation}
The first two equations of (\ref{sei-sedici}) can be written as  
	\begin{equation}
	 \left\{
		\begin{array}{l}
	        \label{sei-diciassette}	
		\displaystyle
		\langle \hat{c}^p\Phi|\psi\rangle=\langle \Phi|ib\frac{\partial}{\partial c^q}\psi\rangle
	        \smallskip \nonumber\\
	        \displaystyle 
	        \langle ib\frac{\partial}{\partial c^q}\Phi|\psi\rangle=\langle \Phi|\hat{c}^p\psi\rangle .
	        \end{array}
		\right.
	\end{equation}
Similarly the last two equations of (\ref{sei-sedici}) are equivalent to 
	\begin{equation}
	 \left\{
		\begin{array}{l}
	        \label{sei-diciotto}	
		\displaystyle
		\langle \hat{c}^q\Phi|\psi\rangle=\langle\Phi|ia\frac{\partial}{\partial c^p}\psi\rangle
	        \smallskip \nonumber\\
	        \displaystyle 
	        \langle ia\frac{\partial}{\partial c^p}\Phi|\psi\rangle=\langle\Phi|\hat{c}^q\psi\rangle.
	        \end{array}
		\right.
	\end{equation}
Inserting (\ref{sei-quindici}) into (\ref{sei-diciassette}), (\ref{sei-diciotto})
we get that for consistency the parameters $a,b$ entering in (\ref{sei-sedici}) must be related as follows
	\begin{equation}
	a=-b \label{sei-diciotto-b}
	\end{equation}
and that, with the choice $g^{00}=1$ (explained in Appendix D), the whole metric turns out to be
	\begin{equation}
	g^{ij}=\pmatrix{1 & 0 & 0 & 0\cr 0 & 0 & -ib & 0\cr 0 & ib & 0 & 0\cr 0 & 0 & 0 & -b^2}.
	\label{sei-diciannove}
	\end{equation}
So this is the metric whose associated scalar product (\ref{sei-quindici}) produces the hermiticity conditions
(\ref{sei-sedici}). The details of these calculations are worked out in Appendix D. Let us first notice that with the
choice $b=-1$ the metric (\ref{sei-diciannove}) reproduces, via (\ref{sei-quindici}), the symplectic scalar product
(\ref{quattro-quattordici}). So we expect that at least one case of (\ref{sei-diciannove}), the symplectic one, will not give a positive
definite scalar product. In general, to check whether (\ref{sei-diciannove}) gives positive definite
scalar products, we should calculate  the eigenvalues of (\ref{sei-diciannove}) and see if they are all positive. It is easy
to work out the four eigenvalues $\lambda_i$ of (\ref{sei-diciannove}) and they are:
	\begin{equation}
	 \left\{
		\begin{array}{l}
	        \label{sei-venti}	
		\displaystyle
		\lambda_1=1
	        \smallskip \nonumber\\
	        \displaystyle 
	        \lambda_2=+b
	        \smallskip \nonumber\\
	        \displaystyle
	        \lambda_3=-b 
	        \smallskip \nonumber\\
	        \displaystyle 
	        \lambda_4=-b^2.
	        \end{array}
		\right.
	\end{equation}
So, being $b$ real, we see that there always are two negative eigenvalues. This ultimately confirms that the scalar product
associated to (\ref{sei-diciannove}) is not positive definite. Let us now turn to another of the nine hermiticity
conditions, in particular the one obtained combining (\ref{sei-dodici-1}) with (\ref{sei-tredici-2}):
	\begin{equation}
	 \left\{
		\begin{array}{l}
	        \label{sei-ventuno}	
		\displaystyle
		\hat{c}^{p\dagger}=ib\hat{\bar{c}}_q
	        \smallskip \nonumber\\
	        \displaystyle 
	        \hat{\bar{c}}_q^{\dagger}=\frac{i}{b}\hat{c}^p
	        \smallskip \nonumber\\
	        \displaystyle
	        \hat{c}^{q\dagger}=e^{i\theta_{\beta}}\hat{c}^q
	        \smallskip \nonumber\\
	        \displaystyle 
	        \hat{\bar{c}}_p^{\dagger}=i\gamma_I^{\prime} \hat{c}^q+e^{-i\theta_{\beta}}\hat{\bar{c}}_p.
	        \end{array}
		\right.
	\end{equation}
It is easy to realize that this choice of hermiticity conditions is not consistent. In fact from the standard commutation
relation $[\hat{c}^q,\hat{\bar{c}}_q]=1$ we get this other one 
	\begin{equation}
	[\hat{c}^{q\dagger},\hat{\bar{c}}_q^{\dagger}]=1 \label{sei-ventidue}
	\end{equation}
and replacing in (\ref{sei-ventidue}) the expression obtained from (\ref{sei-ventuno}) we get 
	\begin{equation}
	\displaystyle [\hat{c}^{q\dagger},\hat{\bar{c}}_q^{\dagger}]=[e^{i\theta_{\beta}}\hat{c}^q,(i/b)\hat{c}^p]=0.
	\end{equation}
This contradicts (\ref{sei-ventidue}). 
We have an analogous problem with the hermiticity conditions obtained combining (\ref{sei-dodici-1}) and (\ref{sei-tredici-3})
	\begin{equation}
	 \left\{
		\begin{array}{l}
	        \label{sei-ventitre}	
		\displaystyle
		\hat{c}^{p\dagger}=ib\hat{\bar{c}}_q
	        \smallskip \nonumber\\
	        \displaystyle 
	        \hat{\bar{c}}_q^{\dagger}=\frac{i}{b}\hat{c}^p
	        \smallskip \nonumber\\
	        \displaystyle
	        \hat{c}^{q\dagger}=e^{i\theta_{\beta}}\hat{c}^q+ia\hat{\bar{c}}_p
	        \smallskip \nonumber\\
	        \displaystyle 
	        \hat{\bar{c}}_p^{\dagger}=e^{-i\theta_{\beta}}\hat{\bar{c}}_p.
	        \end{array}
		\right.
	\end{equation}
In fact from $[\hat{c}^p,\hat{\bar{c}}_p]$=1 we get 
	\begin{equation}
	[\hat{c}^{p\dagger},\hat{\bar{c}}_p^{\dagger}]=1 \label{sei-ventiquattro}
	\end{equation}
but inserting (\ref{sei-ventitre}) in (\ref{sei-ventiquattro}) we obtain 
	\begin{equation}
	[\hat{c}^{p\dagger},\hat{\bar{c}}_p^{\dagger}]=[ib\hat{\bar{c}}_q, e^{-i\theta_{\beta}}\hat{\bar{c}}_p]=0
	\end{equation}
and this contradicts (\ref{sei-ventiquattro}).

Another hermiticity condition which is inconsistent is the one obtained 
combining  (\ref{sei-dodici-2}) with (\ref{sei-tredici-1}). In that case in fact, instead of 
(\ref{sei-ventiquattro}), we would get
	\begin{equation}
	\displaystyle [\hat{c}^{p\dagger},\hat{\bar{c}}_p^{\dagger}]=[e^{i\theta_{\alpha}}\hat{c}^p,
	(i/a)\hat{c}^q]=0.
	\end{equation}
Analogously the hermiticity conditions obtained combining (\ref{sei-dodici-3}) and (\ref{sei-tredici-1}) are inconsistent.
In fact in this case we get
	\begin{equation}
	[\hat{c}^{q\dagger},\hat{\bar{c}}_q^{\dagger}]=[ia\hat{\bar{c}}_p,e^{-i\theta_{\alpha}}\hat{\bar{c}}_q]=0
	\end{equation}
which contradicts (\ref{sei-ventidue}). Let us now analyze the case (\ref{sei-dodici-2})-(\ref{sei-tredici-2}) which gives
as hermiticity conditions
	\begin{equation}
	 \left\{
		\begin{array}{l}
	        \label{sei-ventiquattro-b}	
		\displaystyle
		\hat{c}^{p\dagger}=e^{i\theta_{\alpha}}\hat{c}^p
	        \smallskip \nonumber\\
	        \displaystyle 
	        \hat{\bar{c}}_q^{\dagger}=i\gamma_I\hat{c}^p+e^{-i\theta_{\alpha}}\hat{\bar{c}}_q
	        \smallskip \nonumber\\
	        \displaystyle
	        \hat{c}^{q\dagger}=e^{i\theta_{\beta}}\hat{c}^q
	        \smallskip \nonumber\\
	        \displaystyle 
	        \hat{\bar{c}}_p^{\dagger}=i\gamma_I^{\prime} \hat{c}^q+e^{-i\theta_{\beta}}\hat{\bar{c}}_p.
	        \end{array}
		\right.
	\end{equation}
In Appendix D we will show that, when we make the choice $\theta_{\alpha}=\theta_{\beta}$ and 
$\gamma_I=\gamma_I^{\prime}=0$, this leads to a metric $g^{ij}$ of the form:
	\begin{equation}
	g^{ij}=\pmatrix{0 & 0 & 0 & g^{03}\cr  
	0 & 0 & g^{03}e^{i\theta_{\alpha}} & 0\cr 
	0 & -g^{03}e^{i\theta_{\alpha}} & 0 & 0\cr 
	-g^{03}e^{2i\theta_{\alpha}} & 0 & 0 & 0} \label{sei-venticinque}
	\end{equation}
while, when we make the choice $\gamma_I=-\gamma_I^{\prime}\neq 0$,
we get 
	\begin{equation}
	g^{ij}=\pmatrix{ig^{03}e^{i\theta_{\alpha}}\gamma_I & 0 & 0 & g^{03}\cr
	0 & 0 & g^{03}e^{i\theta_{\alpha}} & 0\cr 
	0 & -g^{03}e^{i\theta_{\alpha}} & 0 & 0\cr
	-g^{03}e^{2i\theta_{\alpha}} & 0 & 0 & 0}. \label{sei-ventisei}
	\end{equation}
The eigenvalue equation associated to this metric is
	\begin{equation}
	\bigl(\lambda^2+(g^{03})^2e^{2i\theta_{\alpha}}\bigr)
	\bigl(\lambda^2-ig^{03}e^{i\theta_{\alpha}}\gamma_I\lambda+(g^{03})^2
	e^{2i\theta_{\alpha}}\bigr)=0.
	\end{equation}
Two of its eigenvalues are
	\begin{equation}
	\lambda=\pm ig^{03}e^{i\theta_{\alpha}} \label{sei-ventisette}
	\end{equation}
and one sees that the choice $\theta_{\alpha}=0$ and $g^{03}=i$ leads to a positive and a negative
eigenvalue for (\ref{sei-ventisette}). This proves that even the scalar product which is associated to the hermiticity
conditions (\ref{sei-ventiquattro-b}) is not positive definite. This case is a generalization of the gauge scalar product 
(\ref{tre-ventotto-x}); in fact it reduces to it with the choice 
$\gamma_I=0, \, g^{03}=-i, \,\theta_{\alpha}=0$. Even the case (\ref{sei-venticinque}) is not positive definite because
it has two eigenvalues which are the same as those in (\ref{sei-ventisette}) relative to the metric (\ref{sei-ventisei}).
Let us now proceed to other cases like for example the one whose hermiticity conditions are the combination 
of (\ref{sei-dodici-2}) and (\ref{sei-tredici-3}). From these relations we get 
	\begin{equation}
	 \left\{
		\begin{array}{l}
	        \label{sei-ventotto}	
		\displaystyle
		\hat{c}^{p\dagger}=e^{i\theta_{\alpha}}\hat{c}^p
	        \smallskip \nonumber\\
	        \displaystyle 
	        \hat{c}^{q\dagger}=e^{i\theta_{\beta}}\hat{c}^q+ia\hat{\bar{c}}_p
	        \smallskip \nonumber\\
	        \displaystyle 
	        \hat{\bar{c}}_q^{\dagger}=i\gamma_I\hat{c}^p+e^{-i\theta_{\alpha}}\hat{\bar{c}}_q
	        \smallskip \nonumber\\
	        \displaystyle 
	        \hat{\bar{c}}_p^{\dagger}=e^{-i\theta_{\beta}}\hat{\bar{c}}_p.
	        \end{array}
		\right.
	\end{equation}
Using these expressions and trying to satisfy the following relations:
	\begin{equation}
	 \left\{
		\begin{array}{l}
	        \label{sei-ventinove}	
		\displaystyle
		[\hat{c}^q,\hat{c}^p]^{\dagger}=0
	        \smallskip \nonumber\\
	        \displaystyle 
	        [\hat{\bar{c}}_q,\hat{\bar{c}}_p]^{\dagger}=0,
	        \smallskip \\
	        \end{array}
		\right.
	\end{equation}
we get that in (\ref{sei-dodici-2}) and (\ref{sei-tredici-3}) we must choose $a=\gamma_I=0$. This choice
makes those hermiticity conditions the same as those which led to the metric (\ref{sei-venticinque}). Therefore also the case
(\ref{sei-dodici-2})-(\ref{sei-tredici-3}) is not positive definite. A similar analysis can be carried out in 
the case where the hermiticity conditions are the combination of (\ref{sei-dodici-3}) and (\ref{sei-tredici-3}).
Also in this case we get that, in order to satisfy the relations (\ref{sei-ventinove}), we must put
$b=\gamma_I^{\prime}=0$ in (\ref{sei-dodici-3}) and (\ref{sei-tredici-2}). 
These relations are then the same as those of
(\ref{sei-dodici-2}) and (\ref{sei-tredici-2}) with $\gamma_I=\gamma_I^{\prime}=0$ 
that led to the metric (\ref{sei-venticinque}) which was not positive definite. 
The last of the nine hermiticity conditions that we have to examine is the
combination of (\ref{sei-dodici-3}) and (\ref{sei-tredici-3}).
This leads to the metric
	\begin{equation}
	g^{ij}=\pmatrix{0 & 0 & 0 & g^{03}\cr
	0 & 0 & g^{03}e^{i\theta_{\alpha}} & 0\cr 
	0 & -g^{03}e^{i\theta_{\alpha}} & 0 & 0\cr
	-g^{03}e^{2i\theta_{\alpha}} & 0 & 0 & -ig^{03}e^{i\theta_{\alpha}}b}. \label{sei-trenta}
	\end{equation}
The details of this derivation are contained in Appendix D. The eigenvalues of this metric are given by the solutions
of the equation
	\begin{equation}
	\bigl(\lambda^2+(g^{03})^2e^{i\theta_{\alpha}}\bigr)\bigl(\lambda^2+
	ig^{03}e^{i\theta_{\alpha}}b\lambda+(g^{03})^2e^{2i\theta_{\alpha}}\bigr)=0.
	\end{equation}
Two solutions are given by
	\begin{equation}
	\lambda=\pm ig^{03}e^{i\theta_{\alpha}} \label{sei-trentuno}
	\end{equation}
and the other two by the solutions of the equation 
	\begin{equation}
	\displaystyle \lambda^2+ig^{03}e^{i\theta_{\alpha}}b\lambda+(g^{03})^2e^{2i\theta_{\alpha}}=0.	
	\label{sei-trentadue}
	\end{equation}
Choosing in (\ref{sei-trentadue}) $g^{03}=i$ and $\theta_{\alpha}=0$ we get 
	\begin{equation}
	\lambda^2-b\lambda-1=0
	\end{equation}
and this equation has two solutions of opposite sign because its solutions $\lambda_1,\lambda_2$ must satisfy 
the relation $\lambda_1\lambda_2=-1$. Also the two solutions (\ref{sei-trentuno}), with $g^{03}=i$ and 
$\theta_{\alpha}=0$, are one positive and one negative
	\begin{equation}
	\lambda_{3,4}=\mp 1.
	\end{equation}
This confirms that, as the metric (\ref{sei-trenta}) can have negative eigenvalues, the scalar product associated to it is
not positive definite. It should be noticed that also (\ref{sei-trenta}) is a generalization of the gauge-scalar product
which is obtained with the following choice of parameters: $b=0$, $\theta_{\alpha}=0$, $g^{03}=-i$.

To summarize what we have done up to now, we can say that the whole set of consistent scalar products under which 
$\widetilde{\cal H}$ is Hermitian is the one associated to the three metrics (\ref{sei-ventisei}),
(\ref{sei-trenta}), (\ref{sei-diciannove}). 
The first two are generalizations of the gauge scalar product and the last is a generalization of the
symplectic case. None of these three generalizations leads to a positive definite scalar product. 
So we have proved that if
$\widetilde{\cal H}$ {\it is Hermitian} then {\it the associated scalar product is not positive definite}. 
As a consequence, based on standard rules of logic, we have that if
{\it the associated scalar product is positive definite} then
$\widetilde{\cal H}$ {\it is not Hermitian}. 
This second case is exemplified in the SvH scalar product. Of course the theorem above holds if we work in the full
Hilbert space.

Even with the generalized scalar products (\ref{sei-diciannove}), (\ref{sei-ventisei}), (\ref{sei-trenta})
it is easy to see that the subspace of positive norm states, closed under time evolution and with the feature of being a
vector space, is isomorphic to the space of zero-forms. We will skip the proof because it is very similar to those
presented in Sec. V.

\section{Conclusions}
\noindent
Despite the detailed mathematical analysis contained in this paper, the reader
may still be puzzled by the results we have
gotten. In fact it is difficult to accept that in classical mechanics, for a
{\it generic potential},
we cannot have at the same time a positive definite
scalar product and a unitary evolution in the space of forms. 
In this section we would like to give some tentative physical explanations
\cite{footnote10} of this result.

Let us for example analyze chaotic systems that are, loosely speaking, those
which have trajectories which fly away
exponentially as time passes by. The variables which describe this behavior
better are the so-called Jacobi fields which
are defined as
	\begin{equation}
	\delta\varphi^a(t,\varphi_0)=
	\varphi_2^a(t,\varphi_0+\delta\varphi_0)-\varphi_1^a(t,\varphi_0)
	\label{sette-uno}
	\end{equation}
where $\varphi_1^a(t)$ and $\varphi_2^a(t)$ are the two trajectories which start
at time $t=0$ very close to each other
respectively in $\varphi_0$ and $\varphi_0+\delta\varphi_0$.
We should notice that the evolution of the Jacobi fields $\delta\varphi^a$ is
the same as that of the Grassmann
variables $c^a$:
	\begin{equation}
	\displaystyle
	\biggl[\delta_b^a\partial_t-\omega^{al}\frac{\partial^2H}{\partial\varphi^l\partial\varphi^b}\biggr]
	\delta\varphi^b=0 .\label{sette-due}
	\end{equation}
The Euclidean square distance between the two trajectories in phase space is
given by
	\begin{equation}
	D(\varphi_0,t)\equiv\|\delta\varphi^a\|^2 \label{sette-tre}
	\end{equation}
and it is a function of $t$ and $\varphi_0$. In more precise mathematical terms
chaotic
systems are defined as those for which the following inequality holds
	\begin{equation}
	\displaystyle \lim_{t\to\infty}\frac{1}{t}\;\ln\int
	d^{2n}\varphi_0D(\varphi_0,t)>0. \label{sette-quattro}
	\end{equation}
One immediately infers from (\ref{sette-quattro}) that the phase space of
chaotic systems has regions of non-zero
measure such that the trajectories which originate from there fly away
exponentially as time passes by.

In Appendix E we will show that, in those regions, also the components
$\psi_q(\varphi_0)$, 
$\psi_p(\varphi_0)$ of any one-form $\psi=\psi_ac^a$ behave as the Jacobi
fields. 
So the norm of these states in the SvH scalar product
	\begin{equation}
	\displaystyle \int d^{2n}\varphi_0\sum_a|\psi_a(\varphi_0,t)|^2
	\label{sette-cinque}
	\end{equation}
diverges exponentially, analogously to (\ref{sette-quattro}).
Let us now take the sum of a zero-form $\psi_0$ and of a one-form
$\psi=\psi_ac^a$
	\begin{equation}
	\widetilde{\psi}\equiv\psi_0+\psi_ac^a.
	\end{equation}
The SvH norm of $\widetilde{\psi}$ is 
	\begin{equation}
	\displaystyle \|\widetilde{\psi}\|^2=\int
	d^{2n}\varphi_0|\psi_0(\varphi_0,t)|^2+\int d^{2n}\varphi_0\sum_a|\psi_a
	(\varphi_0,t)|^2 .\label{sette-sei}
	\end{equation}
Let us now suppose that in the SvH scalar product the $\widetilde{\cal H}$ were
Hermitian. If so then the norm of
$\widetilde{\psi}$ would be conserved under the time evolution. Anyhow we also
know that the second piece in
(\ref{sette-sei}) that is $\displaystyle \biggl(\int d^{2n}\varphi_0\sum_a
|\psi_a|^2\biggr)$ increases for chaotic systems
and 
this implies, for
$\|\widetilde{\psi}\|^2$ to be conserved, that the first term $\displaystyle 
\int d^{2n}\varphi_0|\psi_0|^2$ in (\ref{sette-sei}) cannot be conserved. The
non-conservation of this last piece
implies a violation of the Liouville theorem. 
To put things in simpler terms: if we have chaotic systems, i.e. exponential
behavior, we must be able to
produce, from the operator of evolution $\displaystyle e^{-i\widetilde{\cal
H}t}$, an exponential diverging behavior:
	\begin{equation}
	\displaystyle e^{-i\widetilde{\cal H}t}\;\;\longrightarrow K e^{+Lt}
	\label{sette-sette}
	\end{equation}
with $L$ a real number.
This happens only if $\widetilde{\cal H}$ is not Hermitian and has, as a
consequence, complex eigenvalues which would
produce
something like (\ref{sette-sette}). This same kind of behavior can be produced
in the gauge and symplectic case.
There $\widetilde{\cal H}$ is Hermitian but the scalar product is not positive
definite. In this case
even Hermitian operators can have complex eigenvalues and the proof goes as
follows. Let us start from the hermiticity of
$\widetilde{\cal H}$, i.e.:
	\begin{equation}
	\langle \psi| \widetilde{\cal H}\psi\rangle=\langle \widetilde{\cal H}
	\psi|\psi\rangle \label{sette-otto}
	\end{equation}
where $|\psi\rangle$ is an eigenstate of $\psi$ with eigenvalue $\lambda$; then
(\ref{sette-otto}) can be written as
	\begin{equation}
	\lambda\langle\psi|\psi\rangle=\lambda^*\langle\psi|\psi\rangle.
	\label{sette-nove}
	\end{equation}
From this relation we cannot deduce that $\lambda=\lambda^*$ because, in a
non-positive definite scalar product, the state
$|\psi\rangle$ could be of zero norm and satisfy (\ref{sette-nove}) whatever is
the value, real or complex,
of $\lambda$. 

This analysis explains why for a {\it generic potential}, as we have assumed
throughout this paper, we
have either positive definite Hilbert space or $\widetilde{\cal H}$ Hermitian
but never both of them. 
This last possibility
can
take place only for {\it specific potentials}. For example in the SvH case it
can happen for Hamiltonians of the form
	\begin{equation}
	\displaystyle H=\frac{1}{2}p^2+\frac{1}{2}q^2 
	\end{equation}
for which $\widetilde{\cal H}$ is Hermitian (see Eqs.
(\ref{due-cinquantatre})-(\ref{due-cinquantaquattro})). 
As this is a harmonic
oscillator the reader may be tempted to generalize the result and jump to the
conclusion that, differently than chaotic
systems, for
integrable ones we could have both $\widetilde{\cal H}$ Hermitian and the scalar
product positive definite. It is
actually not so: even among harmonic oscillators only some of them have the
features above as 
explained in Appendix F.
In the same appendix we prove that the Hermiticity condition of ${\cal H}$ is
invariant under 
canonical transformations.

The reader may wonder if all this formalism with Hilbert spaces and scalar
products is actually
needed to study dynamical systems. After all, he may argue, the old fashioned
Hamiltonian formalism
was enough and no scalar product appears there. This is actually not so! In
fact, to study the
ergodic/chaotic properties of a system in the standard Hamiltonian formalism 
one has to introduce by hand the Jacobi fields, like infinitesimal
differences of canonical variables, and
define their {\it length} in order to get the Lyapunov exponents; so a scalar
product, to get the length above, 
somehow would enter also in the standard Hamiltonian approach. In our formalism
instead we have automatically the Jacobi fields since they are related to the 
Grassmann variables and, having a Hilbert space, the introduction of a scalar
product is much more 
explicit than in the standard Hamiltonian formalism. So the reader can see that
the two approaches,
the Hamiltonian and the KvN one, are equivalent because even in the first case
one
has to introduce a scalar product. Moreover via our formalism we have an
interesting  manner to connect the Lyapunov exponents to well-defined
mathematical structures
like the spectral properties of a certain operator \cite{vari}. As we said in
the Introduction
in order to study the spectral properties of an operator it is crucial to know
which
are the features of the Hilbert space on which this operator acts. In particular
it is important 
to know which is the scalar product one uses, the Hermiticity character of
the Hamiltonian and all that. In Ref. \cite{vari} the {\it connection} between
Lyapunov exponents
and spectral properties of the Lie-derivative had been studied but at a rather
{\it formal}
level using the path integral formalism. We felt 
it would be important to check the {\it connection} above at a more {\it
rigorous} 
level using the operatorial
formalism. What we have done in this paper is to prepare 
the mathematical ground for that analysis which will be carried on in later
papers. 
Clearly all the work contained here is of a pure 
{\it mathematical} nature but we feel that, 
combined with the physics studied in Ref. \cite{vari}, it could lead to further 
physical understanding of dynamical systems and of chaotic ones in particular.
Moreover, the {\it mathematical} machinery prepared in this paper has brought to
light
some unexpected phenomena, like the non-Hermiticity of $\widetilde{\cal H}$ for
some
systems. This, we feel, is already a rather interesting result which fully
justifies
the detailed mathematics presented here.

This concludes this paper. In a second one \cite{second} we will analyze the
same issues 
by using an entirely bosonic $\widetilde{\cal H}$ first introduced in Ref.
\cite{Regini}. 
In the same paper
we shall also analyze what happens when we change representations for the
Grassmann variables 
\cite{metaplettico}.

\appendix
\makeatletter
\@addtoreset{equation}{section}
\makeatother
\renewcommand{\theequation}{\thesection\arabic{equation}}


\newpage

\section{}
\noindent
In this appendix we will prove a set of formulas contained in Sec. II.

\noindent First we should remember the usual expression of the Dirac deltas in Grassmannian variables:
	\begin{equation}
	\delta(\alpha-\beta)=\alpha-\beta;\;\;\;\;\;\int d\beta\,\delta(\beta-\alpha)f(\beta)=f(\alpha) \label{i-1}
	\end{equation}
and then we can proceed with the proofs.
\\

\noindent $\bullet$ \underline{\it Derivation of (\ref{due-tredici-a})}: 

\noindent The LHS is
	\begin{eqnarray}
	\displaystyle |\beta+\rangle&=&e^{-\beta c^q}|0+\rangle=e^{-\beta c^q}\bar{c}_q|0-\rangle=
	(1-\beta c^q)\bar{c}_q|0-\rangle=\bar{c}_q|0-\rangle-\beta c^q\bar{c}_q|0-\rangle=\nonumber\\
	&=&\bar{c}_q|0-\rangle-\beta[c^q,\bar{c}_q]_+|0-\rangle=(\bar{c}_q-\beta)|0-\rangle. \label{i-2}
	\end{eqnarray}
The RHS is
	\begin{eqnarray}
	\displaystyle -\int d\alpha \,e^{\alpha\beta}|\alpha-\rangle&=&-\int d\alpha\,
	e^{\alpha\beta}e^{-\alpha\bar{c}_q}|0-\rangle=-\int d\alpha \,e^{\alpha[\beta-\bar{c}_q]}|0-\rangle=\nonumber\\
	&=&-\delta(\beta-\bar{c}_q)|0-\rangle=-(\beta-\bar{c}_q)|0-\rangle=(\bar{c}_q-\beta)|0-\rangle. \label{i-3}
	\end{eqnarray}
As (\ref{i-2}) and (\ref{i-3}) are equal this proves the first relation of Eq. (\ref{due-tredici-a}). 
\\

\noindent $\bullet$ \underline{\it Derivation of (\ref{due-quattordici-a})}.

\noindent Let us evaluate, for example, the scalar product $\biggl(|\alpha -\rangle,|\beta-\rangle\biggr)$
	\begin{eqnarray}
	\displaystyle \biggl(|\alpha -\rangle,|\beta
	-\rangle\biggr)&=&\biggl((1-\alpha\bar{c}_q)|0-\rangle,(1-\beta\bar{c}_q)|0-\rangle\biggr)=\nonumber\\
	&=&\biggl(|0-\rangle,(1+\alpha^*c^q)(1-\beta\bar{c}_q)|0-\rangle\biggr)=\nonumber\\
	&=&\biggl(|0-\rangle,(1-\beta\bar{c}_q+\alpha^*c^q-\alpha^*c^q\beta\bar{c}_q)|0-\rangle\biggr)=\nonumber\\
	&=&\biggl(|0-\rangle,|0-\rangle-\beta\bar{c}_q|0-\rangle-\alpha^*c^q\beta\bar{c}_q|0-\rangle\biggr)=\nonumber\\
	&=&1+\beta\biggl(|0-\rangle,\bar{c}_q|0-\rangle\biggr)-\beta\alpha^*\biggl(|0-\rangle,c^q\bar{c}_q|0-\rangle\biggr)=\nonumber\\
	&=&1+\beta\biggl(c^q|0-\rangle,|0-\rangle\biggr)-\beta\alpha^*\biggl(|0-\rangle,[c^q,\bar{c}_q]|0-\rangle\biggr)=\nonumber\\
	&=&1-\beta\alpha^*=e^{\alpha^*\beta}.
	\end{eqnarray}
Now let us evaluate the scalar product $\biggl(|\alpha^*+\rangle,|\beta-\rangle\biggr)$:
	\begin{eqnarray}
	\displaystyle \biggl(|\alpha^*+\rangle,|\beta
	-\rangle\biggr)&=&\biggl(e^{-\alpha^*c^q}\bar{c}_q|0-\rangle,e^{-\beta\bar{c}_q}|0-\rangle\biggr)=\nonumber\\ &=&
	\biggl((1-\alpha^*c^q)\bar{c}_q|0-\rangle,(1-\beta\bar{c}_q)|0-\rangle\biggr)=\nonumber\\ &=&
	\biggl(|0-\rangle,c^q(1+\alpha\bar{c}_q)(1-\beta\bar{c}_q)|0-\rangle\biggr)=\nonumber\\ &=&
	\biggl(|0-\rangle,c^q(1-\beta\bar{c}_q+\alpha\bar{c}_q)|0-\rangle\biggr)=\nonumber\\ &=&
	\biggl(|0-\rangle,c^q(\alpha-\beta)\bar{c}_q|0-\rangle\biggr)=\nonumber\\ &=&
	\biggl(|0-\rangle,(\beta-\alpha)[c^q,\bar{c}_q]|0-\rangle\biggr)=\nonumber\\ &=& \biggl(|0-\rangle, (\beta-\alpha)
	|0-\rangle\biggr)=\alpha-\beta=\delta(\alpha-\beta)
	\end{eqnarray}
\\

\noindent $\bullet$ \underline{\it Derivation of Eq.(\ref{due-quindici})}.
 
\noindent We should check this formula on all the states. As the basis are $|0+\rangle$ and $|0-\rangle$, let us first prove 
$\displaystyle -\int d\alpha|\alpha+\rangle\langle-\alpha^*|={\bf 1}$ when applied to $|0+\rangle$:
	\begin{eqnarray}
	&&\displaystyle -\int d\alpha|\alpha +\rangle\langle -\alpha^*|0+\rangle=\int d\alpha|\alpha
	+\rangle\delta(\alpha)=\nonumber\\
	&&=\int d\alpha \,\alpha\biggl(|0+\rangle-\alpha|0-\rangle\biggr)=|0+\rangle.
	\end{eqnarray}
Let us now check it on $|0-\rangle$
	\begin{eqnarray}
	&&\displaystyle -\int d\alpha|\alpha +\rangle\langle -\alpha^*|0-\rangle=-\int d\alpha|\alpha
	+\rangle e^{\alpha\cdot 0}=\nonumber\\
	&&=-\int d\alpha \biggl(|0+\rangle-\alpha|0-\rangle\biggr)=|0-\rangle
	\end{eqnarray}
\\

\noindent $\bullet$ \underline{\it Derivation of Eqs. (\ref{due-trentasette-a}), (\ref{due-trentasette-b}),
(\ref{due-trentasette-c})}.

\noindent
Let us start with (\ref{due-trentasette-a}) and use the SvH hermiticity condition $\bar{c}_i^{\dagger}=c_i$:
	\begin{eqnarray}
	&&\biggl(|0+,0-,\ldots,0-\rangle,|0+,0-,\ldots,0-\rangle\biggr)=\nonumber\\
	&&=\biggl(\bar{c}_1|0-,0-,\ldots, 0-\rangle,\bar{c}_1|0-,0-,\ldots, 0-\rangle\biggr)=
	\nonumber\\
	&&=\biggl(|0-,0-,\ldots, 0-\rangle,c_1\bar{c}_1|0-,0-,\ldots, 0-\rangle\biggr)=\nonumber\\
	&&=\biggl(|0-,0-,\ldots,0-\rangle,[c_1,\bar{c}_1]|0-,0-,\ldots,0-\rangle\biggr)=\nonumber\\
	&&=\biggl(|0-,0-,\ldots,0-\rangle,|0-,0-,\ldots, 0-\rangle\biggr)=1.
	\end{eqnarray}
For the Eq. (\ref{due-trentasette-b}) we have 
	\begin{eqnarray}
	&&\biggl(|0-,0+,0-,\ldots,0-\rangle,|0-,0+,0-,\ldots,0-\rangle\biggr)=\nonumber\\
	&&=\biggl(-\bar{c}_2|0-,0-,0-,\ldots,0-\rangle,-\bar{c}_2|0-,0-,0-,\ldots, 0-\rangle\biggr)=\nonumber\\
	&&=\biggl(|0-,0-,0-,\ldots,0-\rangle,c_2\bar{c}_2|0-,0-,0-,\ldots,0-\rangle\biggr)=\nonumber\\
	&&=\biggl(|0-,0-,0-,\ldots,0-\rangle,[c_2,\bar{c}_2]|0-,0-,\ldots,0-\rangle\biggr)=\nonumber\\
	&&=\biggl(|0-,0-,0-,\ldots, 0-\rangle,|0-,0-,\ldots, 0-\rangle\biggr)=1.
	\end{eqnarray}
For the (\ref{due-trentasette-c}) we have 
	\begin{eqnarray}
	&&\biggl(|0+,0-,\ldots, 0-\rangle,|0-,0+,\ldots, 0-\rangle\biggr)=\nonumber\\
	&&=\biggl(\bar{c}_1|0-,0-,\ldots,0-\rangle,-\bar{c}_2|0-,0-,\ldots, 0-\rangle\biggr)=\nonumber\\
	&&=\biggl(|0-,0-,\ldots, 0-\rangle, -c_1\bar{c}_2|0-,0-,\ldots, 0-\rangle\biggr)=\nonumber\\
	&&=\biggl(|0-,0-,\ldots, 0-\rangle,\bar{c}_2c_1|0-,0-,\ldots, 0-\rangle\biggr)=0
	\end{eqnarray}
where in the last step we have used the fact that $\bar{c}_2,c_1$ anticommute and that 
$c_1$ applied on $|0-,0-\ldots,0-\rangle$ gives zero. 
\\

\noindent $\bullet$ \underline{\it Derivation of (\ref{due-trentotto-a})}:
	\begin{eqnarray}
	\displaystyle
	&&\biggl(|\alpha_q-,\alpha_p-\rangle,|\beta_q-,\beta_p-\rangle\biggr)=\nonumber\\
	&&=\biggl((1-\alpha_q\bar{c}_q-\alpha_p\bar{c}_p+\alpha_q
	\bar{c}_q\alpha_p\bar{c}_p)|0-,0-\rangle,
	(1-\beta_q\bar{c}_q-\beta_p\bar{c}_p+\beta_q\bar{c}_q\beta_p\bar{c}_p)|0-,0-\rangle\biggr)=\nonumber\\
	&&=\biggl(|0-,0-\rangle,(1+\alpha_q^*c^q+\alpha_p^*c^p+c^p\alpha_p^*c^q\alpha_q^*)(1-\beta_q\bar{c}_q-\beta_p\bar{c}_p+\beta_q\bar{c}_q
	\beta_p\bar{c}_p)|0-,0-\rangle\biggr)\nonumber\\
	&&=\biggl(|0-,0-\rangle,(1+\alpha_q^*\beta_qc^q\bar{c}_q+\alpha^*_p\beta_pc^p
	\bar{c}_p+c^p\alpha_p^*c^q\alpha_q^*\cdot\beta_q\bar{c}_q\beta_p\bar{c}_p)|0-,0-\rangle\biggr)=\nonumber\\
	&&=(1+\alpha_q^*\beta_q+\alpha_p^*\beta_p+\alpha^*_q\beta_q\alpha_p^*\beta_p)\biggl(|0-,0-\rangle,|0-,0-\rangle\biggr)=
	e^{\alpha_q^*\beta_q+\alpha_p^*\beta_p}
	\end{eqnarray}
\\

\noindent $\bullet$ \underline{\it Derivation of (\ref{due-trentotto-c})}:
	\begin{eqnarray}
	\displaystyle
	&&\biggl(|\alpha_q^*+,\alpha^*_p+\rangle,|\beta_q-,\beta_p-\rangle\biggr)=\nonumber\\
	&&=\biggl(e^{-\alpha_q^*c^q-\alpha_p^*c^p}\bar{c}_p\bar{c}_q|0-,0-\rangle,
	e^{-\beta_q\bar{c}_q-\beta_p\bar{c}_p}|0-,0-\rangle\biggr)=\nonumber\\
	&&=\biggl((1-\alpha_q^*c^q-\alpha_p^*c^p+\alpha_q^*c^q\alpha_p^*c^p)\cdot
	\bar{c}_p\bar{c}_q|0-,0-\rangle,
	(1-\beta_q\bar{c}_q-\beta_p\bar{c}_p+\beta_q\bar{c}_q\beta_p\bar{c}_p)|0-,0-\rangle\biggr)=\nonumber\\
	&&=\biggl( \bar{c}_p\bar{c}_q|0-,0-\rangle+\alpha_q^*\bar{c}_p|0-,0-\rangle-\alpha_p^*\bar{c}_q|0-,0-\rangle
	-\alpha_q^*\alpha^*_p|0-,0-\rangle,\nonumber\\
	&&\quad(1-\beta_q\bar{c}_q-\beta_p\bar{c}_p+\beta_q\bar{c}_q\beta_p\bar{c}_p)|0-,0-\rangle\biggr)=\nonumber\\
	&&=\beta_q\beta_p-\alpha_q\beta_p+\alpha_p\beta_q-\alpha_p\alpha_q=\alpha_q(\alpha_p-\beta_p)-\beta_q(\alpha_p-\beta_p)=\nonumber\\
	&&=\delta(\alpha_q-\beta_q)\delta(\alpha_p-\beta_p).
	\end{eqnarray}
The second equation in (\ref{due-trentotto-c}) can be derived in a similar manner.
\\

\noindent $\bullet$ \underline{\it Derivation of Eq. (\ref{due-quaranta})}.
 
\noindent Let us remember that from the relations
	\begin{equation}
	 \left\{
		\begin{array}{l}
		\displaystyle  
		c|0-\rangle=0,\qquad\qquad\quad \bar{c}|0+\rangle=0
		\smallskip \nonumber\\
		\displaystyle
		\bar{c}|0-\rangle=|0+\rangle\qquad\qquad c|0+\rangle=|0-\rangle,
		\end{array}
		\right.
	\end{equation}
via the hermiticity conditions $\bar{c}^{\dagger}=c, \;\,c^{\dagger}=\bar{c}$, we get 
	\begin{equation}
	 \left\{
		\begin{array}{l}
		\displaystyle  
		\langle -0|\bar{c}=0\qquad\qquad\quad\;\; \langle +0|c=0
		\smallskip \nonumber\\
		\displaystyle
		\langle +0|=\langle -0|c\qquad\qquad \langle -0|=\langle +0|\bar{c}.
	 \smallskip
		\end{array}
		\right.
	\end{equation}
If we now interpreted the following scalar product as: 
	\begin{equation}
	\biggl(|0_q-,0_p+\rangle,|0_q-,0_p+\rangle\biggr)=\langle-0_q,+0_p|\break 0_q-,0_p+\rangle
	\end{equation} 
and not like in (\ref{due-quaranta}), we would get 
	\begin{eqnarray}
	&& \langle -0_q,+0_p|0_q-,0_p+\rangle=\langle -0_q,-0_p|c^p(-\bar{c}_p)|0_q-,0_p-\rangle=\nonumber\\
	&&=-\langle -0_q,-0_p|0_q-,0_p-\rangle=-1=-\biggl(|0_q-,0_p+\rangle,|0_q-,0_p+\rangle\biggr)
	\end{eqnarray}
where we have used the fact that $\biggl(|0_q-,0_p+\rangle,|0_q-,0_p+\rangle\biggr)=1$. This is a contradiction. 
On the other hand it is easy to see that if we interpret the scalar product as
	\begin{equation}
	\biggl(|0_q-,0_p+\rangle,|0_q-,0_p+\rangle\biggr)=\langle+0_p,-0_q|0_q-,0_p+\rangle
	\end{equation}
we do not end up in contradictions. 
\\

\noindent $\bullet$ \underline{\it Derivation of (\ref{due-quarantadue})}.
	\begin{eqnarray}
	&&\int d\alpha_q d\alpha_p|\alpha_q+,\alpha_p+\rangle\langle-\alpha_p^*,-\alpha_q^*|=\\
	&&=\int d\alpha_qd\alpha_p[1-\alpha_q\hat{c}^q-\alpha_p\hat{c}^p+\alpha_q\hat{c}^q 
	\alpha_p\hat{c}^p]|0+,0+\rangle\cdot \langle -0,-0|[1+\alpha_q\hat{c}^q+\alpha_p\hat{c}^p+\alpha_q\hat{c}^q 
	\alpha_p\hat{c}^p]. \nonumber
	\end{eqnarray}
We should remember that $\alpha_q$ and $\alpha_p$ are Grassmannian numbers while $\hat{c}^q,\hat{c}^p$ are 
Grassmannian operators, so they should act on the
states which are either on their left or on their right and we get 
	\begin{eqnarray}
	&&\int d\alpha_q d\alpha_p|\alpha_q+,\alpha_p+\rangle\langle-\alpha_p^*,-\alpha_q^*|=\nonumber\\
	&&=\int d\alpha_qd\alpha_p\biggl[|0+,0+\rangle\langle
	-0,-0|\alpha_q\hat{c}^q\alpha_p\hat{c}^p-\alpha_q\hat{c}^q|0+,0+\rangle\langle-0,-0|\alpha_p\hat{c}^p\nonumber\\
	&&\quad -\alpha_p\hat{c}^p|0+,0+\rangle\langle
	-0,-0|\alpha_q\hat{c}^q+\alpha_q\hat{c}^q\alpha_p\hat{c}^p|0+,0+\rangle\langle-0,-0|\biggr]=\nonumber\\ 
	&&=|0+,0+\rangle\langle -0,-0|\hat{c}^q\hat{c}^p-\hat{c}^q|0+,0+\rangle\langle -0,-0|\hat{c}^p\nonumber\\
	&&\quad +\hat{c}^p|0+,0+\rangle\langle -0,-0|\hat{c}^q+\hat{c}^q\hat{c}^p|0+,0+\rangle\langle-0,-0|=\\
	&&=|0+,0+\rangle\langle +0,+0|+|0-,0+\rangle\langle +0,-0|+|0+,0-\rangle\langle -0,+0|+|0-,0-\rangle\langle -0,-0|.\nonumber
	\end{eqnarray}
\\

\noindent $\bullet$ \underline{\it Derivation of (\ref{due-sessantuno})}. 

\noindent Let us split the derivation in two parts, one for the
bosonic part of $\widetilde{\cal H}$ and the other for the fermionic part. The states we use are the tensor product of the bosonic
and fermionic ones 
	\begin{eqnarray}
	\langle -\bar{c}_{p_j}^*,-\bar{c}^*_{q_j},\lambda_{p_j},\lambda_{q_j}|&=&\langle -\bar{c}^*_{p_j},-\bar{c}_{q_j}^*|\otimes\langle
	\lambda_{p_j},\lambda_{q_j}| \nonumber\\
	|q_j,p_j,c^q_j-,c^p_j-\rangle&=&|q_j,p_j\rangle\otimes |c^q_j-,c^p_j-\rangle.
	\label{i-4}
	\end{eqnarray}
Applying the bosonic part of $\widetilde{\cal H}$ on these states, only the following terms survive:
	\begin{eqnarray}
	\displaystyle 
	&&\langle\lambda_q,\lambda_p|\widetilde{\cal
	H}_{bos}|q,p\rangle=\langle\lambda_q,\lambda_p|\hat{\lambda}_q\hat{p}-\hat{\lambda}_pV^{\prime}(\hat{q})|q,p\rangle=\nonumber\\
	&&=\langle \lambda_q|\hat{\lambda}_q|q\rangle\cdot\langle\lambda_p|\hat{p}|p\rangle-\langle\lambda_q|V^{\prime}(\hat{q})|q\rangle
	\cdot\langle\lambda_p|\hat{\lambda}_p|p\rangle=\nonumber\\
	&&=\lambda_qp\langle \lambda_q|q\rangle\langle \lambda_p|p\rangle-\lambda_pV^{\prime}(q)\langle \lambda_q|q\rangle\langle
	\lambda_p|p\rangle=\nonumber\\
	&&=(\lambda_qp-\lambda_pV^{\prime}(q))e^{-i\lambda_qq}e^{-i\lambda_pp}.
	\label{i-5}
	\end{eqnarray}
For the fermionic part we have 
	\begin{eqnarray}
	\displaystyle 
	&&\langle
	-\bar{c}_p^*,-\bar{c}_q^*,\lambda_p,\lambda_q|i\hat{\bar{c}}_q\hat{c}^p-i\hat{\bar{c}}_pV^{\prime\prime}(\hat{q})\hat{c}^q|
	q,p,c^q-,c^p-\rangle=\nonumber\\ 
	&&=\langle -\bar{c}_q^*|i\hat{\bar{c}}_q|c^q-\rangle\langle
	-\bar{c}_p^*|\hat{c}^p|c^p-\rangle\langle\lambda_q|q\rangle\langle
	\lambda_p|p\rangle
	-\langle-\bar{c}_p^*|i\hat{\bar{c}}_p|c^p-\rangle\langle -\bar{c}_q^*|\hat{c}^q|c^q-\rangle\langle
	\lambda_q|V^{\prime\prime}(\hat{q})|q\rangle\langle \lambda_p|p\rangle=\nonumber\\
	&&=-i\bar{c}_q\langle-\bar{c}_q^*|c^q-\rangle(-c^p)\langle -\bar{c}_p^*|c^p-\rangle 
	e^{-i\lambda_qq}e^{-i\lambda_pp}+i\bar{c}_p
	\langle-\bar{c}_p^*|c^p-\rangle(-c^q)\langle-\bar{c}_q^*|c^q-\rangle V^{\prime\prime}(q)\langle\lambda_q|q\rangle\langle
	\lambda_p|p\rangle\nonumber\\
	&&=i(\bar{c}_qc^p-\bar{c}_pV^{\prime\prime}(q)c^q)e^{\bar{c}_qc^q+\bar{c}_pc^p}e^{-i\lambda_qq-i\lambda_pp}.
	\label{i-6}
	\end{eqnarray}
Combining (\ref{i-5}) and (\ref{i-6}) we get exactly (\ref{due-sessantuno}). 


\newpage

\section{}

\noindent $\bullet$ \underline{\it Derivation of (\ref{tre-diciassette})}.
	\begin{eqnarray}
	\biggl(|\alpha-\rangle,|\beta-\rangle\biggr)&=&\biggl((1-\alpha\bar{c})|0-\rangle,(1-\beta\bar{c})|0-\rangle\biggr)=\nonumber\\
	&=&\biggl(|0-\rangle,(1+\alpha^*\bar{c})(1-\beta\bar{c})|0-\rangle\biggr)=\nonumber\\
	&=&\biggl(|0-\rangle,|0-\rangle+\alpha^*\bar{c}|0-\rangle-\beta\bar{c}|0-\rangle\biggr)=\nonumber\\
	&=&\biggl(|0-\rangle,|0-\rangle\biggr)+\biggl(|0-\rangle,\alpha^*|0+\rangle\biggr)-
	\biggl(|0-\rangle,\beta|0+\rangle\biggr)=\nonumber\\
	&=&\biggl(|0-\rangle,|0-\rangle\biggr)+\alpha^*\biggl(|0-\rangle,|0+\rangle\biggr)-
	\beta\biggl(|0-\rangle,|0+\rangle\biggr)=\nonumber\\
	&=&\alpha^*-\beta=\delta(\alpha^*-\beta).
	\end{eqnarray}
In the last step we used Eq. (\ref{tre-quindici}).
Now let us calculate
	\begin{eqnarray}
	\biggl(|\alpha+\rangle,|\beta-\rangle\biggr)&=&\biggl((1-\alpha c)|0+\rangle,(1-\beta\bar{c})|0-\rangle\biggr)=\nonumber\\
	&=&\biggl(|0+\rangle,(1+\alpha^*c)(1-\beta\bar{c})|0-\rangle\biggr)=\nonumber\\
	&=& \biggl(|0+\rangle,|0-\rangle\biggr)+\biggl(|0+\rangle,\alpha^*c|0-\rangle\biggr)
	-\biggl(|0+\rangle,\beta\bar{c}|0-\rangle\biggr)-\biggl(|0+\rangle,\alpha^*c\beta\bar{c}|0-\rangle\biggr)=\nonumber\\
	&=&1+\alpha^*\beta\biggl(|0+\rangle,|0-\rangle\biggr)=1+\alpha^*\beta=\exp(\alpha^*\beta).
	\end{eqnarray}
The other equations can be easily derived in a similar way.
\\

\noindent $\bullet$ \underline{\it Derivation of (\ref{tre-trentotto})}.
	\begin{eqnarray}
	&&\biggl(|\alpha_q-,\alpha_p-\rangle,|\beta_q+,\beta_p+\rangle\biggr)=\nonumber\\
	&&=\biggl([1-\alpha_q\bar{c}_q-\alpha_p\bar{c}_p+\alpha_q\bar{c}_q\alpha_p
	\bar{c}_p]|0-,0-\rangle,(1-\beta_qc^q-\beta_pc^p+\beta_qc^q\beta_pc^p)|0+,0+\rangle\biggr)=\nonumber\\
	&&=\biggl(|0-,0-\rangle,(1+\alpha_q^*\bar{c}_q+\alpha_p^*\bar{c}_p+\bar{c}_p\alpha_p^*\bar{c}_q\alpha_q^*)\cdot
	(1-\beta_qc^q-\beta_pc^p+\beta_qc^q\beta_pc^p)|0+,0+\rangle\biggr)=\nonumber\\
	&&=\biggl(|0-,0-\rangle,(1-\alpha_q^*\bar{c}_q\beta_qc^q-\alpha_p^*\bar{c}_p\beta_pc^p+\bar{c}_p\alpha_p^*\bar{c}_q\alpha_q^*
	\beta_qc^q\beta_pc^p)|0+,0+\rangle\biggr)=\nonumber\\
	&&=i(1+\alpha_q^*\beta_q+\alpha_p^*\beta_p+\alpha_p^*\alpha_q^*\beta_q\beta_p)=i\cdot
	\exp(\alpha_q^*\beta_q+\alpha_p^*\beta_p)
	\end{eqnarray}

\smallskip

	\begin{eqnarray}
	&&\biggl(|\alpha_q-,\alpha_p-\rangle,|\alpha_q^{\prime}-,\beta_p+\rangle\biggr)=\nonumber\\
	&&=\biggl((1-\alpha_q\bar{c}_q-\alpha_p\bar{c}_p+
	\alpha_q\bar{c}_q\alpha_p\bar{c}_p)|0-,0-\rangle,(c^q-\alpha_q^{\prime}-\beta_pc^pc^q+\alpha_q^{\prime}\beta_pc^p)
	|0+,0+\rangle\biggr)
	=\nonumber\\
	&&=\biggl(|0-,0-\rangle,(1+\alpha_q^*\bar{c}_q+\alpha^*_p\bar{c}_p+\bar{c}_p\alpha^*_p\bar{c}_q\alpha_q^*)\cdot(c^q-\alpha_q^{\prime}
	-\beta_pc^pc^q+\alpha^{\prime}_q\beta_pc^p)|0+,0+\rangle\biggr)=\nonumber\\
	&&=i(-\alpha_q^{\prime}+\alpha_q^*+\alpha_p^*\alpha_q^{\prime}\beta_p-\alpha_p^*\alpha_q^*\beta_p)=\nonumber\\
	&&=i\cdot\delta(\alpha_q^*-\alpha_q^{\prime})\exp(\alpha_p^*\beta_p).
	\end{eqnarray}
In all these derivations we have only used the hermiticity conditions (\ref{tre-due}) of the gauge scalar products and
the scalar
products (\ref{tre-ventinove}). All the other relations in Eq. (\ref{tre-trentotto}), besides the two we have derived here,
can be proved in the same straightforward manner. 
\\

\noindent $\bullet$ \underline{\it Derivation of (\ref{tre-trentanove})}.

\noindent
 Let us rewrite the LHS of the first of the
two resolutions of the identity contained in (\ref{tre-trentanove}) as follows
	\begin{eqnarray}
	&&i\int d\alpha_qd\alpha_p|\alpha_q+,\alpha_p+\rangle\langle +\alpha_p^*,+\alpha^*_q|=\nonumber\\
	&&=i\int d\alpha_q
	d\alpha_p(1-\alpha_qc^q-\alpha_pc^p+\alpha_qc^q\alpha_pc^p)|0+,0+\rangle\cdot\nonumber\\
	&&\quad\cdot\langle +0,+0|(1-c^q\alpha_q-c^p\alpha_p+c^p\alpha_pc^q\alpha_q)=\nonumber\\
	&&=-i\biggl[|0+,0+\rangle\langle+0,+0|c^pc^q+c^q|0+,0+\rangle\langle+0,+0|c^p-\nonumber\\
	&&\quad -c^p|0+,0+\rangle\langle+0,+0|c^q-c^qc^p|0+,0+\rangle\langle+0,+0|\biggr]=\nonumber\\
	&&=-i\biggl[|0+,0+\rangle\langle -0,-0|+|0-,0+\rangle\langle -0,+0|\nonumber\\
	&&\quad -|0+,0-\rangle\langle +0,-0|-|0-,0-\rangle\langle +0,+0|\biggr]
	\end{eqnarray}
So the resolution of the identity we have to check is 
	\begin{eqnarray}
	&&-i\biggl[|0+,0+\rangle\langle -0,-0|+|0-,0+\rangle\langle-0,+0|-|0+,0-\rangle\langle+0,-0|
	-|0-,0-\rangle\langle +0,+0|\biggr]={\bf 1}.\nonumber\\
	\label{i-18-1}
	\end{eqnarray}
Our vector space is spanned by the four states $\bigl\{|0-,0-\rangle$, $|0-,0+\rangle$, 
$|0+,0-\rangle$, $|0+,0+\rangle\bigr\}$. Therefore, to check
the resolution of the identity (\ref{i-18-1}), we just need to verify it on these four states. Let
us first apply the LHS of (\ref{i-18-1}) on $|0+,0+\rangle$:
	\begin{eqnarray}
	&&\label{i-18-3} -i\biggl[|0+,0+\rangle\langle -0,-0|0+,0+\rangle+|0-,0+\rangle\langle-0,+0|0+,0+\rangle\\
	&&-|0+,0-\rangle\langle+0,-0|0+,0+\rangle-|0-,0-\rangle\langle\nonumber
	+0,+0|0+,0+\rangle\biggr]=|0+,0+\rangle .
	\end{eqnarray}
What we get is exactly the RHS of (\ref{i-18-1}) applied on $|0+,0+\rangle$. We have used the scalar products
(\ref{tre-ventinove}) which tell us that only the first term on the LHS of (\ref{i-18-3}) is different from zero. It is
straightforward to verify (\ref{i-18-1}) also on the other three states
$\bigl\{|0-,0-\rangle,|0+,0-\rangle,|0-,0+\rangle\bigr\}$ and we leave it to the reader.  

\newpage

\section{}
\noindent
We confine in this Appendix the mathematical details of Sections IV and V.
\\

\noindent $\bullet$ \underline{\it Derivation of (\ref{quattro-cinque})}.

\noindent
Let us derive the last of the relations in
(\ref{quattro-cinque}):
	\begin{eqnarray}
	&&\biggl(|0-,0-\rangle,|0-,0-\rangle\biggr)=(c^qc^p|0+,0+\rangle,c^qc^p|0+,0+\rangle)=\nonumber\\
	&&=\biggl(|0+,0+\rangle,(c^qc^p)^{\dagger}c^qc^p|0+,0+\rangle\biggr)=
	\biggl(|0+,0+\rangle,\bar{c}_q\bar{c}_pc^qc^p|0+,0+\rangle\biggr)=\nonumber\\
	&&=-\biggl(|0+,0+\rangle,|0+,0+\rangle\biggr)=-1.
	\end{eqnarray}
In the last step we used the commutator structure (\ref{uno-quattordici-b}) of the $c^a,\bar{c}_b$. The other relations
in (\ref{quattro-cinque}) are easily derived in a similar way.
\\

\noindent $\bullet$ \underline{\it Derivation of (\ref{quattro-sette})}. 

\noindent
As an example, let us work out the first of the
relations (\ref{quattro-sette}).
	\begin{eqnarray}
	&&\biggl(|\alpha_q-,\alpha_p-\rangle,|\alpha_q^{\prime}-,\alpha_p^{\prime}-\rangle\biggr)=\nonumber\\
	&&=\biggl((1-\alpha_q\bar{c}_q-\alpha_p\bar{c}_p+\alpha_q\bar{c}_q\alpha_p\bar{c}_p)|0-,0-\rangle,
	(1-\alpha_q^{\prime}\bar{c}_q-\alpha_p^{\prime}\bar{c}_p+\alpha_q^{\prime}\bar{c}_q
	\alpha_p^{\prime}\bar{c}_p)|0-,0-\rangle\biggr)=\nonumber\\
	&&=\biggl(|0-,0-\rangle,(1-i\alpha_q^*c^p+i\alpha_p^*c^q+c^q\alpha_p^*c^p\alpha_q^*)\cdot
	(1-\alpha_q^{\prime}\bar{c}_q-\alpha_p^{\prime}\bar{c}_p+\alpha_q^{\prime}\bar{c}_q\alpha_p^{\prime}\bar{c}_p)
	|0-,0-\rangle\biggr)=\nonumber\\
	&&=-(1-i\alpha_q^*\alpha_p^{\prime}+i\alpha_p^*\alpha_q^{\prime}-\alpha_p^*\alpha_q^*\alpha_q^{\prime}\alpha_p^{\prime})=
	-\exp(-i\alpha_q^*\alpha_p^{\prime}+i\alpha_p^*\alpha_q^{\prime}).
	\end{eqnarray}
The second equation of (\ref{quattro-sette}) can also be easily derived as follow:
	\begin{eqnarray}
	&&\biggl(|\alpha_q-,\beta_p+\rangle,|\beta_q+,\beta_p^{\prime}+\rangle\biggr)=\nonumber\\
	&&=\biggl((1-\alpha_q\bar{c}_q-\beta_pc^p+\alpha_q\bar{c}_q\beta_pc^p)\cdot
	c^q|0+,0+\rangle,(1-\beta_qc^q-\beta_p^{\prime}c^p+\beta_qc^q\beta_p^{\prime}c^p)|0+,0+\rangle\biggr)=\nonumber\\
	&&=\biggl((c^q-\alpha_q-\beta_pc^pc^q+\alpha_q\beta_pc^p)|0+,0+\rangle,(1-\beta_qc^q-\beta_p^{\prime}c^p+
	\beta_qc^q\beta_p^{\prime}c^p)|0+,0+\rangle\biggr)\nonumber\\ 
	&&=\biggl(|0+,0+\rangle,(i\bar{c}_p-\alpha_q^*-\bar{c}_p\bar{c}_q\beta_p^*-i\bar{c}_q\beta_p^*\alpha_q^*)(1-\beta_qc^q
	-\beta_p^{\prime}c^p+\beta_qc^q\beta_p^\prime c^p)|0+0+\rangle
	\biggr)\nonumber\\
	&&=
	i\beta_p^{\prime}-\alpha_q^*+\beta_p^*\beta_q\beta_p^{\prime}-i\beta_p^*\alpha_q^*\beta_q=i\delta(\beta_p^{\prime}+i\alpha_q^*)
	(1+i\beta_q\beta_p^*)
	\end{eqnarray}
and finally let us derive the last of Eq. (\ref{quattro-sette})
	\begin{eqnarray}
	&&\biggl(|\beta_q+,\beta_p+\rangle,|\beta_q^{\prime}+,\beta_p^{\prime}+\rangle\biggr)=\nonumber\\
	&&=\biggl((1-\beta_qc^q-\beta_pc^p+
	\beta_qc^q\beta_pc^p)|0+,0+\rangle,
	(1-\beta_q^{\prime}c^q-\beta_p^{\prime}c^p+\beta_q^{\prime}c^q\beta_p^{\prime}c^p)
	|0+,0+\rangle\biggr)=\nonumber\\
	&&=\biggl(|0+,0+\rangle,(1-i\bar{c}_p\beta_q^*+i\bar{c}_q\beta_p^*+\bar{c}_q\beta_p^*\bar{c}_p\beta_q^*)
	\cdot (1-\beta_q^{\prime}c^q-\beta_p^{\prime}c^p+\beta_q^{\prime}c^q\beta_p^{\prime}c^p)|0+,0+\rangle\biggr)=\nonumber\\
	&&=1+i\beta_q^*\beta_p^{\prime}-i\beta_p^*\beta_q^{\prime}-\beta_p^*\beta_q^*\beta_q^{\prime}\beta_p^{\prime}=
	\exp(i\beta_q^*\beta_p^{\prime}-i\beta_p^*\beta_q^{\prime})
	\end{eqnarray}
\\

\noindent $\bullet$ \underline{\it Derivation of (\ref{quattro-nove-a})-(\ref{quattro-nove-b})}.

\noindent
To prove (\ref{quattro-nove-a}) we
should test it on each of the four states (\ref{due-quarantatre})
which make a basis of our vector space. Let us
first do it on
$|0+,0+\rangle$. We have: 
	\begin{eqnarray}
	&&\int d\alpha_pd\alpha_q|\alpha_q-,\alpha_p-\rangle\biggl(|i\alpha_p^*+,(-i\alpha_q^*)+\rangle,|0+,0+\rangle\biggr)=\nonumber\\
	&&=\int
	d\alpha_pd\alpha_q\biggl(|0-,0-\rangle-\alpha_q|0+,0-\rangle+\alpha_p|0-,0+\rangle+\alpha_q\alpha_p|0+,0+
	\rangle\biggr)\cdot\nonumber\\
	&&\cdot\biggl(|i\alpha_p^*+,(-i\alpha_q^*)+\rangle,|0+,0+\rangle\biggr)=\int d\alpha_pd\alpha_q
	\alpha_q\alpha_p|0+,0+\rangle=|0+,0+\rangle .\label{i-22-1}
	\end{eqnarray}
In the last step we used the scalar products (\ref{quattro-sette}) which give
	\begin{equation}
	\biggl(|i\alpha_p^*+,(-i\alpha_q^*)+\rangle,|0+,0+\rangle\biggr)=1.
	\end{equation}
It is straightforward to repeat the proof
for the other three
states of Eq. (\ref{due-quarantatre}). 
Regarding  (\ref{quattro-nove-b}) let us just test it on the state
$|0-,0-\rangle$ by applying the LHS of (\ref{quattro-nove-b}) on it:
	\begin{eqnarray}
	&&\int
	d\alpha_pd\alpha_q|\alpha_q+,\alpha_p+\rangle\biggl(|(-i\alpha_p^*)-,(i\alpha_q^*)-\rangle,|0-,0-\rangle\biggr)=\nonumber\\
	&&=\int d\alpha_pd\alpha_q\biggl(|0+,0+\rangle-\alpha_q|0-,0+\rangle-\alpha_p|0+,0-\rangle-
	\alpha_q\alpha_p|0-,0-\rangle\biggr)\cdot\nonumber\\
	&&\quad \cdot\biggl(|(-i\alpha_p^*)-,(i\alpha_q^*)-\rangle,|0-,0-\rangle\biggr)=
	\int d\alpha_pd\alpha_q\alpha_q\alpha_p|0-,0-\rangle=|0-,0-\rangle
	\label{i-23-1}
	\end{eqnarray}
where we have used the scalar product (\ref{quattro-sette}) which gives
	\begin{equation}
	\biggl(|(-i\alpha_p^*)-,(i\alpha_q^*)-\rangle,|0-,0-\rangle\biggr)=-1.
	\end{equation}
Eq. (\ref{i-23-1}) proves (\ref{quattro-nove-b}) for the state $|0-,0-\rangle$. It is straightforward to prove it for the other
three states $\bigl\{|0+,0-\rangle, |0-,0+\rangle, |0+,0+\rangle\bigr\}$.
\\

\noindent $\bullet$ \underline{\it Derivation of (\ref{quattro-quattordici-a})}. 

\noindent
If we write $\psi$ as  $\psi=\psi_0+\psi_qc^q+\psi_pc^p+\psi_2c^qc^p$
we have that 
	\begin{equation}
	\hat{c}^q\psi=c^q\psi_0+\psi_pc^qc^p
	\end{equation}
and
	\begin{equation}
	\displaystyle i\hat{\bar{c}}_p\Phi=i\frac{\partial}{\partial c^p}\Phi=i\Phi_p-i\Phi_2c^q.
	\end{equation}
Then, using (\ref{quattro-quattordici}), we get that 
	\begin{equation}
	\langle \Phi|\hat{c}^q\psi\rangle=-i\Phi_p^*\psi_0-\Phi_2^*\psi_p \label{i-25-1}
	\end{equation}
and 
	\begin{equation}
	\langle i\bar{c}_p\Phi|\psi\rangle=(i\Phi_p)^*\psi_0+i(-i\Phi_2)^*\psi_p=-i\Phi_p^*\psi_0-\Phi_2^*\psi_p. \label{i-25-2}
	\end{equation}
So (\ref{i-25-1}) and (\ref{i-25-2}) are equal and this confirms the first of the relations in (\ref{quattro-quattordici-a}). The
others can be obtained in the same  way.
\\

\noindent $\bullet$ \underline{\it Derivation of (\ref{quattro-quindici})}. 

\noindent
To prove (\ref{quattro-quindici}) it is enough to show that in the expression of a generic state
	\begin{equation}
	|\psi\rangle=\psi_0(\varphi)|0+,0+\rangle+\psi_q(\varphi)|0-,0+\rangle+\psi_p(\varphi)|0+,0-\rangle+
	\psi_2(\varphi)|0-,0-\rangle
	\end{equation}
the coefficients of $|0+,0+\rangle$ are the zero-forms. If so then, by choosing
$\||0+,0+\rangle\|^2=\pm1$, we will get the norm of the zero-form to be $\pm\int d\varphi|\psi_0|^2$. The proof that the
coefficients of $|0+,0+\rangle$ are the zero-forms goes as follows. The $\langle bra|$ which are eigenstates of $\hat{c}$ are the
$\langle ic^{p*}+,(-ic^{q*})+|$ (see Eqs. (\ref{quattro-nove-c})-(\ref{quattro-nove-d})). Let us now project the state
$|0+,0+\rangle$ onto $\langle ic^{p*}+,(-ic^{q*})+|$. From (\ref{quattro-sette}) we obtain 
	\begin{equation}
	\langle ic^{p*}+,(-ic^{q*})+|0+,0+\rangle=1.
	\end{equation}
This proves that the coefficient of $|0+,0+\rangle$ is the zero-form $\psi_0(\varphi)$.
\\

\noindent $\bullet$ \underline{\it Derivation of (\ref{cinque-trenta})}. 
	\begin{eqnarray}
	&& \biggl(|0+0+0+0+\rangle,|0+0+0+0+\rangle\biggr)=\nonumber\\
	&&=\biggl(\bar{\xi}_1^*\bar{\xi}_1\bar{\xi}_2^*\bar{\xi}_2|0-0-0-0-\rangle,
	\bar{\xi}_1^*\bar{\xi}_1\bar{\xi}_2^*\bar{\xi}_2|0-0-0-0-\rangle\biggr)=\nonumber\\
	&&=\biggl(|0-0-0-0-\rangle,\xi_2\xi_2^*\xi_1\xi_1^*\bar{\xi}_1^*\bar{\xi}_1\bar{\xi}_2^*\bar{\xi}_2
	|0-0-0-0-\rangle\biggr)=\nonumber\\
	&&=\biggl(|0-0-0-0-\rangle,\xi_2\xi_2^*\xi_1\bar{\xi}_1\bar{\xi}_2^*\bar{\xi}_2[\xi_1^*,
	\bar{\xi_1^*}]|0-0-0-0-\rangle\biggr)=
	\nonumber\\
	&&=\biggl(|0-0-0-0-\rangle,\xi_2\xi_2^*\bar{\xi}_2^*\bar{\xi}_2[\xi_1,\bar{\xi}_1]|0-0-0-0-\rangle\biggr)=\nonumber\\
	&&=-\biggl(|0-0-0-0-\rangle,\xi_2\bar{\xi}_2[\xi_2^*,\bar{\xi}_2^*]|0-0-0-0-\rangle\biggr)=\nonumber\\
	&&=-\biggl(|0-0-0-0-\rangle,[\xi_2,\bar{\xi}_2]|0-0-0-0-\rangle\biggr)=\nonumber\\
	&&=\biggl(|0-0-0-0-\rangle,|0-0-0-0-\rangle\biggr)=1.
	\end{eqnarray}
Note that we have used the indices $(1,2)$ because $n=2$.
\\

\noindent $\bullet$ \underline{\it Derivation of (\ref{cinque-trentuno})}. 

\noindent
In order to have zero-forms with unit norm
it is necessary that $\biggl(|0+0+\ldots0+\rangle,|0+0+\ldots0+\rangle\biggr)=1$. All the other states can be derived from the previous
one by applying $\hat{\xi}$ and $\hat{\xi}^*$ to $|0+0+\ldots 0+\rangle$. Their norms can be obtained using the
anticommutation relations (\ref{cinque-ventuno}) and the hermiticity conditions (\ref{cinque-ventidue}):
	\begin{eqnarray}
	&& \biggl(\hat{\xi}_i|0+0+\ldots 0+\rangle,\hat{\xi}_i|0+0+\ldots 0+\rangle\biggr)=\biggl(|0+0+\ldots
	0+\rangle,\hat{\bar{\xi}}_i\hat{\xi}_i|0+0+\ldots 0+\rangle\biggr)=\nonumber\\
	&& =\biggl(|0+0+\ldots 0+\rangle, [\hat{\bar{\xi}}_i,\hat{\xi}_i]|0+0+\ldots 0+\rangle\biggr)=-1 \label{pao}
	\end{eqnarray}
	
	\begin{eqnarray}
	&& \biggl(\hat{\xi}_i^*|0+0+\ldots 0+\rangle,\hat{\xi}_i^*|0+0+\ldots 0+\rangle\biggr)=\nonumber\\
	&&=\biggl(|0+0+\ldots 0+\rangle, [\hat{\bar{\xi}}_i^*,\hat{\xi}_i^*]|0+0+\ldots 0+\rangle\biggr)=+1. \label{pao1}
	\end{eqnarray}
In general if we apply operators of the type $\hat{\xi}$ $\alpha$ times and $\beta$ times operators
of the type $\hat{\xi}^*$ on $|0+0+\ldots 0+\rangle$ we obtain a state whose
representation contains a number $\alpha$ of variables $\xi$ and a number $\beta$ of variables $\xi^*$ 
and whose norm is $(-1)^{\alpha}$. The reason is that,
via a
calculation analog to that of Eqs. (\ref{pao})-(\ref{pao1}), there will appear $\alpha$ anticommutators of the type
$[\hat{\xi}_i,\hat{\bar{\xi}}_j]=-\delta_{ij}$ and $\beta$ anticommutators of the type
$[\hat{\xi}_i^*,\hat{\bar{\xi}}_j^*]=\delta_{ij}$. This explains the reason why a homogeneous form, containing an odd
number of $\xi$, has negative norm and vice versa. 
As a particular case the state $|0-0-\ldots 0-\rangle$  is obtained from 
$|0+0+\ldots 0+\rangle$ by applying $n$ operators $\hat{\xi}$ and $n$ operators $\hat{\xi}^*$. Therefore its norm is
$(-1)^n$ as expressed by formula (\ref{cinque-trentuno}).
\\

\noindent $\bullet$ \underline{\it Derivation of (\ref{cinque-trentacinque})}.

\noindent
We prove here that the last two terms (\ref{cinque-trentatre}) in
$\widetilde{\cal H}$ annihilate states of the form (\ref{cinque-trentacinque}).
Let us take a generic homogeneous form in (\ref{cinque-trentacinque}) like the 4-form
	\begin{equation}
	\displaystyle \sum_{b,c}\xi_b\xi_{b}^*\xi_c\xi_{c}^* \label{i-27-1}
	\end{equation}
and let us apply on it the last two terms of the $\widetilde{\cal H}$ in (\ref{cinque-trentatre})
	\begin{eqnarray}
	&&\displaystyle \xi_a\partial_a\partial_kH\frac{\partial}{\partial \xi_{k}^*}(\xi_b\xi_{b}^*\xi_c\xi_{c}^*)
	-\xi_{a}^*\bar{\partial}_a\bar{\partial}_kH\frac{\partial}{\partial
	\xi_k}(\xi_b\xi_{b}^*\xi_c\xi_{c}^*)=\nonumber\\
	&&=-\xi_a\partial_a\partial_kH\xi_b\delta^k_b\xi_c\xi_{c}^*-\xi_a\partial_a\partial_kH\xi_b\xi_{b}^*\xi_c\delta^k_c
	-\xi_{a}^*\bar{\partial}_a\bar{\partial}_kH\delta^k_b\xi_{b}^*\xi_c\xi_{c}^*-
	\xi_{a}^*\bar{\partial}_a\bar{\partial}_kH\xi_b\xi_{b}^*\delta^k_c\xi_{c}^*=\nonumber\\
	&&=-\xi_a\partial_a\partial_bH\xi_b\xi_c\xi_{c}^*-\xi_a\partial_a
	\partial_cH\xi_b\xi_{b}^*\xi_c-\xi_{a}^*\bar{\partial}_a\bar{\partial}_bH\xi_{b}^*\xi_c\xi_{c}^*
	-\xi_{a}^*\bar{\partial}_a\bar{\partial}_cH\xi_b\xi_{b}^*\xi_{c}^*=0. 
	\end{eqnarray}
Notice that, for this to be zero, not only it is crucial that each $\xi_j$ in (\ref{i-27-1}) be associated to a
$\xi_{j}^*$, but also that each term in the sum (\ref{i-27-1}) have the same coefficient.
\\

\noindent $\bullet$ \underline{\it{Derivation of (\ref{cinque-trentasei}})}.

\noindent
 Let us verify it on a 4-form
	\begin{eqnarray}
	\displaystyle &&(\hat{\xi}_k\hat{\bar{\xi}}_i+\hat{\xi}^{*}_i\hat{\bar{\xi}}_k^*)(\partial_k\bar{\partial}_iH)
	\biggl[\sum_{j,l}\xi_j\xi^{*}_j\xi_l\xi^{*}_l\biggr]
	=\sum_{j,l}\biggl(-\xi_k\frac{\partial}{\partial \xi_i}+\xi^{*}_i\frac{\partial}{\partial\xi^{*}_k}\biggr)
	(\xi_j\xi^{*}_j\xi_l\xi^{*}_l)\partial_k\bar{\partial}_iH=\nonumber\\
	&&=(-\xi_k\xi^{*}_i\xi_l\xi^{*}_l-\xi_k\xi_j\xi^{*}_j\xi^{*}_i-\xi^{*}_i\xi_k\xi_l\xi^{*}_l-
	\xi^{*}_i\xi_j\xi^{*}_j\xi_k)\partial_k\bar{\partial}_iH=0.
	\end{eqnarray}
Again it was crucial the presence of all the possible pairs $\xi_j$ and $\xi^{*}_j$, summed over
and with the same coefficients.


\newpage

\section{} 

\noindent $\bullet$ \underline{\it{Derivation of (\ref{sei-dodici-1}).}} 

\noindent
In this case, $\beta^*\gamma=1$ and 
from (\ref{sei-sei-a}) and (\ref{sei-sette-b}),
we get 
	\begin{equation}
	\alpha=\delta=0. \label{i-29-1}
	\end{equation}
Moreover from (\ref{sei-cinque}) we have 
	\begin{equation}
	-\beta\gamma=1. \label{i-29-2}
	\end{equation}
Therefore combining the first relation of (\ref{sei-undici}) and (\ref{i-29-2}) we obtain
	\begin{equation}
	(\beta+\beta^*)\gamma=0 \label{i-29-3}
	\end{equation}
The case $\gamma=0$ cannot be a solution otherwise (\ref{i-29-2}) would not be valid, so the only solution of (\ref{i-29-3}) is
	\begin{equation}
	\beta+\beta^*=0
	\end{equation}
which means that $\beta$ is imaginary
	\begin{equation}
	\beta=ib  \label{i-29-4}
	\end{equation}
From (\ref{i-29-2}) we get 
	\begin{equation}
	\displaystyle \gamma=\frac{i}{b}. \label{i-29-5}
	\end{equation}
Putting now together (\ref{i-29-1}), (\ref{i-29-4}), (\ref{i-29-5}) we get that in the case $\beta^*\gamma=1$ the
hermiticity condition (\ref{sei-quattro}) turns out to be 
	\begin{equation}
	 \left\{
		\begin{array}{l}	
		\displaystyle
		\hat{c}^{p\dagger}=ib\hat{\bar{c}}_q
	        \smallskip \nonumber\\
	        \displaystyle 
	        \hat{\bar{c}}_q^{\dagger}=\frac{i}{b}\hat{c}^p
	        \end{array}
		\right.
	\end{equation}
\\

\noindent $\bullet$ \underline{\it{ Derivation of (\ref{sei-dodici-2}) and (\ref{sei-dodici-3})}.} 

\noindent
Let us now analyze the second case
of (\ref{sei-undici}) 
	\begin{equation}
	\beta^*\gamma=0. \label{i-30-1}
	\end{equation}
The solutions are $\beta=0$ or $\gamma=0$. Let us concentrate on the case $\beta=0$. With this value of $\beta$ we get from
(\ref{sei-cinque})-(\ref{sei-sette-b})
	\begin{equation}
	 \left\{
		\begin{array}{l}	
		\displaystyle
		\label{i-30-2-a} |\alpha|=|\delta|=1
	        \smallskip \nonumber\\
	        \displaystyle 
	        \label{i-30-2-b} \alpha\delta=1
	        \smallskip \nonumber\\
	        \displaystyle 
	        \label{i-30-2-c} \alpha\gamma^*+\delta^*\gamma=0.
	        \end{array}
		\right.
	\end{equation}
We can then parametrize $\alpha$ and $\delta$ as $\alpha=e^{i\theta_{\alpha}},\;\;\delta=e^{i\theta_{\delta}}$ and from
(\ref{i-30-2-c}) we obtain
	\begin{equation}	
	(e^{-i\theta_{\alpha}}\gamma)^*=-e^{-i\theta_{\delta}}\gamma \label{i-30-3}
	\end{equation}
and 
	\begin{equation}
	\theta_{\alpha}=-\theta_{\delta}. \label{i-31-1}
	\end{equation}
So (\ref{i-30-3}) gives 
	\begin{equation}
	\gamma^*=-\gamma \label{i-31-2}
	\end{equation}
which means that $\gamma$ is imaginary.
Combining (\ref{i-31-1}), (\ref{i-31-2}) and remembering that this is the case $\beta=0$, we get that the hermiticity
condition (\ref{sei-quattro}) is in this case
	\begin{equation}
	 \left\{
		\begin{array}{l}	
		\displaystyle
		\label{i-31-3}
		\hat{c}^{p\dagger}=e^{i\theta_{\alpha}}\hat{c}^p
	        \smallskip \nonumber\\
	        \displaystyle 
	        \hat{\bar{c}}_q^{\dagger}=i\gamma_I\hat{c}^p+e^{-i\theta_{\alpha}}\hat{\bar{c}}_q.
	        \end{array}
		\right.
	\end{equation}
Let us now  analyze the $\gamma=0$ solution of (\ref{i-30-1}). Then, instead of (\ref{i-30-2-a}),
we would have gotten
	\begin{equation}
	 \left\{
		\begin{array}{l}	
		\displaystyle
		\label{i-31-4}
		|\alpha|=|\delta|=1
	        \smallskip \nonumber\\
	        \displaystyle 
	        \alpha\delta=1
	        \smallskip \nonumber\\
	        \alpha^*\beta+\beta^*\delta=0
	        \end{array}
		\right.
	\end{equation}
from which we obtain
	\begin{equation}
	 \left\{
		\begin{array}{l}	
		\displaystyle
		\label{i-32-1}
		\alpha=e^{i\theta_{\alpha}}
	        \smallskip \nonumber\\
	        \displaystyle 
	        \delta=e^{-i\theta_{\alpha}}
	        \smallskip \nonumber\\
	        \beta=ib 
	        \end{array}
		\right.
	\end{equation}
with $b$ real.
So in this case the hermiticity conditions in (\ref{sei-quattro}) turn out  to be:
	\begin{equation}
	 \left\{
		\begin{array}{l}	
		\displaystyle
		\label{i-32-2}
		\hat{c}^{p\dagger}=e^{i\theta_{\alpha}}\hat{c}^p+ib\hat{\bar{c}}_q
	        \smallskip \nonumber\\
	        \displaystyle 
	        \hat{\bar{c}}_q^{\dagger}=e^{-i\theta_{\alpha}}\hat{\bar{c}}_q
	        \end{array}
		\right.
	\end{equation}
Starting from $\beta^*\gamma=z$ and using (\ref{sei-sei-a})
and (\ref{sei-sette-b}) we get $|\alpha|=|\delta|=\sqrt{1-z}$, so we can write
	\begin{equation}
	 \left\{
		\begin{array}{l}	
		\displaystyle
		\label{i-33-1}
		\alpha=\sqrt{1-z}e^{i\theta_{\alpha}}
	        \smallskip \nonumber\\
	        \displaystyle 
	        \delta=\sqrt{1-z}e^{i\theta_{\delta}}.
	        \end{array}
		\right.
	\end{equation}
Inserting this in (\ref{sei-cinque}) we get 
	\begin{equation}
	\displaystyle e^{i(\theta_{\alpha}+\theta_{\delta})}=\frac{\beta\gamma+1}{1-z} \label{i-33-2}
	\end{equation}
and from (\ref{sei-sei-b}) and (\ref{sei-sette-a}) we obtain 
	\begin{equation}
	 \left\{
		\begin{array}{l}	
		\displaystyle
		\label{i-33-3}
		\beta e^{-i\theta_{\alpha}}=-\beta^*e^{i\theta_{\delta}}
	        \smallskip \nonumber\\
	        \displaystyle 
	        \gamma e^{-i\theta_{\delta}}=-\gamma^*e^{i\theta_{\alpha}}.
	        \end{array}
		\right.
	\end{equation}
Multiplying the LHS of these two equations and doing the same for the RHS we get 
	\begin{equation}
	\beta\gamma e^{-i(\theta_{\alpha}+\theta_{\delta})}=\beta^*\gamma^* e^{i(\theta_{\alpha}+\theta_{\delta})}
	\end{equation}
which implies that the expression
	\begin{equation}
	\beta^*\gamma^* e^{i(\theta_{\alpha}+\theta_{\delta})} \label{i-34-0}
	\end{equation}
is imaginary.
Inserting in this expression the RHS of (\ref{i-33-2}) we get that $\displaystyle \beta^*\gamma^*\frac{\beta\gamma
+1}{1-z}$ is imaginary and so also the numerator $|\beta^*\gamma^*|^2+\beta^*\gamma^*$
is imaginary. 
If we write $\beta^*\gamma^*\equiv a+ib$ we get 
\begin{equation}
|\beta^*\gamma^*|^2+\beta^*\gamma^*=a^2+b^2+a+ib \label{i-34-1}
\end{equation}
Now it is easy to prove that the expression $|\beta^*\gamma^*|^2+\beta^*\gamma^*$ is not only imaginary but also real. In
fact from (\ref{sei-cinque})-(\ref{sei-sette-a})
\begin{equation}
 \left\{
	\begin{array}{l}	
	\displaystyle
	\label{i-34-2}
	\alpha^*\beta+\beta^*\delta=0
        \smallskip \nonumber\\
        \displaystyle 
        \alpha\gamma^*+\delta^*\gamma=0
        \smallskip \nonumber\\
        \alpha\delta-\beta\gamma=1
        \end{array}
	\right.
\end{equation}
we get
\begin{equation}
 \left\{
	\begin{array}{l}	
	\displaystyle
	\label{i-34-3}
	\alpha^*\beta=-\beta^*\delta
        \smallskip \nonumber\\
        \displaystyle 
        \delta^*\gamma=-\alpha\gamma^*
        \end{array}
	\right.
\end{equation}
Multiplying  the LHS and the RHS of the two equations appearing in (\ref{i-34-3}),  we get 
	\begin{equation}
	(\alpha^*\beta\gamma\delta^*)^*=\alpha^*\beta\gamma\delta^*
	\end{equation}
which implies that the expression
	\begin{equation}
	\alpha^*\beta\gamma\delta^* \label{i-34-4}
	\end{equation}
is real.
From the last of Eq. (\ref{i-34-2}) we get that 
	\begin{equation}
	\alpha^*\delta^*=1+\beta^*\gamma^*. \label{i-35-1}
	\end{equation}
and using the RHS of (\ref{i-35-1}) into (\ref{i-34-4}) we obtain that the expression 
$(1+\beta^*\gamma^*)\beta\gamma=|\beta^*\gamma^*|^2+\beta\gamma$ is real. 
Combining this with the fact that (\ref{i-34-0}) is imaginary we get that 
	\begin{equation}
	|\beta^*\gamma^*|^2+\beta^*\gamma^*=0
	\end{equation}
which means 
	\begin{equation}
	a^2+b^2+a+ib=0
	\end{equation}
and this implies $(b=0, a=-1)$ or $(b=0,a=0)$. In the first case as a consequence we have
	\begin{equation}
	\beta^*\gamma^*=-1
	\end{equation}
which, from (\ref{i-33-2}), is the same as 
	\begin{equation}
	e^{i(\theta_{\alpha}+\theta_{\delta})}=0
	\end{equation}
which has no solution. The other case is ($b=0,a=0$) which implies $\beta^*\gamma^*=0$ whose solutions are $\beta=0$ or
$\gamma=0$ and, as a consequence, $\beta^*\gamma=0$ which is exactly the second case of (\ref{sei-undici}). So the last relation of
(\ref{sei-undici}) gives the same hermiticity conditions (\ref{i-31-3}) and (\ref{i-32-2}) given 
by the second case of (\ref{sei-undici}).
\\

\noindent $\bullet$ \underline{\it Derivation of (\ref{sei-diciannove})}. 

\noindent
We have to insert the expression
(\ref{sei-quindici}) into (\ref{sei-diciassette}) and check what we get. The LHS and RHS of the first Eq. of
(\ref{sei-diciassette}) are respectively:
	\begin{eqnarray}
	\displaystyle \langle \hat{c}^p\Phi|\psi\rangle &=& g^{20}\Phi_0^*\psi_0+g^{21}\Phi_0^*\psi_1+g^{22}\Phi_0^*\psi_2+g^{23}\Phi_0^*\psi_3
	\nonumber\\
	& &-g^{30}\Phi_1^*\psi_0-g^{31}\Phi_1^*\psi_1-g^{32}\Phi_1^*\psi_2-g^{33}\Phi_1^*\psi_3
	\label{i-37-1}
	\end{eqnarray}
	\begin{eqnarray}
	\displaystyle \langle \Phi|ib\frac{\partial}{\partial c^q}\psi\rangle &=&
	g^{00}\Phi_0^*ib\psi_1+g^{10}\Phi_1^*ib\psi_1+g^{20}\Phi_2^*ib\psi_1+g^{30}\Phi_3^*ib\psi_1
	\nonumber\\
	& & +g^{02}\Phi_0^*ib\psi_3+g^{12}\Phi_1^*ib\psi_3+g^{22}\Phi_2^*ib\psi_3+g^{32}\Phi_3^*ib\psi_3.
	\label{i-37-2}
	\end{eqnarray}
Equating these two expressions we get 
	\begin{equation}
	 \left\{
		\begin{array}{l}	
		\displaystyle
		\label{i-37-3}
		g^{20}=g^{22}=g^{30}=g^{32}=0
	        \smallskip \nonumber\\
	        \displaystyle 
	        g^{21}=ibg^{00},\;\; g^{23}=ibg^{02}
	        \smallskip \nonumber\\
	        g^{31}=-ibg^{10},\;\; g^{33}=-ibg^{12}.
	        \end{array}
		\right.
	\end{equation}
Doing the same for the second Eq. in  (\ref{sei-diciassette}) we obtain
	\begin{equation}
	 \left\{
		\begin{array}{l}	
		\displaystyle
		\label{i-38-1}
		-ibg^{00}=g^{12};\;\; -ibg^{01}=-g^{13};\;\; g^{02}=g^{03}=0
	        \smallskip \nonumber\\
	        \displaystyle 	
	        -ibg^{20}=g^{32};\;\; -ibg^{21}=-g^{33};\;\;g^{22}=g^{23}=0.
	        \end{array}
		\right.
	\end{equation}
Instead for Eqs. (\ref{sei-diciotto}) we get:
	\begin{equation}
	 \left\{
		\begin{array}{l}
		\displaystyle
		\label{i-38-2}
		g^{10}=0;\;\;g^{11}=0;\;\;g^{12}=iag^{00};\;\;g^{13}=-iag^{01}
	        \smallskip \nonumber\\
	        \displaystyle 
	        g^{30}=0;\;\;g^{31}=0;\;\;g^{32}=iag^{20};\;\;g^{33}=-iag^{21}
	        \smallskip \nonumber\\
	        \displaystyle -iag^{00}=g^{21};\;\;g^{01}=0;\;\;-iag^{02}=g^{23};\;\;g^{03}=0
	        \smallskip \nonumber\\
	        iag^{10}=g^{31};\;\;g^{11}=0;\;\;iag^{12}=g^{33};\;\;g^{13}=0.
	        \end{array}
		\right.
	\end{equation}
From (\ref{i-37-3}), (\ref{i-38-1}) and (\ref{i-38-2}) we deduce  that the only metric components different from zero are
$g^{00}, g^{12}, g^{21}, g^{33}$ and they are given by
	\begin{equation}
	 \left\{
		\begin{array}{l}
		\displaystyle
		\label{i-38-3}
		g^{21}=ibg^{00}=-iag^{00}
	        \smallskip \nonumber\\
	        \displaystyle 
	        g^{12}=-ibg^{00}=iag^{00}
	        \smallskip \nonumber\\
	        \displaystyle 
	        g^{33}=ibg^{21}=-b^2g^{00}.
	        \end{array}
		\right.
	\end{equation}
From this we obtain that, for consistency, $b=-a$ and if we choose $g^{00}=1$ we get the metric (\ref{sei-diciannove}). The
choice $g^{00}=1$ is allowed because the relations (\ref{i-38-3}) do not put any constraint on $g^{00}$. Any other choice
of $g^{00}$ would change all the eigenvalues of (\ref{sei-venti}) only by a common multiplicative factor $g^{00}$.
\\

\noindent $\bullet$ \underline{\it Derivation of (\ref{sei-venticinque})}. 

\noindent
Let us first check whether the commutation relations
(\ref{sei-ventidue}) and (\ref{sei-ventiquattro}) restrict the form of the parameters entering (\ref{sei-ventiquattro-b}).
From (\ref{sei-ventiquattro}) we have
	\begin{equation}
	[\hat{c}^{p\dagger},\hat{\bar{c}}_p^{\dagger}]=[e^{i\theta_{\alpha}}\hat{c}^p,i\gamma_I^{\prime}
	\hat{c}^q+e^{-i\theta_{\beta}}\hat{\bar{c}}_p]=
	e^{i(\theta_{\alpha}
	-\theta_{\beta})}
	\end{equation}
which implies $\theta_{\alpha}=\theta_{\beta}$, so also (\ref{sei-ventidue}) is satisfied
	\begin{equation}
	[\hat{c}^{q\dagger},\hat{\bar{c}}_q^{\dagger}]=[e^{i\theta_{\beta}}\hat{c}^q,i\gamma_{I}\hat{c}^p+
	e^{-i\theta_{\alpha}}\hat{\bar{c}}_q]=1.
	\end{equation}
Let us now check whether $[\hat{\bar{c}}_p^{\dagger},\hat{\bar{c}}_q^{\dagger}]=0$: 
	\begin{eqnarray}
	[\hat{\bar{c}}_p^{\dagger},\hat{\bar{c}}_q^{\dagger}]&=&
	[i\gamma_I^{\prime}\hat{c}^q+e^{-i\theta_{\alpha}}\hat{\bar{c}}_p,i\gamma_I\hat{c}^p+
	e^{-i\theta_{\alpha}}\hat{\bar{c}}_q]=\nonumber\\
	&=& i\gamma_I^{\prime}e^{-i\theta_{\alpha}}+i\gamma_Ie^{-i\theta_{\alpha}}
	\end{eqnarray}
which implies $\gamma_I=-\gamma_I^{\prime}$. Next let us take the matrix element of 
(\ref{sei-ventiquattro-b}) between the bra $\langle\Phi|$ and the ket $|\psi\rangle$:
	\begin{equation}
	\langle \hat{c}^p\Phi|\psi\rangle=\langle\Phi|e^{i\theta_{\alpha}}\hat{c}^p\psi\rangle.
	\label{i-39-1}
	\end{equation}
Proceeding as we did in (\ref{i-37-1})-(\ref{i-37-2}) we get from (\ref{i-39-1})
	\begin{equation}
	 \left\{
		\begin{array}{l}
		\displaystyle
		\label{i-40-1}
		g^{20}=g^{02}e^{i\theta_{\alpha}};\;\;g^{21}=-g^{03}e^{i\theta_{\alpha}};\;\;g^{22}=0;\;\;g^{23}=0
	        \smallskip \nonumber\\
	        \displaystyle 
	        -g^{30}=g^{12}e^{i\theta_{\alpha}};\;\;g^{31}=g^{13}e^{i\theta_{\alpha}};\;\;g^{32}=0;\;\;g^{33}=0.
	        \end{array}
		\right.
	\end{equation}
For simplicity let us choose $\gamma_I=\gamma_I^{\prime}=0$ and then, from the second of Eq.
(\ref{sei-ventiquattro-b}):\break
$\langle \hat{\bar{c}}_q\Phi|\psi\rangle=\langle \Phi| e^{-i\theta_{\alpha}}\hat{\bar{c}}_q|\psi\rangle$, we have:
	\begin{equation}
	 \left\{
		\begin{array}{l}
		\displaystyle
		\label{i-40-2}
		g^{00}=0;\;\; g^{01}=g^{10}e^{-i\theta_{\alpha}}
	        \smallskip \nonumber\\
	        \displaystyle 
	        g^{02}=0;\;\;g^{03}=g^{12}e^{-i\theta_{\alpha}}
	        \smallskip \nonumber\\
	        \displaystyle 
	        g^{20}=0;\;\; g^{21}=g^{30}e^{-i\theta_{\alpha}}.
	        \end{array}
		\right.
	\end{equation}
From the third of Eq. (\ref{sei-ventiquattro-b}) that is
$\langle \hat{c}^q\Phi|\psi\rangle=\langle \Phi|e^{i\theta_{\alpha}}
\hat{c}^q\psi\rangle$ we get
	\begin{equation}
	 \left\{
		\begin{array}{l}
		\displaystyle
		\label{i-40-3}
		g^{11}=0;\;\; g^{13}=0
	        \smallskip \nonumber\\
	        \displaystyle 
	        g^{31}=0;\;\;g^{32}=g^{23}e^{i\theta_{\alpha}}
	        \end{array}
		\right.
	\end{equation}
and from the last one of Eq. (\ref{sei-ventiquattro-b}) that is 
$\langle \hat{\bar{c}}_p\Phi|\psi\rangle=\langle\Phi|
e^{-i\theta_{\alpha}}\hat{\bar{c}}_p
\psi\rangle$ we obtain 
	\begin{equation}
	g^{10}=0;\;\;g^{01}=0;\;\;g^{11}=0 .\label{i-40-4}
	\end{equation}
Combining (\ref{i-40-1})-(\ref{i-40-4}) we get the metric (\ref{sei-venticinque}) in the case 
$\gamma_I=\gamma_I^{\prime}=0$.
The calculations which lead to (\ref{sei-ventisei}) are very similar to the previous ones and we will skip them.
\\

\noindent $\bullet$ {\underline{\it Derivation of (\ref{sei-trenta})}. 

\noindent
From (\ref{sei-dodici-3}) 
and (\ref{sei-tredici-3}) we have that 
	\begin{equation}
	 \left\{
		\begin{array}{l}
		\displaystyle
		\label{i-41-1}
		\hat{c}^{p\dagger}=e^{i\theta_{\alpha}}\hat{c}^p+ib\hat{\bar{c}}_q
	        \smallskip \nonumber\\
	        \displaystyle 
	        \hat{\bar{c}}_q^{\dagger}=e^{-i\theta_{\alpha}}\hat{\bar{c}}_q
	        \smallskip \nonumber\\
	        \displaystyle \hat{c}^{q\dagger}=e^{i\theta_{\beta}}\hat{c}^q+ia\hat{\bar{c}}_p
	        \smallskip \nonumber\\	
	        \displaystyle \hat{\bar{c}}_p^{\dagger}=e^{-i\theta_{\beta}}\hat{\bar{c}}_p
	        \end{array}
		\right.
	\end{equation}
and imposing the conditions 
	\begin{eqnarray}
	&& [\hat{c}^{p\dagger},\hat{\bar{c}}_p^{\dagger}]=1\nonumber\\
	&& [\hat{c}^{q\dagger},\hat{\bar{c}}_q^{\dagger}]=1\\
	&& [\hat{c}^{p\dagger},\hat{c}^{q\dagger}]=0 \nonumber
	\end{eqnarray}
we obtain that 
	\begin{equation}
	 \left\{
		\begin{array}{l}
		\displaystyle
		\label{i-41-2}
		\theta_{\alpha}=\theta_{\beta}
	        \smallskip \nonumber\\
	        \displaystyle 
	        b=-a.
	        \smallskip 
	        \end{array}
		\right.
	\end{equation}
If we choose $b=-a=0$ we end up in the case (\ref{sei-ventiquattro-b}) with $\gamma_I=\gamma_I^{\prime}=0$ 
and get the same metric (\ref{sei-venticinque}).
Now, instead of deriving (\ref{sei-trenta}), we will just verify that the hermiticity conditions (\ref{sei-dodici-3}) 
and (\ref{sei-tredici-3}) are satisfied if the scalar product is calculated using the metric (\ref{sei-trenta}).
Let us, for example, verify one of the hermiticity condition which is:
	\begin{equation}
	\displaystyle \langle \hat{c}^p\Phi| \psi\rangle=\langle \Phi|(e^{i\theta_{\alpha}}\hat{c}^p+ib\hat{\bar{c}}_q)\psi\rangle.
	\label{i-42-1}
	\end{equation}
The LHS of (\ref{i-42-1}), using (\ref{sei-trenta}), is 
	\begin{eqnarray}
	\langle \hat{c}^p\Phi|\psi\rangle&=&\langle 0, 0, \Phi_0, -\Phi_1|\psi_0,\psi_1,\psi_2,\psi_3\rangle=\nonumber\\
	&=&-\Phi_0^*g^{03}e^{i\theta_{\alpha}}\psi_1+\Phi^*_1g^{03}e^{2i\theta_{\alpha}}\psi_0+ig^{03}e^{i\theta_{\alpha}}
	b\Phi_1^*\psi_3. \label{i-42-2}
	\end{eqnarray}
The RHS of (\ref{i-42-1}) is
	\begin{eqnarray}
	&&\label{i-42-3} \langle \Phi|(e^{i\theta_{\alpha}}\hat{c}^p+ib\hat{\bar{c}}_q|\psi\rangle=
	\langle \Phi_0,\Phi_1,\Phi_2,\Phi_3|ib\psi_1, 0, e^{i\theta_{\alpha}}\psi_0+ib\psi_3, -e^{i\theta_{\alpha}}\psi_1\rangle=
	\\
	&&=-\Phi^*_0e^{i\theta_{\alpha}}\psi_1g^{03}+\Phi^*_1g^{03}e^{2i\theta_{\alpha}}\psi_0+ib\Phi_1^*g^{03}
	e^{i\theta_{\alpha}}\psi_3
	-\Phi^*_3g^{03}e^{2i\theta_{\alpha}}ib\psi_1+i\Phi_3^*g^{03}e^{2i\theta_{\alpha}}b\psi_1 \nonumber
	\end{eqnarray}
and it is now easy to see that the RHS of (\ref{i-42-2}) and (\ref{i-42-3}) are equal. In the same manner it is straightforward
to prove that the other hermiticity conditions (\ref{sei-dodici-3}) and (\ref{sei-tredici-3}) are verified using the metric
(\ref{sei-trenta}) and the conditions (\ref{i-41-2}).


\newpage

\section{}
\noindent
In this appendix we show that for those systems (the chaotic ones) where the Jacobi fields (\ref{sette-uno}) diverge
exponentially with time, the same happens with the components $\psi_a$ of the one-forms 

\noindent $\psi=\psi_ac^a$. Let us first
change representation, that means let us turn to a ``momentum" representation
for the Grassmannian variables using $\bar{c}$
in place of $c$. While in the SvH representation that we have been using so far the notation (\ref{due-quarantanove}) was:
	\begin{equation}
	\langle +c^{p*},+c^{q*},\varphi|\psi\rangle=\psi_0+\psi_qc^q+\psi_pc^p+\psi_2c^qc^p \label{i-44-1}
	\end{equation}
it is now natural to use the following other notation for the $\bar{c}$ representation:
	\begin{equation}
	\langle -\bar{c}_p^*,-\bar{c}_q^*,\varphi|\psi\rangle=\psi^0+\psi^q\bar{c}_q+\psi^p\bar{c}_p+\psi^2\bar{c}_q\bar{c}_p.
	\label{i-44-2}
	\end{equation}
In this basis a completeness relation, analog to the first one of (\ref{due-quarantuno}), is then
	\begin{equation}
	\int d\bar{c}_qd\bar{c}_p|\bar{c}_q+,\bar{c}_p+\rangle\langle-\bar{c}_p^*,-\bar{c}_q^*|={\bf 1}
	\end{equation}
and inserting it into the LHS of (\ref{i-44-1}) we get the relation between (\ref{i-44-1}) and (\ref{i-44-2}), i.e.:
	\begin{eqnarray}
	&& \langle +c^{p*},+c^{q*}|\psi\rangle=\int d\bar{c}_qd\bar{c}_p\langle +c^{p*},+c^{q*}|\bar{c}_q+,\bar{c}_p+\rangle
	\cdot \langle -\bar{c}_p^*, -\bar{c}_q^*|\psi\rangle=\nonumber\\
	&& =\int d\bar{c}_qd\bar{c}_p (1+c^q\bar{c}_q+c^p\bar{c}_p-c^qc^p\bar{c}_q\bar{c}_p)\cdot
	(\psi^0+\psi^q\bar{c}_q+\psi^p\bar{c}_p+\psi^2\bar{c}_q\bar{c}_p)=\nonumber\\
	&&=-\psi^2-\psi^pc^q+\psi^qc^p+\psi^0c^qc^p.
	\end{eqnarray}
Comparing this with the RHS of (\ref{i-44-1}) we get that
	\begin{equation}
	 \left\{
		\begin{array}{l}
		\displaystyle
		\label{i-45-1}
		\psi_0=-\psi^2
	        \smallskip \nonumber\\
	        \displaystyle 
	        \psi_q=-\psi^p
	        \smallskip \nonumber\\
	        \displaystyle 
	        \psi_p=\psi^q
	        \smallskip \nonumber\\
	        \displaystyle 
	        \psi_2=\psi^0.
	        \end{array}
		\right.
	\end{equation}
The reason we have introduced the $\bar{c}$ representation is because, as we will show below, the components $\psi^q$ and
$\psi^p$ transform, under time evolution, as the Jacobi fields $\delta q,\delta p$ of (\ref{sette-uno}).
The evolution of the wave functions (\ref{i-44-2}) is given by the Hamiltonian $\widetilde{\cal H}$ expressed
in the ``momentum" representation. The fermionic part of $\widetilde{\cal H}_{ferm}$ is
	\begin{equation}
	\widetilde{\cal H}_{ferm}=i\bar{c}_a\omega^{ab}\partial_b\partial_dHc^d=i\bar{c}_aM^a_dc^d
	\label{i-46-1}
	\end{equation}
where $M^a_d=\omega^{ab}\partial_b\partial_dH$ and its operatorial counterpart is
	\begin{equation}
	\displaystyle \widetilde{\cal H}_{ferm}=i\bar{c}_aM^a_d\frac{\partial}{\partial\bar{c}_d} .\label{i-46-2}
	\end{equation}
The infinitesimal evolution will give:
	\begin{equation}
	\displaystyle \psi^a(\epsilon)\bar{c}_a(\epsilon)=e^{-i\widetilde{\cal H}\epsilon}\,\biggl(\psi^a(0)\bar{c}_a(0)
	\biggr)
	\label{i-46-3}
	\end{equation}
where we have restricted ourselves to the part of the wave functions (\ref{i-44-2}) linear in the $\bar{c}$ variables.
Inserting (\ref{i-46-2}) into (\ref{i-46-3}) we get
	\begin{eqnarray}
	\displaystyle \psi^a(\epsilon)\bar{c}_a(\epsilon)&=&\biggl(1+\epsilon\bar{c}_bM^b_d\frac{\partial}{\partial\bar{c}_d}
	\biggr)\psi^a(0)\bar{c}_a(0)\nonumber\\
	&=& (\psi^a(0)+\epsilon M^a_d\psi^d(0))\bar{c}_a(0). \label{i-46-4}
	\end{eqnarray}
Expanding in $\epsilon$ also the LHS of (\ref{i-46-4}) and comparing terms of the first order in $\epsilon$ and
proportional to $\bar{c}_a(0)$ we get
	\begin{equation}
	\psi^a(\epsilon) =\psi^a(0)+\epsilon M^a_d\psi^d(0). \label{i-47-1}
	\end{equation}
If we now solve, for an infinitesimal time, the equation of motion of the Jacobi fields we get
	\begin{equation}
	\delta\varphi^a(\epsilon)=\delta\varphi^a(0)+\epsilon M^a_d\delta\varphi^d(0). \label{i-47-2}
	\end{equation}
This proves that the variable $\psi^a$ and the Jacobi fields $\delta\varphi^a$ evolve in the same way and if the
latter diverge
exponentially with $t$, the same happens to $\psi^a$. This implies that the behavior with $t$ of the distance
$D(\varphi_0,t)$ of (\ref{sette-tre}) is the same as the one of $\displaystyle \sum_a |\psi_a(\varphi_0)|^2$. This also implies that, if
(\ref{sette-quattro}) holds, also the following inequality holds
	\begin{equation}
	\displaystyle \lim_{t\to\infty}\frac{1}{t}ln\int d^{2n}\varphi_0\sum_a|\psi^a(\varphi_0,t)|^2>0.
	\label{i-47-3}
	\end{equation}
Via then the relations (\ref{i-45-1}) we can replace (\ref{i-47-3}) with 
	\begin{equation}
	\lim_{t\to\infty}\frac{1}{t}ln\int d^{2n}\varphi_0\sum_a|\psi_a(\varphi_0,t)|^2>0.
	\end{equation}
Note that the argument of the logarithm is exactly the SvH norm of the one-form $\psi_a(\varphi_0,t)c^a$. We should point
out that the previous reasoning is correct for those forms which have non-zero components along diverging Jacobi fields.


\newpage

\section{}
\noindent
We noticed in Sec. VII that, with the choice 
	\begin{equation}
	\displaystyle H=\frac{1}{2}p^2+\frac{1}{2}q^2 , \label{i-49-1}
	\end{equation}
the Hamiltonian $\widetilde{\cal H}$ is Hermitian in the SvH scalar product. Let
us now insert the mass and frequency in
$H$ and see what
happens:
	\begin{equation}
	\displaystyle H=\frac{1}{2m}p^2+\frac{1}{2}m\omega^2q^2 .\label{i-49-2}
	\end{equation}
Now let us check if $\widetilde{\cal H}$ is Hermitian under the SvH hermiticity
conditions
	\begin{equation}
	 \left\{
		\begin{array}{l}
		\displaystyle
		\label{i-49-3}
		(\hat{c}^q)^{\dagger}=\hat{\bar{c}}_q
	        \smallskip \nonumber\\
	        \displaystyle 
	        (\hat{c}^p)^{\dagger}=\hat{\bar{c}}_p.
	        \end{array}
		\right.
	\end{equation}
The only part of $\widetilde{\cal H}$ which can encounter problems is
$\widetilde{\cal H}_{ferm}$ which, with the
Hamiltonian (\ref{i-49-2}), turns out to be
	\begin{equation}
	\displaystyle \widetilde{\cal
	H}_{ferm}=i\frac{\hat{\bar{c}}_q\hat{c}^p}{m}-i\hat{\bar{c}}_p\hat{c}^qm\omega^2.
	\label{i-49-4}
	\end{equation}
Its Hermitian conjugate, using (\ref{i-49-3}), is
	\begin{equation}
	\displaystyle \widetilde{\cal H}_{ferm}^{\dagger}=
	i\hat{\bar{c}}_q\hat{c}^pm\omega^2-i\frac{\hat{\bar{c}}_p\hat{c}^q}{m}.
	\label{i-49-5}
	\end{equation}
From (\ref{i-49-4}) and (\ref{i-49-5}) one sees that $\widetilde{\cal H}_{ferm}$
is Hermitian only if 
	\begin{equation}
	\displaystyle \frac{1}{\omega^2}=m^2. \label{i-50-1}
	\end{equation}
So, given the mass of the system, only those harmonic oscillators whose
frequency is given by (\ref{i-50-1}) are
Hermitian. It is easy to see that the oscillators for which relation
(\ref{i-50-1}) holds are those whose trajectories in
phase space are circles (see Fig. 1). Note that in this case the norm of the
Jacobi fields, i.e. 
the arrows in Fig. 1, does not change with time that is
what the Hermiticity of $\widetilde{\cal H}$ under the SvH scalar product 
guarantees. In fact {\it if we use the SvH scalar product
the norm of the Jacobi fields can be put in correspondence with the norm of
one-forms as shown
in Appendix E}. So, as the Hermiticity of $\widetilde{\cal H}$ 
preserves the norm of the one-forms, the same happens for the norm of the Jacobi
fields. In the case in which $\displaystyle \frac{1}{\omega^2}\neq m^2$
the trajectories are ellipses (see Fig. 2). One sees from Fig. 2 that the norm
of the Jacobi fields is not preserved
during the time evolution and this is just a consequence of the non-Hermiticity
of $\widetilde{\cal H}$
under the SvH scalar product. 

The careful reader may object to the previous argument that, even when
$\displaystyle \frac{1}{\omega^2}\neq m^2$, it is
possible to perform the following canonical
tranformation:
\begin{equation}
\displaystyle p=\sqrt{m\omega}\,p^{\prime},\qquad \qquad
q=\frac{1}{\sqrt{m\omega}}\,q^{\prime}
\label{canonical}
\end{equation} 
in order to bring the Hamiltonian (\ref{i-49-2}) into the form:
\begin{equation}
\displaystyle H(q,p)=\frac{1}{2m}p^2+\frac{1}{2}m\omega^2q^2\,\rightarrow\,
H(q^{\prime},p^{\prime})=\frac{1}{2}\omega p^{\prime \,2}
+\frac{1}{2}\omega q^{\prime \,2}. \label{newham}
\end{equation}
In the new phase space labelled by $(q^{\prime},p^{\prime})$ the trajectories of
the 
harmonic oscillator are given by circles and the associated $\widetilde{\cal
H}_{ferm}$
is:
\begin{equation}
\displaystyle \widetilde{\cal
H}_{ferm}=i\omega\hat{\bar{c}}_q^{\,\prime}\hat{c}^{\prime\,p}
-i\omega\hat{\bar{c}}_p^{\,\prime}\hat{c}^{\prime\,q}. \label{hamprimed}
\end{equation}
What can we say about the Hermiticity
of the operator $\widetilde{\cal H}_{ferm}$? Before the canonical transformation
(\ref{canonical}) the Hamiltonian $\widetilde{\cal H}_{ferm}$ in (\ref{i-49-5})
was not Hermitian 
for $\displaystyle \frac{1}{\omega^2}\neq m^2$.
Is this property preserved or not?
To answer this question we must remember that under a canonical transformation,
as well as
under a generic diffeomorphism in phase space, the Grassmann variables of the
theory must
be tranformed like a basis of differential forms and vector fields
respectively, see Eq. (5.7) of the second paper in Ref. [4]:
\begin{equation}
\displaystyle
c^{\prime\,a}=\frac{\partial\varphi^{\prime\,a}}{\partial\varphi^b}c^b,\qquad
\qquad
\bar{c}_a^{\,\prime}=\frac{\partial\varphi^b}{\partial\varphi^{\prime\,a}}\bar{c}_b.
\label{transf}
\end{equation}
In our particular case the transformations on the Grassmann variables
induced by (\ref{canonical}) are given by:
\begin{equation}
\left\{
\begin{array}{l}
\displaystyle c^{\prime\,q}=\sqrt{m\omega}\,c^q,\qquad\qquad
c^{\prime\,p}=\frac{1}{\sqrt{m\omega}}\,c^p,\\
\displaystyle \bar{c}_q^{\,\prime}=\frac{1}{\sqrt{m\omega}}\,\bar{c}_q,
\qquad\qquad \bar{c}_p^{\,\prime}=\sqrt{m\omega}\,\bar{c}_p. \label{enrico}
\end{array}
\right.
\end{equation}
To check the consistency of the formalism we can note that, by inserting the
inverse of (\ref{enrico})
into (\ref{i-49-4}) one obtains just the Hamiltonian in the primed variables
(\ref{hamprimed}).
Furthermore the SvH scalar product (\ref{i-49-3}) in the 
primed variables becomes:
\begin{equation}
\left\{
\begin{array}{l}
\displaystyle
(\hat{c}^{\prime\,q})^{\dagger}=m\omega\,\hat{\bar{c}}_q^{\,\prime}\medskip \\ 
\displaystyle
(\hat{c}^{\prime\,p})^{\dagger}=\frac{1}{m\omega}\,\hat{\bar{c}}_p^{\,\prime}.
\label{newscpr}
\end{array}
\right. 
\end{equation}
Since we are interested in the case $m\omega\neq 1$ we have that, after the
canonical 
transformation (\ref{canonical})-(\ref{enrico}), the original SvH scalar product
(\ref{i-49-3})
changed its explicit form. Under this scalar product 
(\ref{newscpr}) the Hermitian conjugate of the Hamiltonian (\ref{hamprimed}) is:
\begin{equation}
\displaystyle \widetilde{\cal H}^{\dagger}_{ferm}=
i\omega^3m^2\hat{\bar{c}}_q^{\,\prime}\hat{c}^{\prime\,p}
-\frac{i}{m^2\omega}\hat{\bar{c}}_p^{\,\prime}
\hat{c}^{\prime\,q}. \label{hamprimeddag}
\end{equation}
From (\ref{hamprimed}) and (\ref{hamprimeddag}) one sees that the operator
$\widetilde{\cal H}_{ferm}$
is not Hermitian if $\displaystyle \frac{1}{\omega^2}\neq m^2$. Therefore the
property of 
non-Hermiticity of $\widetilde{\cal H}_{ferm}$ is preserved by the canonical 
transformation\footnote{We want to stress that there exists a particular scalar
product under which the Hamiltonian
$\widetilde{\cal H}_{ferm}$ of (\ref{hamprimed}) is Hermitian: it is the SvH
scalar product in the primed variables
$(\hat{c}^{\prime\,a})^{\dagger}=\hat{\bar{c}}_a^{\,\prime}$;
anyway this scalar product is different from the SvH one in the unprimed
variables 
(\ref{i-49-3}) that we originally imposed upon our Hilbert space. 
Once we fix a particular scalar product there exists only one
particular harmonic oscillator whose Hamiltonian $\widetilde{\cal H}_{ferm}$ is
Hermitian.} $(q,p,c^q,c^p)\rightarrow
(q^{\prime},p^{\prime},c^{\prime\, q},c^{\prime\,p})$.

Vice versa let us suppose we consider the oscillator $\displaystyle 
H=\frac{1}{2}q^2+\frac{1}{2}p^2$ whose associated $\widetilde{\cal H}_{ferm}$ is
Hermitian with the 
SvH scalar product (\ref{i-49-3}) and let us perform the following canonical
transformation with $\alpha\neq 1$:
$\displaystyle q=\frac{1}{\alpha}\,q^{\prime},\;\; p=\alpha \,p^{\prime}$, which
induces the following 
transformations on the $c$ and $\bar{c}$:
\begin{equation}
\left\{
\begin{array}{l}
\displaystyle c^q=\frac{1}{\alpha}\,c^{\prime\,q},\qquad\quad\; c^p=\alpha
\,c^{\prime\,p},
\medskip\\
\displaystyle \bar{c}_q=\alpha\,\bar{c}_q^{\,\prime},\qquad\qquad
\bar{c}_p=\frac{1}{\alpha}\,\bar{c}_p^{\,\prime}.
\end{array}
\right.
\end{equation}
Under these transformations the Hamiltonian $H$ becomes
$\displaystyle H(q^{\prime},p^{\prime})=\frac{1}{2\alpha^2}q^{\prime\,
2}+\frac{1}{2}\alpha^2p^{\prime \,2}$, 
while the fermionic part of $\widetilde{\cal H}$ turns out to be:
\begin{equation}
\displaystyle \widetilde{\cal
H}_{ferm}=i\bar{c}_q^{\,\prime}c^{\prime\,p}\alpha^2
-i\bar{c}_p^{\,\prime}c^{\prime\,q}\frac{1}{\alpha^2} \label{giogio}
\end{equation}
and the scalar product (\ref{i-49-3}) becomes:
\begin{equation}
\left\{
\begin{array}{l}
\displaystyle
(\hat{c}^{\prime\,q})^{\dagger}=\alpha^2\,\hat{\bar{c}}_q^{\,\prime} \medskip \\
\displaystyle
(\hat{c}^{\prime\,p})^{\dagger}=\frac{1}{\alpha^2}\,\hat{\bar{c}}_p^{\,\prime}.
\label{pav}
\end{array}
\right.
\end{equation}
With this scalar product it is easy to prove that the Hamiltonian (\ref{giogio})
is still 
Hermitian and this confirms that the canonical
transformations do not spoil the Hermiticity of the Hamiltonian $\widetilde{\cal
H}$. 
This proof can be extended to the most general canonical 
transformation. In fact in the primed variables given by (\ref{transf}) 
the fermionic Hamiltonian associated to $H=\frac{1}{2}q^2+\frac{1}{2}p^2$
becomes:
\begin{equation}
\displaystyle \widetilde{\cal H}_{ferm}=i\frac{\partial\varphi^{\prime
b}}{\partial q}
\frac{\partial p}{\partial \varphi^{\prime d}}\hat{\bar{c}}_b^{\prime}
\hat{c}^{\prime d}-i\frac{\partial \varphi^{\prime b}}{\partial p}
\frac{\partial q}{\partial \varphi^{\prime
d}}\hat{\bar{c}}^{\prime}_b\hat{c}^{\prime d}. \label{rus}
\end{equation}
The SvH scalar product, which gives $\hat{c}^{a \dagger}=\hat{\bar{c}}_a$, in
the primed variables
becomes:
\begin{equation}
\displaystyle \delta_k^a(\hat{c}^{\prime d})^{\dagger}=\frac{\partial
\varphi^{\prime
b}}{\partial \varphi^k}\frac{\partial
\varphi^{\prime d}}
{\partial \varphi^{a}}\hat{\bar{c}}_b^{\prime}, \qquad \qquad 
\delta_b^k(\hat{\bar{c}}_d^{\prime})^{\dagger}
=\frac{\partial \varphi^k}{\partial\varphi^{\prime
d}}\frac{\partial\varphi^b}{\partial \varphi^{\prime f}}
\hat{c}^{\prime f}
\end{equation}
and with this scalar product it is straightforward to prove that the Hamiltonian
(\ref{rus}) is still Hermitian. 
This implies that the Hermiticity of $\widetilde{\cal H}$ is an
intrinsic (i.e. canonical invariant) property of the system which can only
change with the 
scalar product we choose.

Now the careful reader can ask: in the case described by (\ref{giogio}) the
Hamiltonian of the system 
is Hermitian but the norm of the Jacobi fields is not conserved since the
trajectories in the
phase space labelled by $(q^{\prime},p^{\prime})$ are ellipses. So in this case
the Hermiticity of the Hamiltonian
$\widetilde{\cal H}$ cannot imply that the Jacobi fields are conserved in time. 
The reason is easily understood if we consider that the matrix $g^{ij}$ 
which produces the scalar product (\ref{pav}) in the primed variables is given
by:
\begin{equation}
g^{ij}=\pmatrix{1 & 0 & 0 & 0\cr
0 & \alpha^2 & 0 & 0\cr 0 & 0 & 1/\alpha^2 & 0\cr 0 & 0 & 0 & 1}.
\end{equation}
Consequently the norm of the state $\psi=\psi_{\scriptscriptstyle
0}+\psi_qc^{q\,\prime}+\psi_pc^{p\,\prime}$ 
is:
\begin{equation}
\displaystyle \langle \psi|\psi\rangle =\int
d\varphi\biggl(|\psi_{\scriptscriptstyle 0}(\varphi)|^2+
\alpha^2|\psi_q(\varphi)|^2+\frac{1}{\alpha^2}|\psi_p(\varphi)|^2\biggr).
\label{c-18}
\end{equation}
Remember that the piece $\displaystyle \int d\varphi |\psi_{\scriptscriptstyle
0}(\varphi)|^2$
in (\ref{c-18}) is conserved because $\psi_{\scriptscriptstyle 0}(\varphi)$
evolves
with the Liouvillian which is a Hermitian operator. Therefore the Hermiticity of
$\widetilde{\cal H}_{ferm}$
implies immediately that the leftover piece in (\ref{c-18}), which is
$\displaystyle
\int d\varphi\biggl(
\alpha^2|\psi_q(\varphi)|^2+\frac{1}{\alpha^2}|\psi_p(\varphi)|^2\biggr)$, is
conserved in time. 
But since $\alpha\neq 1$ this does not imply that the norm of the Jacobi fields
is conserved.
In fact the square of the length of the Jacobi field is given by $
|\delta q|^2+|\delta p|^2$ and, according to what we proved in Appendix E, 
this quantity can be put in correspondence with
$|\psi_q|^2+|\psi_p|^2$.
Therefore the length of the Jacobi field can be associated with the contribution
to the 
norm of $\psi$ given by the one-form only if $\alpha=1$, i.e. only if we
consider the SvH scalar 
product in the primed variables
$\hat{c}^{\prime\,\dagger}=\hat{\bar{c}}^{\,\prime}$. 
With all the other scalar products,
including (\ref{pav}) which is canonically ``equivalent" to the original SvH one
(\ref{i-49-3}), 
the Hermiticity of
the Hamiltonian $\widetilde{\cal H}$ does not imply that the length of the
Jacobi fields
is conserved. The same thing can be phrased as follows: if we want that the
condition of
Hermiticity/non-Hermiticity of $\widetilde{\cal H}$ precisely signals the
conservation/non-conservation
of the length of the Jacobi fields, we have to use always the SvH 
scalar product and not those canonically equivalent to it. So, for example, if
we start with the SvH 
scalar product and those phase space coordinates in which the trajectories are
ellipses, the $\widetilde{\cal H}$
will not be Hermitian. The same feature will be inherited by the
canonically-transformed $\widetilde{\cal H}$ 
in which the trajectories are circles but this feature (the non-Hermiticity)
will appear only if we use the canonically
transformed SvH scalar product. If we instead use the original SvH scalar
product only the $\widetilde{\cal H}$
of the circles in the original (unprimed) variables will be Hermitian.

To conclude this Appendix we can note that the relation (\ref{i-50-1}) can be
disrupted not only by
a canonical transformation but also by changing the system of units which we
use in measuring $\omega$ and $m$. But we should note that, if we change the
units, we have to change also the hermiticity 
conditions (\ref{i-49-3}) for
dimensional reasons.
Under the new hermiticity conditions $\widetilde{\cal H}$ is again Hermitian.
The reason why, by
changing the system of units, we have to change the hermiticity conditions
(\ref{i-49-3}) can be explained as follows.
Looking at the Lagrangian (\ref{uno-tredici}) we notice that the kinetic term of
the action is $\displaystyle \int
dt\,\bar{c}_q\dot{c}^q$ and so the dimension of $c^q$ is the inverse of
$\bar{c}_q$:
	\begin{equation}
	[c^q]=[\bar{c}_q]^{-1}. \label{i-52-1}
	\end{equation}
The same for the $c^p$:
	\begin{equation}
	[c^p]=[\bar{c}_p]^{-1}. \label{i-52-2}
	\end{equation}
From the interaction term $\displaystyle \int
dt\,\bar{c}_p\partial_q\partial_qHc^q$ of the action associated to the
$\widetilde{\cal L}$
of (\ref{uno-tredici}) we get that
	\begin{equation}
	\displaystyle [c^p]=\frac{M}{T}[c^q]. \label{i-52-3}
	\end{equation}
These are the dimensional relations among the Grassmann variables.
So the hermiticity conditions (\ref{i-49-3}) should be written as
	\begin{equation}
	(\hat{c}^q)^{\dagger}=\hat{\bar{c}}_q\cdot {\bf 1}_q,\;\;\;\;\;\;\;
	(\hat{c}^p)^{\dagger}=\hat{\bar{c}}_p\cdot 
	{\bf 1}_p
	\end{equation}
where ${\bf 1}_q, {\bf 1}_p$ are dimensionful quantities. Of course we could
choose $c^q$ to be dimensionless at all, and
so ${\bf 1}_q$ could be a number like 1. But if $c^q$ is dimensionless then
$c^p$, because of (\ref{i-52-3}), must have
dimension. ${\bf 1}_p$ would then have the dimension of $[c^p]^2$ and so
it is a number which changes with the system of units.
Let us for example choose the S.I. system of units and the hermiticity
conditions 
	\begin{equation}
	 \left\{
		\begin{array}{l}
		\displaystyle
		\label{i-53-1}
		(\hat{c}^q)^{\dagger}=\hat{\bar{c}}_q
	        \smallskip \nonumber\\
	        \displaystyle 
	        (\hat{c}^p)^{\dagger}=\hat{\bar{c}}_p.
	        \end{array}
		\right.
	\end{equation}
The harmonic oscillators which are Hermitian are those for which 
	\begin{equation}
	 \left\{
		\begin{array}{l}
		\displaystyle
		\label{i-53-2}
		m=\alpha\cdot Kg
	        \smallskip \nonumber\\
	        \displaystyle 
	        \omega=\frac{1}{\alpha} sec^{-1}.
	        \end{array}
		\right.
	\end{equation}
If we now pass to the CGS system the relation (\ref{i-53-2}) becomes 
	\begin{equation}
	 \left\{
		\begin{array}{l}
		\displaystyle
		\label{i-53-3}
		m=\alpha\cdot10^3 g
	        \smallskip \nonumber\\
	        \displaystyle 
	        \omega=\frac{1}{\alpha} sec^{-1}
	        \end{array}
		\right.
	\end{equation}
and $m\omega=10^3\neq 1$. But the hermiticity relations (\ref{i-53-1}), in the
new units, become
	\begin{equation}
	 \left\{
		\begin{array}{l}
		\displaystyle
		\label{i-53-4}
		(\hat{c}^q)^{\dagger}=\hat{\bar{c}}_q
	        \smallskip \nonumber\\
	        \displaystyle 
	        (\hat{c}^p)^{\dagger}=10^6\,\hat{\bar{c}}_p
	        \end{array}
		\right.
	\end{equation}
and with these new hermiticity conditions $\widetilde{\cal H}$ is Hermitian even
under the condition 
(\ref{i-53-3}).


\newpage
\noindent {\LARGE \underline{Figure Caption}}

\bigskip
\bigskip
\bigskip

\noindent {\large Figure  1}: Phase space trajectories for a harmonic oscillator
with $\displaystyle \frac{1}{\omega^2}=m^2$.\\

\noindent {\large Figure 2}: Phase space trajectories for a harmonic oscillator
with $\displaystyle \frac{1}{\omega^2}\neq m^2$.

\newpage

\begin{figure}
\centering
\includegraphics{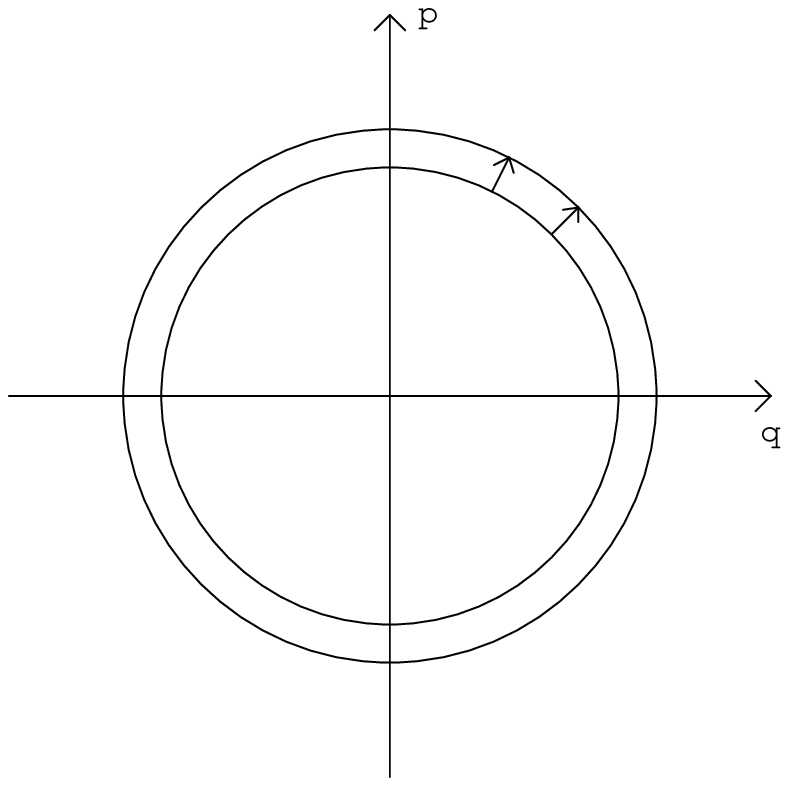}
\caption{Phase space trajectories for a harmonic oscillator with $\displaystyle
\frac{1}{\omega^2}=m^2$.} \label{circus}
\end{figure}

\bigskip
\bigskip
\bigskip
\bigskip

{\large Figure 1. Authors: E. Deotto, E. Gozzi and D. Mauro}\\
\newpage

\begin{figure}
\centering
\includegraphics{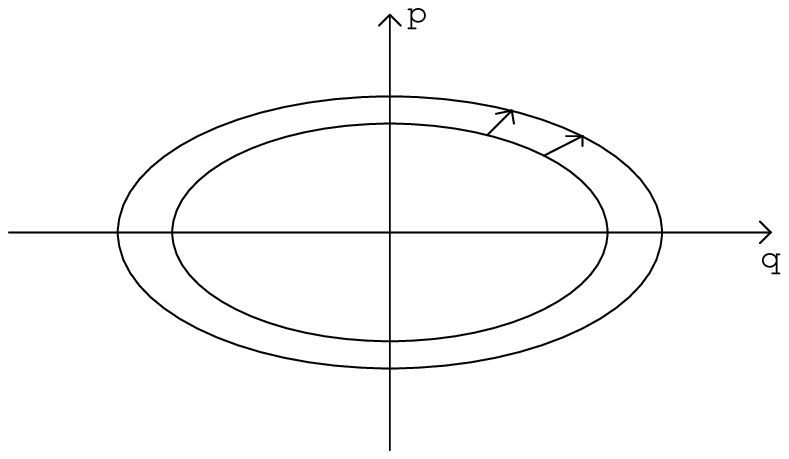}
\caption{Phase space trajectories for a harmonic oscillator with $\displaystyle
\frac{1}{\omega^2}\neq m^2$.}
\label{ellipse}
\end{figure}

\bigskip
\bigskip
\bigskip
\bigskip

{\large	 Figure 2. Authors: E. Deotto, E. Gozzi and D. Mauro}\\


\begin{thebibliography}{99}
\bibitem{footnote1}
$[\quad]_-$ and $[\quad]_+$ are respectively the commutator and the anticommutator.
\bibitem{footnote2}
Regarding dimensional problems which arise from the hermiticity conditions see Appendix F.
\bibitem{footnote3}
Properly speaking ours is not yet a Hilbert but just a vector space.
\bibitem{footnote5}
We call it ``configurational" even if we use phase
space variables because the $(q,p)$ are the basic variables in the CPI while
the $\lambda_{a}$ are the associated momenta.
\bibitem{footnote6}
In the formula below we have suppressed the index ``$a$" on
$(\varphi^a,\lambda_a,c^a,\bar{c}_a)$.
\bibitem{footnote7}
The proof that they are eigenstates 
of $\hat{c}$ and $\hat{\bar{c}}$ is the same as in the case of the SvH scalar
product.
\bibitem{footnote8}
We have used the variable
$c$ instead of $\alpha$ in the ket $|\alpha -\rangle$ to indicate that this last one is an eigenstate of $\hat{c}$
with eigenvalue $c$. The same we did for $|\alpha+\rangle$ replacing it with $|\bar{c} +\rangle$.
\bibitem{footnote8bis}
The reader may be puzzled by the ``$i$" which 
appears in front of the RHS of (\ref{tre-quarantadue-b}) and which did not appear in the SvH. 
The reason is that, because of the normalization we have chosen, the $K(f|i)$ in Eq. (\ref{tre-quarantuno})
is not ``${\bf 1}$" for $t_f=t_i$ and $f=i$ but it is equal to ``$i$".
\bibitem{footnote9}
For $\varphi$ and $\lambda$ we will stick to the standard hermiticity 
conditions (\ref{due-uno}).
\bibitem{footnote10}
The content of this section has been generated by many very helpful discussions with M. Reuter whom we wish to warmly 
thank for his generosity in sharing with us many of his ideas and insights on these topics.
\end{thebibliography}
\end{document}